\documentclass[amsmath,amssymb,aps,10pt,prd,letterpaper,nofootinbib,balancelastpage,notitlepage,superscriptaddress,twocolumn,floatfix]{revtex4-2}
\usepackage{graphicx}	
\usepackage{ragged2e}	
\usepackage{bm}		    
\usepackage[sort&compress]{natbib}	 	
	\setcitestyle{square,numbers,comma}	
\usepackage[colorlinks=true,urlcolor=blue,linkcolor=black,citecolor=red]{hyperref}	
\newcommand{\up}[1]{\textsuperscript{#1}}			
\newcommand{\tabref}[2][]{Tab{#1}.~\ref{#2}}		
\newcommand{\figref}[2][]{Fig{#1}.~\ref{#2}}		
\newcommand{\secref}[2][]{Sec{#1}.~\ref{#2}}		
\newcommand{\appref}[2][x]{Appendi{#1}~\ref{#2}}	
\renewcommand{\eqref}[2][]{Eq{#1}.~(\ref{eq:#2})}	
\newcommand{\eqrefRange}[2]{Eqs.~(\ref{eq:#1})--(\ref{eq:#2})}		
\newcommand{\citeR}[2][]{Ref{#1}.~\cite{#2}}		
\newcommand{\orcid}[1]{\href{https://orcid.org/#1}{\,\includegraphics[width=8px]{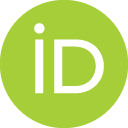}}}
\newcommand{\lb}{\ensuremath{\left}}					
\newcommand{\rb}{\ensuremath{\right}}					
\newcommand{\nl}{\nonumber \\ & \quad }					
\newcommand{\tsc}[1]{\textsc{#1}}
\newcommand{\LL}{\mathcal{L}}

\newcommand{\phihat}{\bm{\hat{\phi}}}
\newcommand{\thetahat}{\bm{\hat{\theta}}}
\newcommand{\rhat}{\bm{\hat{r}}}
\newcommand{\xhat}{\bm{\hat{x}}}
\newcommand{\yhat}{\bm{\hat{y}}}
\newcommand{\zhat}{\bm{\hat{z}}}
\newcommand{\Hz}{\,\textrm{Hz}}
\newcommand{\eV}{\,\textrm{eV}}

\newcommand{\kin}[1]{({#1}_{\textsc{k}})}
\newcommand{\primekin}[1]{({#1}_{\textsc{k}}^\prime)}
\newcommand{\mass}[1]{({#1}_{\textsc{m}})}
\newcommand{\primemass}[1]{({#1}_{\textsc{m}}^\prime)}
\newcommand{\inter}[1]{({#1}_{\textsc{i}})}
\newcommand{\primeinter}[1]{({#1}_{\textsc{i}}^\prime)}

\begin{document}

\title{Earth as a transducer for dark-photon dark-matter detection}
\date{\today}
\author{Michael A.~Fedderke\orcid{0000-0002-1319-1622}}
\email{mfedderke@jhu.edu}
\affiliation{Department of Physics and Astronomy, The Johns Hopkins University, Baltimore, MD 21218, USA}
\affiliation{Stanford Institute for Theoretical Physics, Department of Physics, Stanford University, Stanford, CA 94305, USA}
\author{Peter W.~Graham\orcid{0000-0002-1600-1601}}
\email{pwgraham@stanford.edu}
\affiliation{Stanford Institute for Theoretical Physics, Department of Physics, Stanford University, Stanford, CA 94305, USA}
\author{Derek F.~Jackson Kimball\orcid{0000-0003-2479-6034}}	
\email{derek.jacksonkimball@csueastbay.edu}
\affiliation{Department of Physics, California State University -- East Bay, Hayward, CA 94542, USA}
\author{Saarik Kalia\orcid{0000-0002-7362-6501}\,}	
\email{saarik@stanford.edu}
\affiliation{Stanford Institute for Theoretical Physics, Department of Physics, Stanford University, Stanford, CA 94305, USA}

\begin{abstract}
We propose the use of the Earth as a transducer for ultralight dark-matter detection.
In particular we point out a novel signal of kinetically mixed dark-photon dark matter: a monochromatic oscillating magnetic field generated at the surface of the Earth.
Similar to the signal in a laboratory experiment in a shielded box~(or cavity), this signal arises because the lower atmosphere is a low-conductivity air gap sandwiched between the highly conductive interior of the Earth below and ionosphere or interplanetary medium above.
At low masses~(frequencies) the signal in a laboratory detector is usually suppressed by the size of the detector multiplied by the dark-matter mass.
Crucially, in our case the suppression is by the radius of the Earth, and not by the~(much smaller) height of the atmosphere.
We compute the size and global vectorial pattern of our magnetic field signal, which enables sensitive searches for this signal using unshielded magnetometers dispersed over the surface of the Earth.
In principle, the signal we compute exists for any dark photon in the mass range $10^{-21}\,\text{eV}\lesssim m_{A'} \lesssim 3\times 10^{-14}\,\text{eV}$.
We summarize the results of our companion paper~\cite{Fedderke:2021qva}, in which we detail such a search using a publicly available dataset from the SuperMAG Collaboration: we report no robust signal candidates and so place constraints in the (more limited) dark-photon dark-matter mass range $2\times 10^{-18}\,\textrm{eV} \lesssim m_{A'} \lesssim 7\times 10^{-17}\,\textrm{eV}$ (corresponding to frequencies $6\times 10^{-4}\,\text{Hz}\lesssim f \lesssim 2\times 10^{-2}\,\text{Hz}$).
These constraints are complementary to existing astrophysical bounds.
Future searches for this signal may improve the sensitivity over a wide range of ultralight dark-matter candidates and masses.
\end{abstract}

\maketitle
\tableofcontents

\section{Introduction}
\label{sec:introduction}

The nature of dark matter remains one of the great mysteries in fundamental physics.
Myriad dark-matter candidates exist, spanning a wide range of allowed masses.
Excitingly, there has recently been significant progress in the exploration of more of this dark-matter parameter space.
In this work, we hone in on the `ultralight' portion of the allowed dark-matter mass range, and propose a new detection technique for the dark photon, a well-motivated dark-matter candidate.

The dark photon is a new $U(1)$ gauge boson coupled to the Standard Model (SM) through a kinetic mixing with the ordinary SM photon~\cite{Holdom:1985ag}.
Dark-photon dark matter~\cite{Nelson:2011sf} is generically produced from inflation~\cite{Graham:2015rva,Ahmed:2020fhc,Kolb:2020fwh} and can also be produced in other, model-dependent ways (e.g.,~\citeR[s]{Arias:2012az,Agrawal:2018vin,dror2019parametric,bastero2019vector,ema2019production,co2019dark,long2019dark,Nakai:2020cfw,nakayama2020gravitational,Salehian:2020asa,Bastero-Gil:2021wsf,co2021gravitational}).
Several new experimental approaches aiming at dark-photon detection have recently been developed, and there is significant ongoing interest in this field (see, e.g.,~\citeR[s]{Wagner:2010mi, Redondo:2010dp, Bahre:2013ywa, Graham:2014sha, Chaudhuri:2014dla, Phipps:2019cqy, TheMADMAXWorkingGroup:2016hpc, Baryakhtar:2018doz, Lawson:2019brd, Gelmini:2020kcu, Jaeckel:2013sqa,Horns:2012jf,Suzuki:2015sza,Andrianavalomahefa:2020ucg,Cantatore:2020obc,Su:2021jvk}).

Several existing direct detection experiments that are sensitive to the electromagnetic effects of dark-photon dark matter, such as ADMX~\cite{Wagner:2010mi} and DM Radio~\cite{Chaudhuri:2014dla}, employ a highly sensitive magnetometer with an electromagnetic resonator (e.g., a cavity or lumped-element circuit) inside a shielded region.
While these approaches are powerful, their sensitivity falls off at lower dark-photon masses because the signal size is parametrically suppressed by $\sim m_{A'} L$ where $m_{A'}$ is the dark-photon mass and $L$ is the characteristic linear size of the shielded region~\cite{Chaudhuri:2014dla}.
If the electromagnetically shielded region is $L \sim 1 \, \text{m}$, then the measurable signal decreases for masses $m_{A'} \lesssim 10^{-7} \, \text{eV}$ (frequencies below approximately 30\,MHz).%
\footnote{\label{ftnt:hbarc1}%
    Throughout this paper we work in natural units where $\hbar = c = 1$.
    The conversion to frequency from mass is $f = m_{A'}/(2\pi)$ in natural units; that is, $f \sim 24 \,\text{mHz} \times ( m_{A'} / 10^{-16}\,\text{eV})$.
} %

We propose a new way to detect dark-photon dark matter at much lower dark-photon masses.
In contrast to these existing detection approaches, we propose removing the human-made shield and using a sensitive magnetometer exposed to the ambient electromagnetic environment.
While at first it might appear that we have removed the significant $\sim m_{A'} L$ suppression, in fact an `unshielded' magnetometer on Earth is necessarily still surrounded by naturally occurring shields; indeed, these natural shields are essential in generating the signal we consider in this work.
Given the low-mass dark photons of interest to us, corresponding to signal frequencies $f \lesssim  10^{-2}\,$Hz (though still well above $\text{yr}^{-1}$), the Earth itself behaves as an excellent conductor, and acts to damp the interacting component of the photon--dark-photon system in exactly the same fashion as a shield.
Similarly, while the ionosphere surrounding the Earth may or may not (we will consider both cases) constitute a sufficiently thick layer of good conductor to qualify as a good natural shield, it is certainly true that the interplanetary medium permeating the Solar System beyond behaves as an almost collisionless plasma with a large plasma frequency and is amply thick to damp the interacting component of the photon--dark-photon system.
Interestingly though, it turns out that the lower few kilometers of the atmosphere are a marginal or poor conductor.
For the relevant frequencies then, we thus effectively have a naturally shielded, vacuum-like air-gap region near the surface of the Earth.
But the natural shields at play here have very large characteristic sizes, and we consequently expect enhanced sensitivity to lower-mass dark photons as compared to any conceivable experiment employing laboratory-scale magnetic shields.

In this paper, we calculate the signal of dark-photon dark matter that is expected in a magnetometer that is exposed to the ambient electrical environment near the surface of the Earth, by modeling the naturally shielded atmospheric `cavity' as bounded below by the conducting Earth and bounded above by either (a) the conducting ionosphere, or (b) the plasma of the interplanetary medium. 
In the former case (a), the cavity is a simple thin spherical shell sandwiched between two good conductors, and the computation proceeds straightforwardly: dark-photon dark matter can drive oscillating charge motion at the interfaces of the Earth and ionosphere with the air gap of the lower atmosphere, and these surface currents give rise to a leading-order magnetic field in the lower atmosphere.
In the latter case (b), there is a more complicated electrical environment between the natural shields, and we examine how our results from the former case are modified to more realistically account for the complexities of the electrical environment in the vicinity of the Earth.
Perhaps surprisingly, we find that the same leading-order magnetic field is expected in either case.
Importantly, our computation resolves a crucial question (see, e.g., \citeR{Dubovsky:2015cca}) regarding the length-scale $L$ that enters the geometrical $\sim m_{A'} L$ suppression: is it the height of the atmosphere $L \sim h \lesssim 10^2\,$km, or the radius of the Earth, $L\sim R\sim 6\times 10^3\,$km?
Perhaps counter-intuitively, we show that it is the larger radius of the Earth which enters the suppression factor, which is much more favorable for the signal.
This makes possible a sensitive search for low-mass dark-photon dark matter.

Of course, with a magnetometer lacking a human-made shield, we must ask whether ambient electromagnetic noise will swamp any possible signal, making a sensitive experiment impossible. 
Rather than trying to estimate all possible noise sources, we have instead carried out a full analysis of an existing dataset from a global network of unshielded, geographically dispersed, three-axis magnetometers that have been operating for decades for the purposes of geophysical metrology~\cite{SuperMAGwebsite,Gjerloev:2009wsd,Gjerloev:2012sdg}.
We present the results of this analysis in summary form in this work; technical details of the analysis are presented in a companion paper~\cite{Fedderke:2021qva}. 
As we report no robust signal candidates in this analysis, we present limits on the parameter space for dark-photon dark matter.
These limits augment existing astrophysical constraints applicable in this dark-photon dark-matter mass range that arise from bounds on gas heating in various environments (see,  e.g.,~\citeR[s]{Dubovsky:2015cca,Bhoonah:2018gjb,McDermott:2019lch,Wadekar:2019xnf,Kovetz:2018zes}).
Our search results, arising from significantly different measurements, are competitively complementary to these existing constraints.
Future searches for this signal hold promise to significantly expand the reach of this approach beyond existing astrophysical bounds, particularly at higher frequencies.

The rest of this paper is structured as follows.
In \secref{sec:problemStatement} we present an overview of the relevant physics of the kinetically mixed photon--dark-photon system, and a review of the electrical (conductivity) environment near the Earth.
\secref{sec:signal} describes our actual signal.
We begin \secref{sec:signal} with a simple toy example to illustrate the origin of the dark-photon dark-matter signal we propose to search for and highlight an important point regarding the geometrical suppression factor in a shielded region (\secref{sec:toyTheory}).
We follow on from this toy example by presenting our calculation of the dark-photon signal near the Earth under two different sets of assumptions about the damping thickness of the ionospheric conductivity layers near the top of the atmosphere: first assuming that the ionosphere is an effective shield (\secref{sec:earthTheory}), and then assuming it is not (\secref{sec:generalEarthTheory}).
The results of the experimental analysis that is detailed in our companion paper~\cite{Fedderke:2021qva} are presented in summary form in \secref{sec:experimentalSearch}.
We conclude in \secref{sec:conclusion}.
We present supplemental material in a number of appendices: \appref{app:systemBehavior} gives an in-depth review of the dynamics of the photon--dark-photon system; 
\appref{app:finiteConductivty} gives a treatment of our signal without assuming infinite-conductivity (or infinite plasma frequency) boundary conditions near the Earth;
\appref{app:fullCoeffModel1} gives the full forms of some lengthy expressions whose limiting forms we present in the main text;
and \appref{app:vectorSphericalHarmonics} gives our conventions for the vector spherical harmonics.

\section{Preliminaries}
\label{sec:problemStatement}

The behavior of the kinetically mixed photon--dark-photon system in the vicinity of an ordinary electromagnetically (EM) conducting interface exhibits a rich and non-trivial phenomenology.
We consider the case of dark-photon dark matter, for which there exists a background, non-relativistic dark-photon field. 
Observable electromagnetic effects are generated by this background field~\cite{Graham:2014sha,Chaudhuri:2014dla}.
Most physically, one can think about these effects as arising due to ordinary electric charges acquiring an effective millicharge under the dark $U(1)$ gauge group (in the so-called `mass basis').
The action of the dark-photon field then causes surface currents to be driven at a conducting interface, and those currents in turn source observable electromagnetic fields on the non-conducting side of the interface.

More abstractly, one can consider the observable fields to arise from an abrupt change at the conductor--vacuum interface in the relationship between the vacuum propagation eigenstates and the interaction eigenstates of the mixed photon--dark-photon system; this abrupt change gives rise to neutrino-oscillation-like phenomena in the photon--dark-photon system on the non-conducting side of the interface that lead to the generation of an interacting component of the photon--dark-photon system away from the interface.

In this section, we will first give a short qualitative theory review of the behavior of the photon--dark-photon system, both in vacuum and in conductors or plasmas, and discuss implications for phenomenology; we defer technical details and derivations to \appref{app:systemBehavior}.
With this theoretical motivation in place, we will then review the electromagnetic environment near the surface of the Earth in order to demonstrate that, in some range of frequencies, the lower atmosphere constitutes precisely the kind of environment in which we expect the generation of observable EM signals due to the effects noted above.
Specifically, we discuss how the lower atmosphere constitutes a low-conductivity gap sandwiched between two layers in which the active mode of the photon--dark-photon system is effectively damped: (1) the ground, which acts as a good conductor; and (2) either (a) the ionosphere, which as a relatively thin conductive layer may or may not be thick enough to damp the interacting mode, or (b) the interplanetary medium beyond, which acts as a collisionless plasma with a high plasma frequency and which is amply thick enough to achieve the necessary damping.

\subsection{Overview of photon--dark-photon phenomenology}
\label{sec:phenomenologyOverview}

In this work, we consider a massive dark photon $\primekin{A}$ kinetically mixed with the SM $U(1)$ photon $\kin{A}$, described by the Lagrangian
\begin{align}
    \LL &\supset - \frac{1}{4} \kin{F}_{\mu\nu}\kin{F}^{\mu\nu} - \frac{1}{4} \primekin{F}_{\mu\nu} \primekin{F}^{\mu\nu} \nl
    + \frac{\varepsilon}{2} \kin{F}_{\mu\nu} \primekin{F}^{\mu\nu} + \frac{1}{2} m_{A'}^2 \primekin{A}_\mu \primekin{A}^{\mu} \nl
    - J_{\textsc{em}}^\mu \kin{A}_\mu. \qquad \text{\footnotesize[kinetically mixed basis]}
    \label{eq:kineticallyMixedLagrangian}
\end{align}
Here, $(F_{\textsc{k}}^{(\prime)})_{\mu\nu} \equiv \partial_\mu (A_{\textsc{k}}^{(\prime)})_\nu - \partial_\nu (A_{\textsc{k}}^{(\prime)})_\mu$ is the field strength tensor for the ordinary (respectively, dark) photon, and $J_{\textsc{em}}^\mu$ is the usual SM $U(1)$ electromagnetic current.
We assume that the kinetic mixing parameter $\varepsilon$ is \linebreak small: $\varepsilon\ll1$.

While this basis is convenient for making explicit the `vector portal' nature of the mixing (see, e.g., \citeR[s]{graham2016dark,safronova2018search} for discussion of `portal' phenomenology), it is more convenient for our purposes to perform a field re-definition and work in the so-called interaction basis; see \appref{app:systemBehavior} for a detailed discussion of basis choices and the relationships between various choices.
Making the substitutions $A_{\tsc{k}} \rightarrow A_{\tsc{i}}$ and $A'_{\tsc{k}} \rightarrow A'_{\tsc{i}} + \varepsilon A_{\tsc{i}}$ in \eqref{kineticallyMixedLagrangian}, followed by dropping terms at $\mathcal{O}(\varepsilon^2)$, we have%
\footnote{\label{ftnt:notationChange}%
    For the moment, we will keep the subscript-$\tsc{i}$ on the interaction basis states.
    Later, in \secref{sec:signal}, where we work solely in the interaction basis to make field computations, we will drop these subscripts and identify $A_{\tsc{i}}\equiv A$, and $A_{\tsc{i}}' \equiv A'$ for notational simplicity.
} %
\begin{align}
    \LL &\supset - \frac{1}{4} \inter{F}_{\mu\nu}\inter{F}^{\mu\nu} - \frac{1}{4} \primeinter{F}_{\mu\nu} \primeinter{F}^{\mu\nu}\nl
    + \frac{1}{2} m_{A'}^2 \primeinter{A}_\mu \primeinter{A}^{\mu} 
    + \varepsilon m_{A'}^2 \primeinter{A}^{\mu} \inter{A}_\mu \nl
    - J_{\textsc{em}}^\mu \inter{A}_\mu. \qquad \text{\footnotesize[interaction basis, } \mathcal{O}(\varepsilon)\text{]}
    \label{eq:interactionBasisLagrangianLinear}
\end{align}

It is apparent from \eqref{interactionBasisLagrangianLinear} that, of the interaction-basis states, only the `interacting state' $A_{\tsc{i}}$ (sometimes also called the `active mode') couples to EM charges; the `sterile state' $A'_{\tsc{i}}$ does not. 
On the other hand, the presence of the mass-mixing term $m_{A'}^2 A_{\tsc{i}}A_{\tsc{i}}'$ makes clear that these interaction eigenstates are not the propagation (momentum) eigenstates in vacuum (we denote these $A_{1,2}$).
In particular, in vacuum, these are related at $\mathcal{O}(\varepsilon)$ by
\begin{align}
    \begin{pmatrix} A_1 \\ A_2 \end{pmatrix}
    &= \begin{pmatrix}
            1 & -\varepsilon \\
            +\varepsilon & 1
        \end{pmatrix}
        \begin{pmatrix} A_{\textsc{i}} \\ A_{\textsc{i}}' \end{pmatrix};
   \qquad \text{\footnotesize[vacuum, } \mathcal{O}(\varepsilon)\text{]}
    \label{eq:vacuumPropagationRelation}
\end{align}
see the detailed discussion in \appref{app:systemBehavior}.

This mismatch of the interaction and propagation eigenstates and, in particular, the changes in the relationships between those eigenstates that occur as one moves from one medium to another can give rise to phenomena directly analogous~\cite{Graham:2014sha} to neutrino oscillations.
For example, suppose that a field configuration is such that, on some physical boundary $\Sigma$ to a region of vacuum, we have that (1) the interacting field component is vanishing, $A_{\tsc{i}}|_{\Sigma} =0$ (as happens, e.g., at a conductive interface).
Suppose also that (2) the sterile field component is non-vanishing, $A'_{\tsc{i}}|_{\Sigma}\neq0$.
Because of the misalignment of the interaction and momentum eigenbases, these conditions require that the momentum eigenstates $A_{1,2}$ have a certain fixed relationship with each other: under the assumptions here, (1) imposes that $(A_1 + \varepsilon A_2)|_{\Sigma} \approx 0$, while (2) imposes $(-\varepsilon A_1 + A_2)|_{\Sigma} \approx A'_{\tsc{i}}|_{\Sigma}$.
At leading order then, $A_2|_{\Sigma} \approx A'_{\tsc{i}}|_{\Sigma}$, while $A_1|_{\Sigma} \approx -\varepsilon A'_{\tsc{i}}|_{\Sigma}$.
Now consider a test charge located at $\bm{x}$, within the vacuum region; it will be sensitive to the local interacting field component $A_{\tsc{i}}(\bm{x}) \approx A_1(\bm{x}) + \varepsilon A_2(\bm{x})$.
If $\bm{x}$ were located on the surface $\Sigma$, then condition (1) would by construction cause this linear combination to vanish and the test charge would experience no effect.
However, because the momentum eigenstates $A_{1,2}$ have different momentum eigenvalues $k_{1,2}$ and hence different phase evolution $\sim \exp\lb[ i\bm{k}_i\cdot \bm{x} \rb]$ under translations, we generically have $A_1(\bm{x}) \not\approx -\varepsilon A_2(\bm{x})$ when $\bm{x}$ is not on the surface $\Sigma$.
As such, the interacting field component $A_{\tsc{i}}(\bm{x})$ will not vanish at $\bm{x}$ away from the boundary $\Sigma$, and a test charge at $\bm{x}$ will thus be accelerated.
In other words, in this language, simple vacuum propagation of this coupled system causes a measurable interacting field that is constrained to be zero on some boundary, to be re-generated some distance away from the boundary.

The interface between vacuum and a good conductor supplies a natural location for the interacting component $A_{\tsc{i}}$ of the field to vanish.
As we discuss in detail in \appref{app:systemBehavior}, in a good conductor (conductivity $\sigma \gg m_{A'}^2/\omega$ where $\omega$ is the angular frequency of interest for the field oscillation), the large self-energy for the interacting mode $A_{\tsc{i}}$ leads to a close alignment of the interaction and momentum eigenstates in the conductor. 
Moreover, in this limit, the interacting component $A_{\tsc{i}}$ rapidly decays on the skin-depth length-scale $\delta \sim \sqrt{2/(\omega \sigma)} \ll m_{A'}^{-1}$ (see, e.g., \tabref{tab:earthConductivity} and \secref{sec:conductivityNearEarth}).
On the other hand, the sterile state $A_{\tsc{i}}'$ has the dispersion relation $k^2 \approx \omega^2 - m_{A'}^2$ up to highly suppressed corrections: it behaves as a particle with a mass $m_{A'}$ and is barely impacted by the medium at all.
Therefore, deep in a conducting medium (i.e., any more than a few skin-depths from any interface), any non-zero field configuration must be purely in the sterile state $A_{\tsc{i}}'$, which is itself essentially unaffected by the presence of the medium.

If we specialize to the case of dark-photon dark matter, then in order to match astrophysical and cosmological observations, the dark photons must be non-relativistic (i.e., $\omega\approx m_{A'}$).
Consider a region of space characterized by a high conductivity, $\sigma \gg m_{A'}$, and assume that this region in space is large compared to the skin-depth $\delta$ for the interacting state in the conductor. 
To excellent approximation, the dark-matter field in that region will then be purely in the sterile state, with no interacting admixture.

Specifically, the dark-photon dark-matter field in the vicinity of the Earth is a coherently oscillating vector field with a random initial polarization state, which can be written as (the real part of)
\begin{align}
    \bm{A}'_\textsc{I}(\bm{x},t) \approx \frac{\sqrt{2\rho_{\textsc{dm}}}}{m_{A'}} e^{-im_{A'}t} \times\sum_{i=1}^3  \xi_i(\bm{x},t) \bm{\hat{n}_i} e^{i\phi_i(\bm{x},t)},
    \label{eq:dmvec}
\end{align}
where $\bm{\hat{n}_i}$ ($i=1,2,3$) are a set of orthonormal Cartesian basis vectors fixed in an inertial frame.
The~$\xi_i(\bm{x},t)$ are $\mathcal{O}(1)$ functions and the $\phi_i(\bm{x},t)$ are phases; together these fix the dark-photon polarization state.
Compared to the leading $e^{-im_{A'}t}$ phase evolution, the functions $\xi_i$ and $\phi_i$ all vary slowly, on length-scales $L_{\text{coh}} \sim \lambda_{\text{de Broglie}} \sim 2\pi/(m_{A'}v_{\textsc{dm}})$ and timescales $T_{\text{coh}}\sim L_{\text{coh}}/v_{\textsc{dm}} \sim 2\pi/(m_{A'}v_{\textsc{dm}}^2)$, owing to the dispersion of DM velocities $v_{\textsc{dm}}\sim 10^{-3}$ in the Milky Way.%
\footnote{\label{ftnt:coherence}%
    Indeed, one can arrive at \eqref{dmvec} by integrating a set of plane waves with random phase offsets and vectorial orientations, and phase evolution governed by $\exp\lb[ -i \lb( \omega t - \bm{k}\cdot\bm{x} \rb) \rb]$ where $\omega = m_{A'} \sqrt{ 1  + \bm{v}^2 }$ and $\bm{k} = m_{A'}\bm{v}$, over the standard galactic-rest-frame Maxwell--Boltzmann DM velocity distribution for $\bm{v}$.
} %
Note that the dark-photon polarization state~$\propto \sum_i \xi_i \bm{\hat{n}_i}e^{i\phi_i}$ is generally elliptical: the field is not generally simply oscillating back and forth along a real 3-vector direction with its magnitude passing back and forth through zero.
Moreover, at leading order, the direction of the DM velocity vector (including any net DM wind) is not relevant for setting the polarization state of the DM field (i.e., the vectorial orientation of $\bm{A}_{\textsc{i}}'$).

If there is a cavity hollowed out within the high-conductivity region of space mentioned above, such that in the cavity we have $\sigma \ll m_{A'}$, then the interfaces between the cavity and the conducting material will, of course, be surfaces on which the interacting state must vanish, while the sterile state simply takes the same non-zero value at the interface that it does just inside the conductor.
This setup is precisely that required to give rise to the oscillation phenomenon discussed above, and an interacting, detectable field will be generated inside the cavity~\cite{Graham:2014sha} (similar observations in the context of light-shining-through-walls experiments appear in \citeR[s]{Okun:1982xi,Ahlers:2007rd,Jaeckel:2007ch}).
Specifically, in the limit where the geometrical dimension $R$ of the cavity, as measured transverse to the axis on which the polarization vector of the sterile field oscillates, satisfies the condition $m_{A'}R \ll 1$, it can be shown that the dominant field generated within the cavity is an oscillating magnetic field with a magnitude $B \sim \varepsilon (m_{A'}R) \sqrt{\rho_{\textsc{dm}}}$~\cite{Chaudhuri:2014dla} near the walls of the cavity.
This is the origin of the signal being searched for by, e.g., the DM Radio experiment~\cite{Chaudhuri:2014dla,Silva-Feaver:2016qhh,Phipps:2019cqy}. 

Because the only role played in the preceding few paragraphs by the conductive medium was to supply boundary conditions for the interacting state in the cavity (or, more physically, to supply charges that could be accelerated to generate surface currents that allow the net parallel electric field to be canceled exactly at the cavity walls), similar conclusions also naturally apply to the case where the medium surrounding the cavity is instead a nearly collisionless plasma with a high plasma frequency; i.e., $\omega_p \gg \omega \sim m_{A'} \gg \nu$, where $\omega_p$ and $\nu$ are the plasma and collision frequencies, respectively.
For this case, the approximate replacement rule in the discussion above about the active mode damping length is $\sigma \rightarrow \omega_p^2 / \omega$;
see \appref{app:systemBehavior} for detailed discussion.

In this work, we will apply these observations to a natural physical system of experimental interest: the Earth.
As it turns out, the Earth acts as an excellent conductor in the dark-photon mass range of interest to us.
On the other hand, in the same mass range, the conductivity of the lower atmosphere is poor. 
Above the lower atmosphere, the electrical environment near the Earth is complicated, with possible effects from both the ionosphere/magnetosphere (which may act as a layer of good conductivity, but also may not) and the interplanetary medium beyond the Earth's magnetosphere (which acts as a collisionless plasma with a high plasma frequency).
In either scenario, however, the lower atmosphere constitutes a large `vacuum' gap%
\footnote{\label{ftnt:dielectricEffectsOK}%
    Dielectric effects of the atmospheric medium do not spoil this, as such effects would enter only via the relative permittivity, which is approximately unity.
} %
sandwiched between two media which efficiently damp the active component of the photon--dark-photon system.
Thus, we expect to find an observable magnetic field in the gap. 
A similar observation that such a signal may exist was made briefly in \citeR{Dubovsky:2015cca}, but we disagree with the brief comments made therein on the size of the possible suppression of the effect.
However, before turning to the computation of the size of the expected signal (see \secref{sec:signal}), we first discuss the electrical environment near the Earth in more detail.

\subsection{Electrical environment near the Earth}
\label{sec:conductivityNearEarth}

\begin{figure*}[t]
\includegraphics[width=0.85\textwidth]{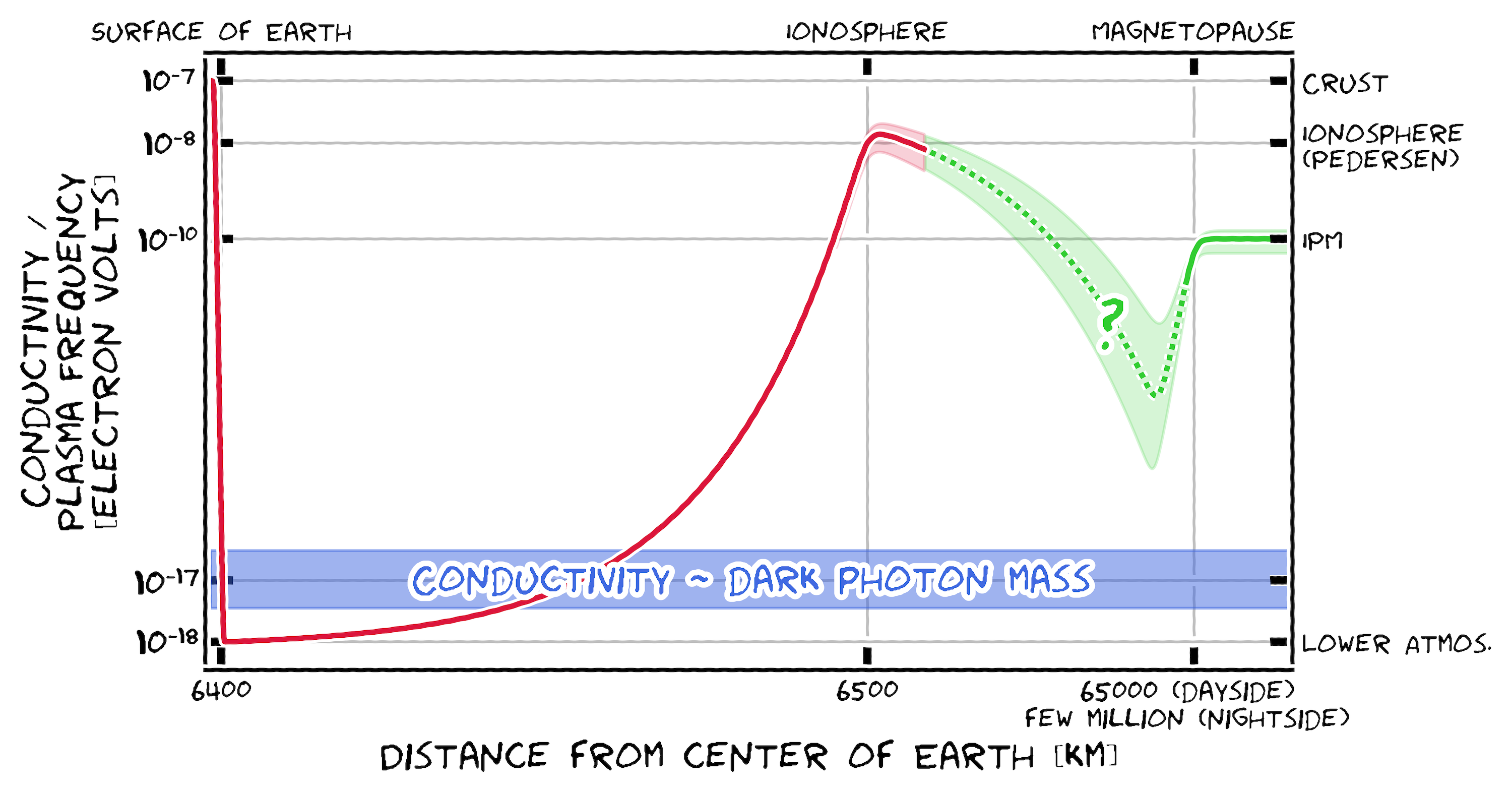}
\caption{\label{fig:conductivityProfile}%
    Sketch of the conductivity (or plasma frequency, where applicable) profile in the vicinity of the Earth's surface in (1) the case where the ionosphere is thick enough to damp active photon modes [solid red], and (2) the case where it is not and we must consider the electrical environment out to the interplanetary medium (IPM) [dotted and solid green].
    In the former case, we terminate the sketch some distance into the ionosphere, as the details above that altitude are irrelevant.
    In the latter case, we note that the conductivity or plasma frequency in the magnetospheric region (i.e., between the ionosphere and the magnetopause) can be quite complicated, and we have indicated this by dotting the green line, and including the shaded band and the question mark: we show in \secref{sec:generalEarthTheory} that, under reasonable assumptions, the details of the profile here do not matter for our signal.
    Note that the magnetopause has an aspherical tear-drop shape which is much larger in spatial extent in the downstream direction of the solar wind (i.e., the nightside of the Earth) than in the upstream direction (dayside).
    The vertical extent of the blue band indicates the range of dark-photon masses $m_{A'}$ we consider, in correct relation to the relative sizes of (a) the conductivity in the lower atmosphere, (b) the conductivity in the Earth's crust/upper part of the mantle (representative value labeled on the plot), (c) the (Pedersen) conductivity of the ionosphere, and (d) the plasma frequency of the interplanetary medium beyond.
    While we stress that this plot is highly schematic, it correctly captures that, in the upper atmosphere (interplanetary medium) and the Earth's crust, the conductivity (plasma frequency) exceeds the dark-photon mass, while the opposite is true in the lower atmosphere.
    Representative numerical values shown are very rough, and the reader is referred to the text of \secref{sec:conductivityNearEarth} for detailed discussion and caveats associated with this sketch.
    } 
\end{figure*}

In this subsection, we discuss in detail the electrical conductivity environment in the vicinity of the Earth's surface. 
The main purpose of this discussion is to establish that, for an interesting range of dark-photon masses, this environment approximates a poor conductivity gap---the lower atmosphere---sandwiched between two layers which effectively damp any active component of the mixed photon--dark-photon system: (1) at the inner edge of the gap, the interior of the Earth; and (2) at the outer edge of the gap, either (a) the interplanetary medium, or (b) the ionosphere.
Readers who are interested mainly in the conclusions of this section can refer to \figref{fig:conductivityProfile} for a rough sketch of the conductivity profile near the Earth's surface and continue to \secref{sec:conductivitySummary} for a brief summary of the discussion.

Although the signal we find in this work would in principle be present for any dark photon in a wide range of masses $10^{-21}\,\text{eV} \lesssim m_{A'} \lesssim 3\times 10^{-14}\,\text{eV}$ (see \secref{sec:conductivitySummary}), by way of calibration for the present discussion, the range of dark-photon masses of practical interest in this work (see \secref{sec:experimentalSearch}) and \citeR{Fedderke:2021qva} is $2\times 10^{-18}~\textrm{eV} \lesssim m_{A'} \lesssim 7\times 10^{-17}~\textrm{eV}$, corresponding to oscillation frequencies $6\times 10^{-4}~\textrm{Hz} \lesssim f_{A'} \lesssim 2\times10^{-2}~\textrm{Hz}$ and Compton wavelengths $8\times10^4 \gtrsim \lambda_{A'}/R \gtrsim 3\times10^3$, where $R$ is the Earth radius (the dark-photon de Broglie wavelengths are $10^3$ times larger since $v_{\textsc{dm}}\sim10^{-3}$).

\subsubsection{Surface and interior of the Earth}
\label{sec:conductivityEarthInterior}
At the frequencies of interest to our work, it is common practice (see, e.g., \citeR[s]{Simoes:2012asf,Jackson}) to approximate the Earth as a highly conductive spherical ball. 
This can easily be justified by examining the representative conductivities for the various layers of the surface and interior of the Earth which are shown in \tabref{tab:earthConductivity}, along with the corresponding skin-depths for the interacting mode of the photon--dark-photon system (see \appref{app:systemBehavior}).

\begin{table*}[t]
\begin{ruledtabular}
\caption{\label{tab:earthConductivity}%
	  Representative values for the conductivity of various parts of the bulk of the Earth.
	  We give a description, approximate depth below Earth's surface, reference conductivity $\sigma$ (or range of conductivities) in both SI and natural units,%
        \footnote{\label{ftnt:conductivityUnits}%
            Recall: $1\,\text{S/m} \equiv 1/(\Omega\text{m}) \approx 7.4\times 10^{-5}\,\text{eV}\approx 10^{11}\,\text{s}^{-1}$. 
            In older literature, units of `e.m.u.' are sometimes used: $1\,\text{S/m}=10^{-11}\,$e.m.u.
            } %
      active-mode skin-depth $\delta \sim (\sigma \omega / 2)^{-1/2}$ (see \appref{app:systemBehavior}) for $\omega \sim \omega_\star \equiv 10^{-18}\,$eV given the reference conductivity (or range), and references for the conductivity values quoted.
      The specific numbers quoted here are less important than the following general conclusion: the active-mode skin-depths for the lower mantle and deeper layers are all some orders of magnitude smaller than the thicknesses of those layers, making the Earth an excellent conductor that damps the interacting component efficiently at a radius that is $\mathcal{O}(1)$ of the full radius of the Earth.
	 }
\begin{tabular}{llllll}
Description & Depth [km] & $\sigma$ [S/m] &  $\sigma$ [eV] & $\delta(\omega_\star)$ [km] & Ref(s). \\ \hline
Surface/crust ($f\lesssim 30\,$kHz)\footnote{Conductivity varies by geographical location (local ground composition)~\cite{atlas}.} & 0--30 & $10^{-4}$--$10^{-2}$ & $7\times 10^{-9}$--$7\times 10^{-7}$ & 3200--320 & \cite{atlas}\\
Oceans ($f\lesssim 30\,$kHz)          & 0--10 & $\sim 4$ & $\sim 3\times 10^{-4}$ & $\sim 16$ & \cite{atlas} \\
Upper mantle    & 30--500    &   $\sim 10^{-2}$ & $\sim 7\times 10^{-7}$ & 320 & \cite{1970QJRAS..11...23P,Hutton_1976} \\
Lower mantle (upper)    & 500--1000 &  $1$--$10$ & $7\times 10^{-4}\,$--$7\times 10^{-3}$ & 30--10 & \cite{1970QJRAS..11...23P,Hutton_1976} \\
Lower mantle (core--mantle boundary) & $\sim 2900$ & $\sim 10^2$ & $\sim 7\times 10^{-3}$ & $\sim 3$ & \cite{Peyronneau1989} \\
Outer core\footnote{Conductivity inferred from inner core values and comments in \citeR{Pozzo_2012}.} & $2900$--$5200$ & $(1.2$--$1.3)\times 10^6$    & $\sim 90$--$95$   & $3.0\times10^{-2}$ & \cite{Pozzo_2012}\\
Inner core & $5200$--$6400$ & $(1.5$--$1.6)\times 10^6$    & $110-120$   & $2.5\times10^{-2}$ & \cite{Pozzo_2012}
\end{tabular}
\end{ruledtabular}
\end{table*}

The low-frequency ($f\lesssim 30\,$kHz) conductivity of the crust of the Earth exhibits fairly large local fluctuations near the surface owing to the presence of oceans and varying solid ground composition~\cite{atlas}, and is typically insufficiently thick to be damping for an active mode with $\omega \sim 10^{-18}$\,eV.
However, the crust is only a few tens of kilometers thick.
The mostly molten layer immediately below the crust, the upper mantle, can be approximated as a bulk layer with an approximately uniform, isotropic conductivity, and is already thick enough to be moderately damping for the interacting component of the photon--dark-photon system even for $\omega \sim 10^{-18}\,$eV.
By the depth of the lower mantle, roughly $500$--$1000$\,km below the surface, the conductivity has increased sufficiently that the lower mantle, and outer and inner cores are all some orders of magnitude thicker than the active-mode skin-depth for $\omega \gtrsim 10^{-18}\,$eV (and down to lower frequency modes as well, although this is not relevant for our work).

Given these observations, a highly conductive spherical ball model for the Earth, with the thickness and conductivity required to completely damp any interacting component of the photon--dark-photon system, is justified down to at least $\omega \sim 10^{-18}\,$eV (and onward to much lower frequencies too).
It is also clear that the radius of the spherical ball at which the interacting component can be assumed to be completely damped is in the worst case only $\sim (\text{few})\times 10^{2}\,$km--$10^3$\,km below the surface; given that the Earth radius is $\sim 6.4 \times 10^3\,$km, this sufficiently highly conductive ball has a radius that is $\mathcal{O}(1)$ of that of the Earth.
We will show in \secref{sec:generalEarthTheory}, however, that the exact assumed radius of this conductive ball will not affect our leading order result: regardless of the details of the conductivity profile of the Earth, the relevant length-scale that will appear in our magnetic field signal will be the radius at which the magnetic field is measured, which will be the radius of the Earth $R$.

\subsubsection{Lower atmosphere}

As might be expected, however, the gaseous lower atmosphere presents a vastly different electrical conductivity environment as compared to the solid/liquid environments found in the interior of the Earth.
The low free-charge densities (near the ground, induced mainly by ground radioactivity and radioactive gases~\cite{Handbook20.1}) and short collision lengths in the lower atmosphere guarantee that the lower few kilometers of the atmosphere are a fairly poor conductor: typical values of conductivity just above the ground are in the range 
$\sigma(h=0) \sim 1$--$3\times 10^{-14}\,\text{S/m}\sim 7\times 10^{-19}$--$2\times 10^{-18}$\,eV~\cite{Handbook20.1}.%
\footnote{\label{ftnt:conductivity}%
    We note that the upper end of this range lies near the lower end of our range for $m_{A'}$.  
    We demonstrate in \appref{app:finiteConductivty} that our calculation is still valid even when $\sigma\sim m_{A'}$ in the lower atmosphere.
} %
With increasing altitude $h$, the density of neutral atoms falls (leading to longer collision times) and the (cosmogenic) small-ion charge density increases~\cite{Handbook20.1}, which both act to cause the conductivity to then rise exponentially: $\sigma(h) \approx \sigma(h=0) e^{h/h_\sigma}$, with the scale height $h_\sigma \sim 5$--$6\,$km~\cite{Handbook20.1}; this expression is valid until $h\sim 60$--$90$\,km.

Because the delineation between good and poor conductor behavior for the photon--dark-photon system is (see \appref{app:systemBehavior}) $\sigma_{\star} \sim m_{A'}^2/\omega \sim m_{A'}$ for $\omega \sim m_{A'}$, there is thus a thin layer in the lower atmosphere---from just above the ground, to a few to perhaps tens of kilometers of altitude---that acts as a relatively poor conductor for the photon--dark-photon system with $\omega \sim m_{A'} \gtrsim (\text{few}) \times 10^{-18}\,$eV. 
Moreover, the active-mode damping length, $\delta$, in the lower atmosphere is enormous compared to either the thickness of the atmospheric layer or size of the Earth: 
\begin{align}
    \delta &\sim \sqrt{\frac{2}{m_{A'}\sigma}} \nonumber \\
    &\sim 1.3\,\text{AU}\times \sqrt{\frac{10^{-18}\,\text{eV}}{m_{A'}}} \times \sqrt{ \frac{ 3\times 10^{-14}\,\text{S/m}}{\sigma }}.
    \label{eq:dampAtmos}
\end{align} 

We generally assume a homogeneous, time-invariant lower-atmospheric conductivity. 
Weather phenomena would, of course, cause largely stochastic, short-lived, and relatively local (even for the largest storm systems) fluctuations to the lower atmospheric conductivity environment, for instance via rain, clouds, and/or lightning.
For example, there can be an associated increase in the (mostly negative polar) electricity conductivity of the atmosphere during certain types of heavy rainfall, but the effect appears to be at the level of a factor-of-3 increase~\cite{atmos11111195}; in view of the damping length estimate at \eqref{dampAtmos}, this would not dramatically alter whether any unshielded magnetometer station would, for instance, end up being effectively shielded during a heavy rainfall.
Moreover, the signal we will derive in \secref[s]{sec:earthTheory}~and~\ref{sec:generalEarthTheory} is a long-lived, global magnetic field signal.
As such, while weather-related atmospheric conductivity phenomena would undoubtedly cause additional local signal features, we do not expect them to be able to significantly modify the leading global signal that we report.
See also further discussion, including comments on the Schumann Resonances and Ionospheric Alfv\'en Resonator, in \secref{sec:generalEarthTheory}.

\subsubsection{Ionosphere}
\label{sec:ionosphere}

The simple conductivity model for the atmosphere mentioned in the previous subsection begins to break down at the ionospheric layers that surround the Earth, at altitudes $\sim (0.9$--$\text{few})\times 10^2$\,km. 
The ionosphere is a complicated, multi-layer, anisotropic conductive medium whose properties depend sensitively on altitude and geographical location, and which exhibits both daily and longer-period modulation~\cite{Simoes:2012asf,Takeda:1985hcf,GM118}.

The highest concentration of ionized electrons and ions that are found to occur in the ionospheric layers, $n\sim 10^{6}\,\text{cm}^{-3}$ under optimal conditions, would in principle be sufficient to support isotropic conductivities on the order of $\sigma \sim 10^{1}$--$10^{2}\,$\,S/m~\cite{Takeda:1985hcf,GM118}.
However, charge motion in the ionosphere is subject to non-negligible effects of the Earth's magnetic field $\bm{B}_{\oplus}$, and this significantly modifies the conductivity properties of the medium, particularly in directions perpendicular to the magnetic field lines~\cite{Takeda:1985hcf,GM118}.

The `parallel conductivity' (i.e., that which applies for charge motion in response to an electric field applied along the direction of $\bm{B}_{\oplus}$ field lines) is effectively the same as the isotropic conductivity one would obtain absent the $\bm{B}_{\oplus}$ field: it rises to $\sigma_{\shortparallel} \sim 1\,\text{S/m} \sim 7\times 10^{-5}$\,eV by an altitude of 120\,km (ionospheric E layer), and continues to rise as high as $\sigma_\shortparallel \sim 10^2\,\text{S/m}\sim 7\times 10^{-3}$\,eV at an altitude of $\sim 300\,$km (F layer)~\cite{Takeda:1985hcf,GM118}.
Moreover, the high altitude (upper F layer) parallel conductivity varies temporally by less than an order of magnitude over daily or solar cycle periods, and remains in the $\sigma_{\shortparallel} \sim 1$--$10$\,S/m range~\cite{Takeda:1985hcf}.
If this were an isotropic conductivity, the associated characteristic active-mode skin-depth
\begin{align}
    \delta &
    \sim \sqrt{\frac{2}{m_{A'}\sigma}}
    \sim 2\,\text{km}\times \sqrt{\frac{10^{-18}\,\text{eV}}{m_{A'}}} \times \sqrt{\frac{10^{2}\,\text{S/m}}{\sigma}}
\end{align}
would easily be short enough to completely damp the interacting mode within the ionosphere.

However, the conductivity relevant for charge motion in the direction of an electric field applied perpendicular to the $\bm{B}_{\oplus}$ field lines, the so-called Pedersen conductivity $\sigma_{\textsc{p}}$~\cite{Takeda:1985hcf,GM118}, behaves very differently from the parallel conductivity.%
\footnote{\label{ftnt:Hall}%
    There is also a third conductivity, the Hall conductivity $\sigma_{\textsc{h}}$, which characterizes charge motion perpendicular to both applied electric field and $\bm{B}_{\oplus}$.
    Qualitatively, the Hall conductivity behaves broadly similarly to the Pedersen conductivity: they have similar peak values, and both peak in the ionospheric layers and then drop at higher altitude, but there are important differences with regard to the details of their altitude profiles~\cite{Takeda:1985hcf,GM118}.
    It is not clear that a Hall conductivity is relevant to questions of active-mode damping, as Joule energy loss is $\propto \bm{J}\cdot\bm{E}$, and $\bm{J}_{\textsc{h}} \perp \bm{E}$.
    However, even if it is, its effects would be qualitatively similar to the Pedersen conductivity; as a result, the Hall conductivity will not modify our qualitative arguments in the text regarding the ionospheric layer thicknesses vis \'a vis the active-mode damping length.
} %

Characteristic values for the Pedersen conductivity around noon at mid-latitude locations during medium solar activity 
(Wolf number%
\footnote{\label{ftnt:WolfNumber}%
    The Wolf number $R_{\textsc{Wolf}}$ measures the number of sunspots, and varies from $R_{\textsc{Wolf}}\sim 0$ at solar minimum to $R_{\textsc{Wolf}} \sim 100$--$200$ at solar maximum, on the $\sim 11$-year solar cycle (see, e.g., \citeR{Clette_2014}).
} %
$R_{\textsc{Wolf}} \sim 70$) 
are $\sigma_{\textsc{p}} \sim (\text{few})\times 10^{-4}\,\text{S/m}\sim (\text{few}) \times 10^{-8}\,$eV at an altitude of $\sim 120\,$km (E layer), falling to $\sigma_{\textsc{p}} \sim (\text{few})\times 10^{-5}\,\text{S/m} \sim (\text{few}) \times 10^{-9}\,$eV by an altitude of $\sim 160\,$km (lower F layer), and remaining there until an altitude of $\sim 250\,$km.
The Pedersen conductivity then falls exponentially with increasing altitude, reaching $\sigma_{\textsc{p}} \sim 10^{-7}\,\text{S/m} \sim 7\times 10^{-12}\,$eV around $\sim 500\,$km (upper F layer). 
These values, however, exhibit significant daily and longer-term (solar cycle) modulations~\cite{Takeda:1985hcf}: at times of low solar activity ($R_{\textsc{Wolf}}\sim 35$), night-time Pedersen conductivities are up to 2 orders of magnitude smaller in the E layer than during the day, and approximately an order of magnitude smaller in the F layer~\cite{Takeda:1985hcf}.
At times of peak solar activity ($R_{\textsc{Wolf}}\sim 200$), there are regions where the Pedersen conductivity remains $\sigma_{\textsc{p}} \sim (\text{few})\times 10^{-4}\,$S/m at all hours of the day, although the altitude and thickness of this layer varies: it is in the E layer at $\sim 100\,$km during the day, and in the lower F layer at $\sim 200\,$km at night~\cite{Takeda:1985hcf}.
Although approximate and quite variable, these characteristic values are all very high compared to the dark-photon mass range of interest $\sigma \gg m_{A'}$.

However, it is clear that the layer of high Pedersen conductivity is only $\sim (\text{few} \times 10^1)$--$10^2$\,km thick.
By contrast, a \emph{homogeneous, isotropic} conductor with \emph{homogeneous, isotropic} conductivity values on the order of the peak Pedersen conductivity would exhibit an active-mode damping length of order 
\begin{align}
    \delta &\sim \sqrt{\frac{2}{m_{A'}\sigma}} \nonumber \\
    &\sim 1300\,\text{km}\times \sqrt{\frac{10^{-18}\,\text{eV}}{m_{A'}}} \times \sqrt{\frac{3\times10^{-4}\,\text{S/m}}{\sigma}}.
\end{align}
Although this is not strictly the correct comparison (i.e., damping in an isotropic conductor with isotropic conductivity of order $\sigma_{\textsc{p}}$ is not the same as damping in an anisotropic conductor with the smallest conductivity of order $\sigma_{\textsc{p}}$), the fact that this characteristic damping length exceeds (or, depending on $m_{A'}$, is comparable to) the thickness of the relevant ionospheric layer where the Pedersen conductivity has such large values, makes it questionable whether the interacting mode will damp within the ionospheric layer in our dark-photon mass range of interest. 

The upshot of this discussion is that the ionosphere always has high characteristic anisotropic conductivities $\sigma_{\{\shortparallel,\textsc{p},\textsc{h}\}} \gg m_{A'}$, within some thickness.
However, only the parallel conductivity $\sigma_\shortparallel$ attains values sufficiently large that an isotropic conductor with the same conductivity would result in guaranteed damping of the interacting mode within the thickness of the ionospheric layers throughout the whole mass range in which our signal computation is valid; see \secref{sec:conductivitySummary}.
On the other hand, for $m_{A'}\lesssim (\text{few})\times 10^{-16}\,$eV, an isotropic conductivity of the same size as typical mid-solar-cycle peak Pedersen conductivity would not necessarily be sufficient to significantly damp the interacting mode within the thickness of the ionosphere; see again the discussion in \secref{sec:conductivitySummary}. 
As a result, we will hedge our modeling of the ionosphere and consider two possible cases: (a) the ionosphere does act to completely damp the interacting mode within its thickness; and (b) it does not, so we must consider the medium beyond the ionosphere.

\subsubsection{Earth's magnetosphere}
\label{sec:magnetosphere}

The ionosphere is only a constituent part of the larger magnetosphere, the region of space where the magnetic field is dominated by the Earth's own (mostly dipolar) field.
This is a complicated and highly dynamic environment, which in addition to the ionosphere, contains other distinctive features.
Just above the ionosphere is the so-called plasmasphere (some sources define the ionosphere as being the lower part of the plasmasphere), a region of cold charged plasma (mostly originating from the solar wind) which can extend up to a few Earth radii from the surface.
The outer edge of this region is defined by a steep decline in plasma density, dubbed the plasmapause~\cite{LaaksoJarva}.
In addition, the magnetosphere contains the two Van Allen radiation belts, which are regions of highly energetic electrons and protons trapped by the Earth's magnetic field.
The inner belt, located at 1--3 Earth radii, is relatively stable, while the outer belt, located at 3--7 Earth radii, can vary significantly in response to solar activity~\cite{Ganushkina}.
Finally, the boundary of the magnetosphere, the magnetopause, marks the outset of the interplanetary medium (see next subsection), where the dominant magnetic field is that of the Sun.
The magnetopause has a location and shape that is highly variable and depends on the prevailing state of the solar wind; generally, it takes a highly aspherical tear-drop-like shape that extends up to 10 Earth radii in the upstream direction of the solar wind (i.e., toward the Sun) and up to 200 Earth radii in the downstream direction (i.e., away from the Sun)~\cite{ShueSong,SibeckLin}.
For the purposes of this current work, we do not attempt to explicitly account for this environmental complexity; instead, we will argue that the relevant part of the signal we have found should be independent of these details (see \secref{sec:generalEarthTheory}).

\subsubsection{Interplanetary medium}
\label{sec:interplanetaryMedium}

Beyond the Earth's magnetopause lies the interplanetary medium which permeates the Solar System.
The interplanetary medium consists of a hot collisionless plasma consisting of fast-moving electrons, and ions streaming outward from the Sun at a few hundred km/s.
This plasma will also damp low-frequency interacting photon modes.

The interplanetary medium electron number density in the vicinity of the Earth is, on average,%
\footnote{\label{ftnt:SolarStorms}%
    Large upward transitory excursions by factors of $\sim 10$ are of course seen during solar storm events, such as flares or coronal mass ejections~\cite{ISSAUTIER20052141,celias-mtofARTICLE, celias-mtofDATASET, celias-mtofURL}.
}%
\up{,}%
\footnote{\label{ftnt:IPM}%
    Voyager mission measurements indicate that the interplanetary medium maintains an electron and ion density $n \gtrsim 10^{-3}\,\text{cm}^{-3}$~\cite{Gurnett_2019} all the way out to the heliopause, some $\sim 100\,$AU from Earth.
} %
$n_e\sim5\,\text{cm}^{-3}$,
while the electron temperature is $T_e\sim10^5\,\text K$~\cite{ISSAUTIER20052141, kallenrode2004space, russell}.
This implies an electron-ion collision frequency of roughly~\cite{Dubovsky:2015cca}
\begin{align}
    \nu=\frac{4\sqrt{2\pi}\alpha^2n_e}{3\sqrt{m_eT_e^3}}\ln\Lambda_C\sim10^{-20}\,\text{eV},
\end{align}
where $\alpha$ is the fine structure constant and the Coulomb logarithm can be estimated as~\cite{Dubovsky:2015cca}
\begin{align}
    \ln\Lambda_C=\frac12\ln\left(\frac{4\pi T_e^3}{\alpha^3n_e}\right)\approx27.
\end{align}
This collision frequency lies below the dark-photon mass range of interest to us in this work; the plasma can thus be treated as collisionless.

The ionic solar wind flowing out from the Sun carries with it solar magnetic field lines~\cite{1958ApJ...128..664P}, leading to a characteristic magnetic field in the vicinity of the Earth (outside the magnetopause) of around $B_{\odot}\sim5\,\text{nT}$~\cite{russell}.
This implies a cyclotron frequency for the electrons of
\begin{align}
    \omega_c=\frac{eB_{\odot}}{m_e}\sim6\times10^{-13}\,\text{eV},
\end{align}
which lies far below the characteristic electron plasma frequency of%
\footnote{\label{ftnt:ionPlasmaFreq}%
    The charged ion plasma frequency is of course a factor of $\sqrt{m_p/m_e} \sim \sqrt{1800}$ smaller. 
} %
\begin{align}
    \omega_p=\sqrt{\frac{4\pi n_e\alpha}{m_e}}\sim10^{-10}\,\text{eV},
\end{align}
or $f_p \approx 20\,\text{kHz}$.
Therefore the effects of the magnetic field can be neglected as well.

The primary effect of the plasma will thus be to add an effective mass $\omega_p \gg m_{A'}$ to the dispersion relation of interacting modes in the interplanetary medium (see \appref{app:systemBehavior} for more nuanced discussion).
Interacting modes in the medium with frequencies below $\omega_p$ will not propagate; they will instead be damped over the characteristic scale $\delta\sim1/\omega_p\sim2\,\text{km} \ll L_{\text{Solar System}}$. 
As this is an extremely short length-scale compared to characteristic distances in the Solar System, it is thus safe to assume that the interacting mode of the photon--dark-photon system throughout our entire mass range of interest is effectively damped out completely within the interplanetary medium.

\subsubsection{Summary}
\label{sec:conductivitySummary}

Here we summarize the relevant features of the near-Earth environment discussed in this section, and outline the mass range of validity for our models of the environment used in \secref{sec:signal}.
As we are considering the effects of ultralight dark-photon dark matter, our discussion will be restricted to masses $m_{A'}\gtrsim 10^{-21}\,\text{eV}$ (i.e., $f\sim 2.5 \times 10^{-7}\,$Hz) which are sufficiently large to allow for observed small-scale dark-matter structure~\cite{Hui:2016ltb,Kobayashi:2017jcf,Irsic:2017yje,Nadler:2020prv}.
On the other hand, the signal derived in \secref{sec:signal} crucially relies on the Compton wavelength of the dark matter being larger than the radius of the Earth, so we will also restrict to masses $m_{A'}\lesssim3\times10^{-14}\,\text{eV}$ (i.e., $f\lesssim 7\,$Hz).
Throughout this whole mass range, the innermost layers of the Earth, which are $\mathcal O(1000\,\text{km})$ deep (see \tabref{tab:earthConductivity}), are sufficiently conductive and thick to damp the active photon mode.
The lower atmosphere, on the other hand, acts as a relatively poor conductor throughout this range in the sense that the active-mode skin-depth greatly exceeds the radius of the Earth: the lower atmosphere thus contributes negligible damping to photon modes.
The effects of the ionosphere present a more complicated situation however, as the ionospheric layers have a highly anisotropic conductivity.
For masses $m_{A'}\gtrsim (\text{few}) \times 10^{-16}\,\text{eV}$ [i.e., $f\gtrsim (\text{few}) \times 10^{-2}\,$Hz], the ionospheric layers are thick enough that the active photon mode would be efficiently damped within the ionosphere, even using a conservative skin-depth estimate based on the Pedersen conductivity.
However, for masses $m_{A'}\lesssim (\text{few}) \times 10^{-16}\,\text{eV}$, the Pedersen conductivity becomes sufficiently low that the anisotropy of the ionosphere must be accounted for and the damping of active photon modes is not guaranteed.
Finally, in this case, the interplanetary medium beyond the ionosphere acts as a plasma with high plasma frequency for all relevant masses; it will thus damp the active photon mode for the entire mass range $10^{-21}\,\text{eV}\lesssim m_{A'}\lesssim 3\times10^{-14}\,\text{eV}$.

In summary then, for $(\text{few}) \times 10^{-16}\,\text{eV}\lesssim m_{A'}\lesssim 3\times10^{-14}\,\text{eV}$, the atmospheric gap between the Earth and the ionosphere represents a cavity between two active-mode-damping layers; on the other hand, for $10^{-21}\,\text{eV}\lesssim m_{A'}\lesssim (\text{few}) \times 10^{-16}\,\text{eV}$, the damping effects of the ionosphere are not guaranteed, but the gap between the Earth and the Earth's magnetopause represents a cavity between two active-mode-damping layers.
The situation right around $m_{A'} \sim (\text{few}) \times 10^{-16}\,\text{eV}$ may be fairly complicated; however, this possibly complicated region of parameter space lies above the mass range we consider explicitly in our search for this signal in this work (see \secref{sec:experimentalSearch}) and \citeR{Fedderke:2021qva}.

\section{Signal}
\label{sec:signal}

In this section, we derive the observable magnetic field signal which the dark photon sources near the Earth's surface in the atmospheric cavity bounded by the Earth itself below, and by either the ionosphere or the interplanetary medium above.

As discussed in \secref{sec:conductivityNearEarth}, the Earth may be treated as a good conductor in which the active mode of the photon--dark-photon system is efficiently damped, while the lower atmosphere is a region of relatively poor conductivity where the active mode propagates almost without attenuation.
However, the effects of the ionosphere above are more complicated, as this layer may or may not be thick enough to act as an adequate shield for the active mode.
The interplanetary medium beyond this, however, can be considered a plasma with a high plasma frequency (i.e., much above our frequency range of interest) and essentially infinite extent, and thus a good shield for the active mode.
Therefore, in order to remain agnostic about the effect of the ionosphere, in this section we compute the expected signal considering two different idealized models for the environment near the Earth.

In both models, we idealize the Earth as a perfect conductor and the lower atmosphere as a vacuum.
In light of the long active-mode damping length, we show in detail in \appref{app:finiteConductivty} that even having the conductivity as large as $\sigma \sim m_{A'}$ in the lower atmosphere does not spoil the assumption that this gap is effectively vacuum.
For the first model, we take the outer boundary of our geometry to be the ionosphere, which we assume to be a perfectly conducting spherical layer (i.e., a layer of sufficient thickness to completely damp the active mode of the photon--dark-photon system).
That is, we take the vacuum atmospheric air gap to be sandwiched between two perfect spherical conductors separated by a gap (the height of the atmosphere) much less than the radius of the Earth.
For the second model, we ignore the ionosphere and magnetospheric environment, and take the outer boundary to be the Earth's aspherical magnetopause, assuming that the interplanetary plasma medium beyond acts to completely damp the active mode of the photon--dark-photon system at the location of the magnetopause.

In both cases, we find the same signal at leading order: a monochromatic magnetic field signal with the spatial dependence of a particular vector spherical harmonic (VSH) [see \appref{app:vectorSphericalHarmonics} for VSH conventions] at the surface of the Earth.
In the aspherical case, additional magnetic field contributions appear, but they are in different VSH components which can easily be distinguished from the one of interest.

A key feature of our result is the characteristic length-scale that determines the suppression of the dark-photon signal. 
Similar to many other dark-photon observables, our signal is suppressed by $\varepsilon m_{A'}$~\cite{Chaudhuri:2014dla,Dubovsky:2015cca,Bhoonah:2018gjb,McDermott:2019lch,Wadekar:2019xnf,Kovetz:2018zes}; on dimensional grounds, the factor of $m_{A'}$ comes along with a length-scale.
In either model, our cavity has two such scales: the radius of the Earth $R$, and the characteristic size of the gap between the Earth and the outer boundary (either the ionosphere or magnetopause) $h$.
In the case where the ionosphere functions as our outer boundary, the latter is far smaller than the former.
\emph{A priori} one may expect that the suppression would be determined by the shortest length-scale of the cavity, which in the case where the ionosphere functions as the shield would be the  height of the atmosphere, $h \ll R$ (see, e.g., comments in \citeR{Dubovsky:2015cca}). 
However, we show that in both models the observable magnetic field generated by the dark-photon field is, in fact, suppressed by $m_{A'}R$, not by $m_{A'}h$.

In this section we proceed as follows:
First, to motivate the appearance of the $m_{A'}R$ dependence, as well as to introduce some features of our Earth calculation, we calculate the effect of a dark photon in a simple toy example of a wide and squat cylindrical cavity hollowed out of a perfect conductor; see also Appendix A.b of \citeR{Chaudhuri:2014dla}.
Second, we calculate the magnetic field signal in the vicinity of the Earth, for the case of the first model with a spherical, perfectly conducting outer boundary at the ionosphere.
Finally, we compute our signal in the second model with an aspherical outer boundary of the magnetosphere.

\subsection{Toy example: cylindrical cavity}
\label{sec:toyTheory}

\begin{figure}[!t]
\includegraphics[width=0.8\columnwidth]{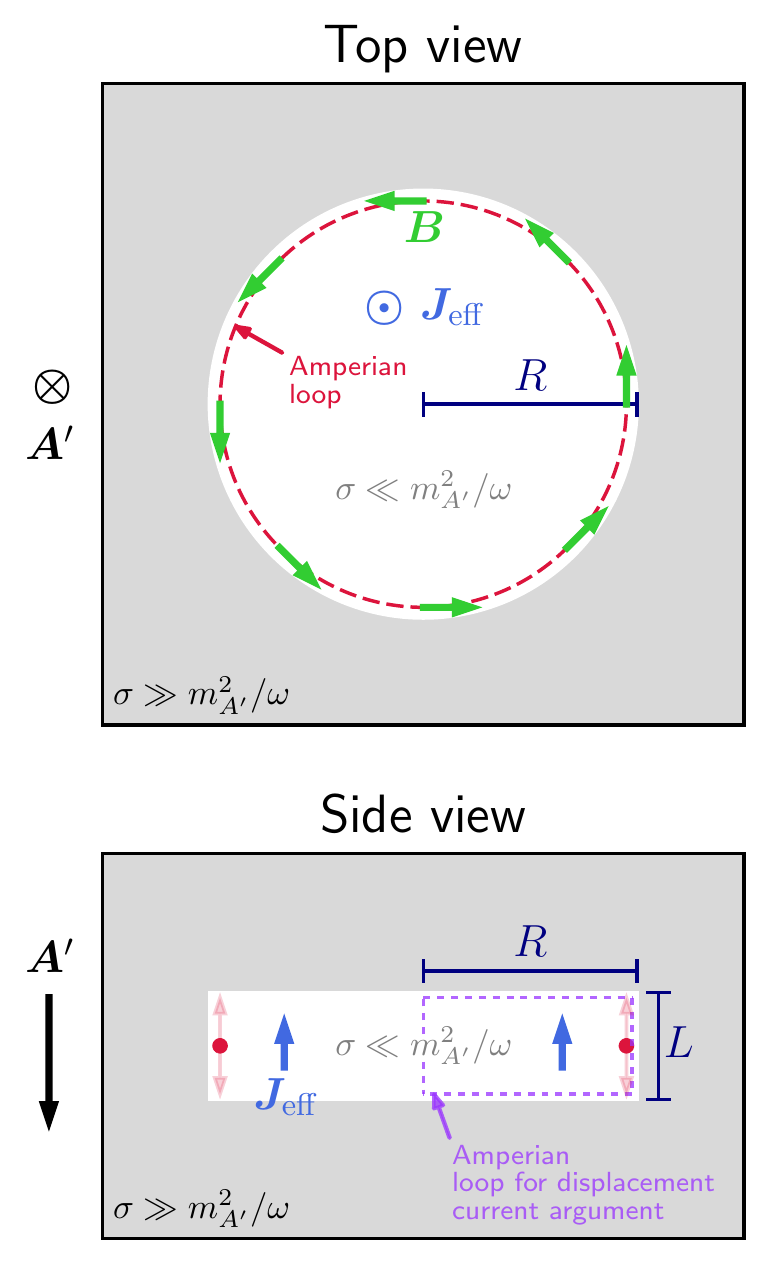}
\caption{\label{fig:cavity}%
    Schematic view of the toy example cylindrical cavity of poorly conducting material (white) of radius $R$ and height $L\ll R$ hollowed out of a good conductor (gray).
    Also shown are (1) the Amp\`erian loop discussed in the text (red dashed in top view; red markers in side view), which can be located anywhere within the vertical height of the cavity along the short dimension (as indicated by the pink arrows in the side view); (2) the effective current $\bm{J}_{\text{eff}} \propto -\bm{A'}$ [blue indicator (arrows) in top (side) view], assuming here for the purposes of this argument that the dark-photon field $\bm{A}'$ [black indicator (arrow) in top (side) view] happens to be oriented to be perpendicular to the top and bottom surfaces of the cavity (see discussion in text of \secref{sec:toyTheory} for generalization); (3) the induced observable magnetic field $\bm{B}$ (green arrows in top view); and (4) the Amp\`erian loop (dotted purple in the side view) that we consider for the argument which we advance in footnote \ref{ftnt:displacementCurrent} that the displacement current can be ignored (this loop runs along a straight ray emanating from the center of the cavity in the top view).
    Note that the direction of $\bm{J}_{\text{eff}}$ is set by the direction of $-\bm{A}'$ [see \eqref{JeffAprime}]: see discussion around \eqref{dmvec} for how the direction of $\bm{A}'$ is set for the case of realistic dark-photon dark matter, and note that the DM velocity is not relevant at leading order in setting the direction of $\bm{A}'$.%
    } 
\end{figure}

Consider a cylindrical cavity of radius $R$ and height $L$ whose walls have infinite conductivity, in the presence of a dark-photon field oriented along the axial symmetry axis of the cylinder (which we will take to be the $z$-axis).
We will demonstrate that the magnetic field sourced by the dark-photon field does not depend on the dimension  of the cavity $L$ that is longitudinal to the dark-photon field, but rather only on the transverse dimension $R$, even if $L\ll R$.

Before proceeding, we note that unless otherwise specified, from this point onward in this paper, the terminology `dark photon' or 'dark-photon field' refers to the sterile state in the interaction basis, $A'_{\tsc{i}}$.
Also, because we work solely in the interaction basis from this point onward, we will for notational simplicity drop the\linebreak  {subscript-$\tsc{i}$} on both the sterile and interaction states in the interaction basis: i.e., for the remainder of the body of the paper, $A_{\tsc{i}} \equiv A$ and $A_{\tsc{i}}' \equiv A'$.

Our calculation will be based on the `effective background current' approach for treating the effects of the dark-photon field, which we will briefly outline here; see \citeR[s]{Graham:2014sha,Chaudhuri:2014dla} and \appref{app:effectiveCurrent} for careful treatments.
Because the sterile dark-photon field itself is unaffected to leading order in $\varepsilon$ by the presence of a conductor or by the presence of an interacting component (see \appref{app:systemBehavior}), it is consistent when $\varepsilon\ll1$ to neglect back-reaction on the dark-photon field and treat $A'$ as a background field. 
In the interaction-basis Lagrangian, \eqref{interactionBasisLagrangianLinear}, the terms $\LL \supset -(J^\mu_{\tsc{em}}-\varepsilon m_{A'}^2A^{\prime\, \mu}  )A_\mu $ appear.
It is thus clear that with $A'$ treated as a background field at leading order, it acts to source observable electromagnetic fields in a manner indistinguishable from an effective current given by
\begin{align}
    \bm{J}_\text{eff}(\bm{x},t)=-\varepsilon m_{A'}^2\bm{A}'(\bm{x},t).
    \label{eq:JeffAprime}
\end{align}
Note that we have written only spatial components here because, in the non-relativistic limit $\omega \gg k$, the effective charge density vanishes,  $J_{\text{eff}}^0 \rightarrow 0$ (see \appref{app:effectiveCurrent} for a detailed discussion).

Before explicitly computing the electric and magnetic field solutions, let us first explore a simple argument to see why the result will depend on $R$ but not $L$.
Consider a circular Amp\`erian loop that runs around the inner circumference of the cavity (parallel to the top and bottom faces of the cavity); see \figref{fig:cavity}.
Assume the dark-photon field $\bm{A}'$ is aligned with the axial symmetry axis of the cavity (i.e., perpendicular to the top and bottom surfaces).
The magnetic field inside the conductor is sourced by the effective background current density $\bm{J}_\text{eff}$ and must be axial on symmetry grounds.
By the Amp\`ere--Maxwell law, the integrated magnetic field along this loop, $\oint \bm{B}\cdot\bm{dl} \sim BR$, is equal to the current flux through the surface it bounds,%
\footnote{\label{ftnt:displacementCurrent}%
    Since we operate in the quasi-static limit $m_{A'} R \ll 1$, the displacement current term in the Amp\`ere--Maxwell law can be ignored at leading order.
    Consider the integral form of Faraday's law applied on the purple Amp\`erian loop shown in the side view in \figref{fig:cavity}.
    Boundary conditions require that $E_z$ vanishes at the wall, and that the radial electric field must be zero near the top and bottom cavity surfaces; we thus have $\oint \bm{E} \cdot\bm{dl} \sim E L$, where $E$ is the value of the vertical electric field near the center of the cavity (which is similar to the generic value for $E_z$ not in the vicinity of the cavity walls).
    The Gaussian surface integral that appears on the RHS of Faraday's law is $\iint \partial_t \bm{B} \cdot \bm{dA} \sim m_{A'} B R L$, where $B$ is a representative value of the axial magnetic field.
    Therefore, $E \sim (m_{A'}R) B$.
    The additional displacement current term in the Amp\`ere--Maxwell law that we ignored in the main text would thus be $\iint \partial_t \bm{E} \cdot\bm{dA} \sim m_{A'} R^2 E \sim R ( m_{A'} R)^2 B$, which clearly only modifies the $B$ field result at sub-leading order in $m_{A'}R$.
} %
$\iint \bm{dA} \cdot \bm{J}_\text{eff} \sim \varepsilon m_{A'}^2R^2 A'$. 
Therefore, we expect $B \sim \varepsilon m_{A'}^2R A'$.
If we normalize $A'$ to be all of the dark matter, $A' \sim \sqrt{\rho_{\textsc{dm}}}/m_{A'}$, it follows that the $B$ field will be $B \sim \varepsilon m_{A'}R \sqrt{\rho_{\tsc{dm}}}$; cf.~the result at \eqref{mag_cyl}, and the discussion immediately following.
Note that this argument does \emph{not} depend on where in the short geometrical dimension of the cavity the Amp\`erian loop is located: the answer is independent of $L$~\cite{Chaudhuri:2014dla}.

Note that if $\bm{A}'$ were not perfectly aligned with the symmetry axis $\bm{\hat{s}}$ of the cylinder, the above parametric argument would still go through [up to $\mathcal{O}(1)$ geometrical factors], with one exception: there would be an additional angular suppression $B \propto \bm{\hat{A}}'\cdot \bm{\hat{s}}$.
Unless the background dark-photon field is nearly perpendicular to the symmetry axis of the cylinder ($\bm{\hat{A}}'\cdot \bm{\hat{s}} \lesssim L/R$), the axial magnetic field is therefore still parametrically larger than an estimate suppressed by the small length-scale $L$.

The intuitive lesson to draw from this discussion is that the magnetic field amplitude depends on the separation distance between the surfaces in which the screening currents that lie along the direction of the would-be dark-photon electric field run.

Let us now find the quantitative solution to see that this parametric argument holds.
Given the effective current described above, the full electric field solution must satisfy
\begin{align}
    (\nabla^2-\partial_t^2)\bm{E}&=\partial_t\bm{J}_\text{eff},&\nabla\cdot\bm{E}&=0.
\end{align}
Assuming some boundary conditions for our problem, we can decompose the full solution as
\begin{align}
    \bm{E}=\bm{E}_\text{inh}+\bm{E}_\text{hom},
\end{align}
where $\bm{E}_\text{inh}$ is chosen to satisfy
\begin{align}
    (\nabla^2-\partial_t^2)\bm{E}_\text{inh}=\partial_t\bm{J}_\text{eff},
\end{align}
and $\bm{E}_\text{hom}$ is chosen to fulfill the boundary conditions on the full solution, while satisfying
\begin{align}
    (\nabla^2-\partial_t^2)\bm{E}_\text{hom}=0.
    \label{eq:hom}
\end{align}
Both contributions must also satisfy $\nabla\cdot\bm{E}_\text{inh/hom}=0$.

Neglecting the velocity of the dark photon, we may write its effective current density as
\begin{align}
    \bm{J}_\text{eff}(\bm{x},t)=-\varepsilon m_{A'}^2A'_0e^{-im_{A'}t}\zhat.
\end{align}
Then we may take our inhomogeneous solution to be
\begin{align}
    \bm{E}_\text{inh}(\bm{x},t)=i\varepsilon m_{A'}A'_0e^{-im_{A'}t}\zhat.
    \label{eq:inh}
\end{align}
In accordance with the symmetries of the problem, we will write our homogeneous solution as a linear combination
\begin{align}
    \bm{E}_\text{hom}(\bm{x},t)=\big[aJ_0(m_{A'}r)+bY_0(m_{A'}r)\big]e^{-im_{A'}t}\zhat,
    \label{eq:hom_cyl}
\end{align}
for some constants $a$ and $b$.
Using the properties of the (cylindrical) Bessel functions $J_n$ and $Y_n$, it is straightforward to show that \eqref{hom_cyl} satisfies \eqref{hom}.

The two boundary conditions at the cavity walls determine $a$ and $b$.
Since the walls of the cavity are assumed to have infinite conductivity, the $z$-component of the electric field must vanish at a radius $r=R$ (recall, we are working in the interaction basis).
Moreover, the electric field must be regular at the origin $r=0$.  The latter condition forces $b=0$, and the former then requires
\begin{align}
    a=-\frac{i\varepsilon m_{A'}A'_0}{J_0(m_{A'}R)}.
\end{align}
This means that the full solution for the electric field inside the cavity is~\cite{Chaudhuri:2014dla}
\begin{align}
    \bm{E}(\bm{x},t)=i\varepsilon m_{A'}A'_0\left(1-\frac{J_0(m_{A'}r)}{J_0(m_{A'}R)}\right)e^{-im_{A'}t}\zhat.
    \label{eq:elec_cyl}
\end{align}
The corresponding magnetic field is~\cite{Chaudhuri:2014dla}
\begin{align}
    \bm{B}(\bm{x},t)&=-\frac i{m_{A'}}\nabla\times\bm{E} \label{eq:BfromE}
    \\&=-\varepsilon m_{A'}A'_0\frac{J_1(m_{A'}r)}{J_0(m_{A'}R)}e^{-im_{A'}t}\phihat.
    \label{eq:mag_cyl}
\end{align}
Near the cavity walls and in the limit $m_{A'}R\ll 1$, this axial magnetic field oscillates with magnitude $|B|=\varepsilon m_{A'}^2 R A_0' / 3$.
Normalizing $A_0'$ to be all of the DM, this result has the exact same parametric scalings as the simple Amp\`erian-loop argument advanced above.

Note that neither \eqref{elec_cyl} nor \eqref{mag_cyl} depend explicitly on the dimension of the cavity $L$ along the direction of $\bm{A}'$.
This means that even if the cylinder is very squat (i.e., $L\ll R$), the observable fields inside the cavity will suffer no additional suppression.
This effect is not particular to this geometry.
For instance, for a rectilinear cavity of side lengths $L_x$, $L_y$, and $L_z$, it can be shown that the magnetic field sourced by a dark photon oriented along the $z$-direction is%
\footnote{\label{ftnt:rectilinear}%
        This result is not derived using the above approach of breaking down the electric field into homogeneous and inhomogeneous contributions.
        Rather, it is derived using a cavity mode decomposition (cf.~Appendix A.c of \citeR{Chaudhuri:2014dla}).
        A similar approach can be applied to the cylindrical cavity and will give an equivalent result to \eqref{mag_cyl}, but in the form of a more complicated sum.
} %
\begin{widetext}
\begin{align}
    \bm{B}=-16\varepsilon m_{A'}^2A'_0\sum_{p,q\text{ odd}}\frac{\frac p{L_x}\cos\left(\frac{\pi px}{L_x}\right)\sin\left(\frac{\pi qy}{L_y}\right)\yhat-\frac q{L_y}\sin\left(\frac{\pi px}{L_x}\right)\cos\left(\frac{\pi qy}{L_y}\right)\xhat}{\pi pq\left(m_{A'}^2-\frac{\pi^2p^2}{L_x^2}-\frac{\pi^2q^2}{L_y^2}\right)}e^{-im_{A'}t}.
    \label{eq:Bbox}
\end{align}
\end{widetext}
Again, this expression does not depend on $L_z$, so that even if $L_z\ll L_x,L_y$, the magnetic field will not be suppressed by the shortest length-scale of the cavity.
In fact, this is generically true regardless of the dark-photon orientation: typically, the $z$-component of the dark-photon field will be nonzero, and the magnetic field contribution generated by the $z$-component of $\bm{A}'$ will still take the form of \eqref{Bbox}, but with $A_0' \rightarrow \bm{A}'\cdot\bm{\hat{z}}$.
In order to suppress this field contribution by an amount equivalent to making the geometrical suppression factor $\sim m_{A'}L_z$ as opposed to $\sim m_{A'}\times\min\{L_x,L_y\}$ would require close alignment between $\bm{A}'$ and the $xy$-plane, to within an angle of $\mathcal{O}( L_z / \min\{L_x,L_y\}) \ll 1$.

\subsection{Earth model 1: Ionosphere as boundary}
\label{sec:earthTheory}

We now consider the computation of the dark-photon signal in our first idealized model of the electrical environment near the Earth: a vacuum cavity bounded between two concentric spherical walls.
This is the physical situation in the vicinity of the Earth if, in fact, the ionosphere acts as an effective shield for the active mode of the photon--dark-photon system.

For the purposes of this computation, we approximate the lower atmosphere as a cavity of zero conductivity bounded by an inner spherical wall of radius $R\ll 1/m_{A'}$ (the Earth's surface) and an outer spherical wall of radius $L=R+h$ (the ionosphere), where $h\ll R$ (see also Sec.~8.9 of \citeR{Jackson} for a similar model for discussing the Schumann resonances~\cite{Schumann+1952+149+154}).
We will take both the ground and the ionosphere to have infinite conductivity in our calculation; see \appref{app:finiteConductivty} for a discussion of modifications to this picture if finite conductivity effects are included.

\begin{figure}[t]
\includegraphics[width=0.75\columnwidth]{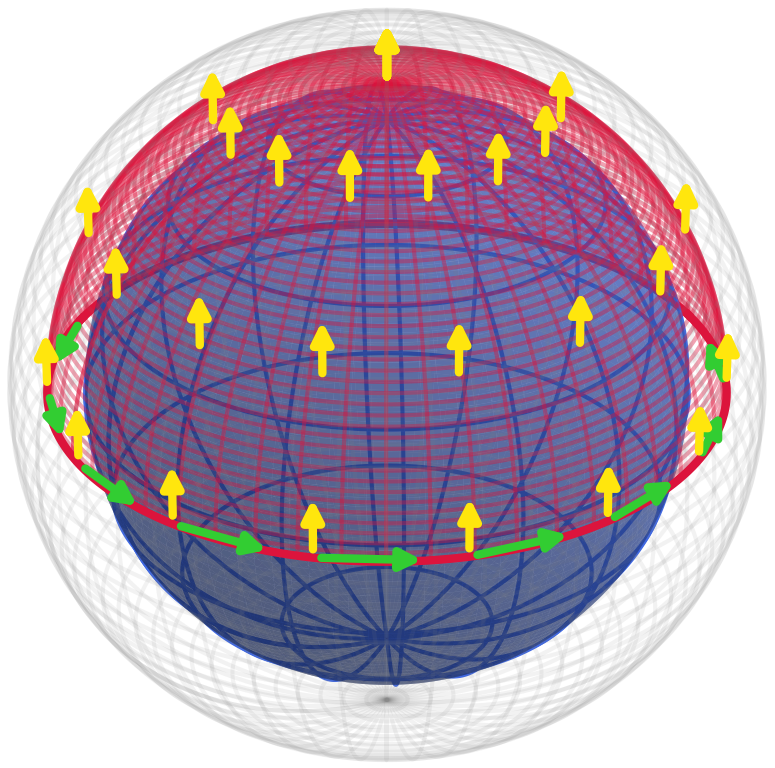}
\caption{\label{fig:earth}%
    Sketch of the Amp\`erian loop setup for the Earth (not to scale).
    The inner conducting sphere of the Earth (radius $R$) is shown as the blue sphere.
    The solid green arrows represent the axial magnetic field on the Equator.
    The thick solid red line at the Equator is the Amp\`erian loop discussed in the text, with the red hemisphere being the Gaussian surface spanned by the loop through which the effective current $\bm{J}_{\text{eff}}$ (yellow arrows), here assumed to point along the Earth's rotational axis, is integrated.
    The assumed conducting ionospheric layer is represented by the outer enveloping gray sphere, a distance $h\ll R$ above the surface of the Earth sphere.
    The Amp\`erian loop could be located anywhere in the gap between the Earth sphere and the ionospheric conductive layer without modifying the leading-order result of the Amp\`erian loop argument given in the text [i.e., this would only induce corrections $\propto (h/R)^n$ to the leading order $\bm{B}$ field].
    Note that the dark-photon-induced effective current is shown here as aligned with the rotational axis of the Earth strictly for the sake of visualization; our search (see \secref{sec:experimentalSearch}) marginalizes over the spatial orientation (i.e., polarization state) of the dark-photon field.
    } 
\end{figure}

Before proceeding to the calculation, we reiterate the point that the result will depend only on $R$, and not $h$, with another simple argument based on the Amp\`ere--Maxwell law.
As above, we will treat the dark photon as an effective background current.
Suppose for simplicity that $\bm{A}'$ is oriented along the rotational axis of the Earth (which we take to be the $z$-axis).
Consider the Gaussian surface that covers the Northern Hemisphere of the Earth (but lies just outside the inner conductive sphere); see the red hemisphere in \figref{fig:earth}.
The boundary of this surface is an Amp\`erian loop in the plane of the Earth's equator.
By the Amp\`ere--Maxwell law, the integrated magnetic field along this loop, once again $\oint \bm{B}\cdot\bm{dl} \sim BR$, is equal to the current flux through this surface,%
\footnote{\label{ftnt:noDisplacement2}%
    Again, we ignore the higher-order-in-$(m_{A'}R)$ displacement current by virtue of an argument very similar to that advanced in footnote \ref{ftnt:displacementCurrent}, modified as required to account for the different geometry here.
} %
which is given parametrically by $\iint \bm{J}_{\text{eff}} \cdot \bm{dA} \sim R^2 J_{\text{eff}} \sim \varepsilon m_{A'}^2R^2 A' \sim  \varepsilon m_{A'} R^2 \sqrt{\rho_{\textsc{dm}}}$,
The latter expression is obtained assuming the dark photon is all of the dark matter, and throughout this series of estimates we neglected $\mathcal{O}(1)$ geometric factors, and corrections $\sim h/R$.
Clearly we once again arrive at the conclusion that $B \sim \varepsilon m_{A'}^2RA' \sim  \varepsilon m_{A'}R \sqrt{\rho_{\textsc{dm}}}$ at leading order, up to $\mathcal{O}(1)$ factors.
The leading order answer is independent of $h$, the height of the atmosphere. 
Thus, if the height of the atmosphere is varied, it will not have any effect on the strength of the magnetic field at the equator.
In particular, we emphasize that \emph{the field is not suppressed by $m_{A'}h\ (\ll m_{A'}R)$.}

Note that the intuition developed in \secref{sec:toyTheory} regarding the relevant length-scale that enters the geometrical suppression factor holds up here too, albeit with one minor modification.
Previously we argued that the relevant length scale is the separation between the surfaces on which the screening currents that lie along the direction of the would-be dark-photon electric field run.
In the geometry here, screening currents run in opposite directions in the inner and outer shielding layers, so one should not consider the gap $h$ between the inner and outer shields to be the relevant separation distance, as the magnetic field contributions from those opposite current directions will constructively superpose in the gap.
Rather, the relevant separation distance is that between \emph{like-sense} screening currents; here, that is approximately the radius of the Earth $R$ [up to $\mathcal{O}(h)$ corrections], which is indeed the length-scale entering the suppression factor.

Let us proceed with the quantitative calculation, which will confirm the foregoing parametric argument.
Because we assume spherical symmetry of the Earth's surface and ionospheric layer, whether the Earth is rotating is irrelevant for the purposes of computing the fields at a fixed location in absolute, inertial coordinates (i.e., coordinates fixed to the locations of the average positions of a set of distant stars).
To begin with then, we work in inertial spherical coordinates with the $z$-axis aligned to the Earth's rotational axis, and compute the signal at a fixed inertial position near the Earth's surface. 
In this case, the inertial spherical coordinate $\theta$ corresponds to a fixed latitude on the Earth's surface, but the geographical longitude to which the inertial spherical co-ordinate $\phi$ corresponds on the Earth's surface evolves as the Earth rotates in the inertial co-ordinate frame.
Accounting for this to find the signal at a fixed location on the Earth's surface (i.e., at a fixed location in the body-fixed rotating frame) will, however, be trivial once we have the signal in inertial coordinates, and we defer this correction to the end of the computation.

We will take the orientation of the dark-photon vector potential $\bm{A}'$ to be generic and, for convenience, introduce the notation
\begin{align}
	A'_+ &= -\tfrac{1}{\sqrt{2}} \lb(A'_x - iA'_y \rb), \label{eq:Aplus}\\
	A'_- &= +\tfrac{1}{\sqrt{2}} \lb(A'_x + iA'_y \rb), \label{eq:Aminus}\\
	A'_0 &= A'_z, \label{eq:Azero}
\end{align}
where $A'_x$, $A'_y$, and $A'_z$ are the Cartesian components of the dark-photon vector potential in the inertial frame.
Because we assume that the dark photon is non-relativistic, the $A'_i$ are constant over the whole surface of the Earth; i.e., the dark-photon de Broglie wavelength $\lambda_{\text{dB}} = 2\pi/ (m_{A'}v_{\textsc{dm}})$ is much larger than the radius of the Earth: $3 \times 10^6 \lesssim \lambda_{\text{dB}} / R \lesssim 8\times 10^7$ in our mass range of interest.
With this notation, we can then use the VSH identities at \eqrefRange{xharm}{zharm} to write the effective background current corresponding to the dark photon in terms of VSH as%
\footnote{\label{ftnt:VSHnotation}%
    We follow convention and label the VSH with degree $\ell$ and order $m$; the order symbol $m$ should not be confused with the dark-photon mass, which we label $m_{A'}$.
} %
\begin{align}
    \bm{J}_\text{eff}=-\sqrt{\frac{4\pi}3}\varepsilon m_{A'}^2\sum_{m=-1}^1A'_m(\bm{Y}_{1m}+\bm{\Psi}_{1m})e^{-im_{A'}t}, \label{eq:Jeff}
\end{align}
where we have employed the notation $A'_{\pm1}\equiv A'_\pm$.
This form of the effective current is applicable everywhere in the cavity gap between the surface of the Earth and the ionosphere.
Note that the appearance in \eqref{Jeff} of the form $\bm{V}_{m} \equiv \bm{Y}_{1m}+\bm{\Psi}_{1m}$ is easily understood: $\bm{V}_0 \propto \bm{\hat{z}}$, and $\bm{V}_{\pm1} \propto \mp ( \bm{\hat{x}} \pm i \bm{\hat{y}})$; see \appref{app:vectorSphericalHarmonics}.

As in \secref{sec:toyTheory}, we proceed by computing the homogeneous and inhomogeneous contributions to the electric field inside the cavity.
The inhomogeneous contribution will simply be
\begin{align}
    \bm{E}_\text{inh}=\sqrt{\frac{4\pi}3}i\varepsilon m_{A'}\sum_{m=-1}^1A'_m(\bm{Y}_{1m}+\bm{\Psi}_{1m})e^{-im_{A'}t}.
    \label{eq:inhom_sph}
\end{align}
In terms of the VSH, the homogeneous contribution can be decomposed into a `transverse electric' (TE) and a `transverse magnetic' (TM) contribution~\cite{Jackson}:%
\footnote{\label{ftnt:TEmode}%
    These modes are `transverse' in the sense that their electric [\eqref{ETE}] and magnetic [\eqref{BTM}] fields are, respectively, tangent to the sphere: $\bm{\hat{r}}\cdot\bm E_\text{TE}=\bm{\hat{r}}\cdot\bm B_\text{TM}=0$.
} %
\begin{align}
    \bm E_\text{hom}=\bm E_\text{TE} + \bm E_\text{TM},
\end{align}
where
\begin{align}\label{eq:ETE}
    \bm E_\text{TE}&\equiv\sum_{\ell m}f_{\ell m}(m_{A'}r)\bm\Phi_{\ell m}e^{-im_{A'}t},\\
    \bm E_\text{TM}&\equiv\sum_{\ell m}\frac1{m_{A'}}\nabla\times\Big[g_{\ell m}(m_{A'}r)\bm\Phi_{\ell m}\Big]e^{-im_{A'}t}\nonumber\\&=\sum_{\ell m}\lb[\begin{array}{l}-\dfrac{\ell(\ell+1)g_{\ell m}(m_{A'}r)}{m_{A'}r}\bm Y_{\ell m}\\[2ex]-\left(g'_{\ell m}(m_{A'}r)+\dfrac{g_{\ell m}(m_{A'}r)}{m_{A'}r}\right)\bm\Psi_{\ell m}\end{array}\rb]\nl\qquad\quad \times e^{-im_{A'}t},
    \label{eq:ETM}
\end{align}
and where the VSH Laplacian properties \eqrefRange{Ylaplace}{Philaplace} can easily be used to show that $f_{\ell m}$ and $g_{\ell m}$ must each be linear combinations of spherical Bessel functions $j_\ell$ and spherical Neumann functions $y_\ell$, in order to satisfy \eqref{hom}.%
\footnote{\label{ftnt:sBsNfuncs}%
    Recall that for $F_\ell \in \{j_\ell,\,y_\ell\}$ we have~\cite{ArfkenWeber}
    \[ x^2 F_\ell'' + 2xF_\ell' + \lb(x^2 - \ell(\ell+1)\rb)F_\ell = 0. \]%
} %

Using the VSH curl properties \eqrefRange{Ycurl}{Phicurl}, the corresponding magnetic fields can be computed to be
\begin{align}\label{eq:BTE}
    \bm B_\text{TE}&=-i\sum_{\ell m}\frac1{m_{A'}}\nabla\times\Big[f_{\ell m}(m_{A'}r)\bm\Phi_{\ell m}\Big]e^{-im_{A'}t}\nonumber\\&=-i\sum_{\ell m}\lb[\begin{array}{l}-\dfrac{\ell(\ell+1)f_{\ell m}(m_{A'}r)}{m_{A'}r}\bm Y_{\ell m}\\[2ex]-\left(f'_{\ell m}(m_{A'}r)+\dfrac{f_{\ell m}(m_{A'}r)}{m_{A'}r}\right)\bm\Psi_{\ell m}\end{array}\rb]\nl\qquad\quad \times e^{-im_{A'}t},\\
    \bm B_\text{TM}&=-i\sum_{\ell m}g_{\ell m}(m_{A'}r)\bm\Phi_{\ell m}e^{-im_{A'}t}.
    \label{eq:BTM}
\end{align}
Note that $\bm B_\text{inh}$ vanishes under our approximations since $\bm E_\text{inh}$ points in a fixed direction and is constant throughout the atmospheric air gap. 

Because \eqref{inhom_sph} contains no $\bm{\Phi}_{\ell m}$ components and the boundary geometry is spherical, it is clear that only the $\ell=1$ TM homogeneous components will be relevant for our computation in this section.
That is, $g_{\ell m}=0$ for $\ell\neq1$, and $f_{\ell m}=0$ for all $\ell,m$.
Let us then write
\begin{align}
    g_{1m}(x)&=a_mj_1(x)+b_m x_0^3 y_1(x) &
    [x_0 &\equiv m_{A'}R]
    \label{eq:fmDefn}
\end{align}
for $m=0,\pm1$; here $a_m$ and $b_m$ are constants, and we have introduced a dimensionless scale factor $x_0$ for later convenience (see footnote \ref{ftnt:powerCounting}).

Electromagnetic boundary conditions enforce that the total electric field tangent to a perfectly conducting boundary must vanish (recall, we work in the interaction basis): that is, we must set $\bm{E}_\parallel=\bm{0}$ on the ground and at the ionosphere.
Since $\bm{Y}_{1m} \propto \bm{\hat{r}}$ points only radially [cf.~\eqref{VSHdef}], the boundary conditions as applied to the field expansions we have developed are such that the coefficient of $\bm{\Psi}_{1m}$ in the total electric field must vanish both at $r=R$, and at $r=R+h$.%
\footnote{\label{ftnt:meaningOfY}%
    Of course, a non-zero component of $\bm{E} \propto \bm{Y}_{\ell m} \propto \bm{\hat{r}}$ merely indicates the presence of an induced surface charge density at the conductive boundaries.
} %
Imposing these conditions yields algebraically complicated expressions for $a_m$ and $b_m$ in terms of the spherical Bessel and Neumann functions; see \eqref[s]{amFull} and (\ref{eq:bmFull}) in \appref{app:fullCoeffModel1}.
Since we will be interested in the limits $m_{A'}R \ll 1$ and $h \ll R$, we may use the small-$x$ limits of the spherical Bessel and Neumann functions, $j_1(x)\sim(x/3)-(x^3/30)$ and $y_1(x)\sim-x^{-2}-1/2$, respectively, to expand \eqref[s]{amFull} and (\ref{eq:bmFull}).
Retaining leading terms and the first few corrections yields%
\footnote{\label{ftnt:orderexpansion}%
    To compute the leading-order magnetic field, we require only the first term $\propto (m_{A'}R)^0$ in $a_m$ in \eqref{am}, and we can set $b_m=0$.
    For completeness, we have kept those higher-order terms here which would be required to calculate the leading-order piece of the electric field that is $\propto \bm{\Psi}_{1m}$ \emph{and} have it satisfy the boundary conditions approximately.
}%
\up{,}%
\footnote{\label{ftnt:powerCounting}%
    The $x_0^3$ that we explicitly factored out in \eqref{fmDefn} preserves a common small-parameter power counting in $(m_{A'}R)$ for $a_m$ and $b_m$: since, parametrically, $y_1(x) \sim x^{-3} j_1(x)$ at small $x$, it follows that for $x \approx x_0 \ll 1$, we have $f_m(x_0) \sim x_0 ( a_m /3 - b_m) + \cdots$.
    Therefore, like powers of $x_0 = m_{A'}R$ appearing in $a_m$ and $b_m$ contribute at the same order to $f_m(x\approx x_0)$.
} %
\begin{align}
    a_m &= \sqrt{3\pi}\, i\varepsilon m_{A'}A'_m \lb[ 1 + \frac{1}{3} \lb(m_{A'}R\rb)^2 \lb( 1 + \frac{h}{R} + \frac{2h^2}{3R^2} \rb) \rb],
    \label{eq:am}\\
    b_m &= -\frac{4\sqrt{\pi}}{15\sqrt{3}}\,i\varepsilon m_{A'}A'_m \lb(m_{A'}R\rb)^2 \lb( 1 + \frac{5h}{2R} + \frac{5h^2}{3R^2} \rb).
    \label{eq:bm}
\end{align}
Substituting \eqref[s]{am} and (\ref{eq:bm}) into \eqref{BTM}, we find that, to leading order in $m_{A'}R$, the magnetic field at $\Omega=(\theta,\phi)$ is
\begin{align}
    \bm{B}(\Omega,t)=\sqrt{\frac\pi3}\varepsilon m_{A'}^2R\sum_{m=-1}^1A'_m\bm{\Phi}_{1m}(\Omega)e^{-im_{A'}t}.
    \label{eq:pre_signal}
\end{align}
Note that, as advertised, the magnetic field signal is suppressed not by $m_{A'}h$, but rather by $m_{A'}R$.
Note also that it has exactly the parametric scaling advanced by the simple Amp\`erian loop argument above ($\Phi_{10}\propto \phihat$ on the Equator at $\theta =\pi/2$).

It remains to account for the rotation of the Earth; see also \citeR{Caputo:2021eaa} for recent discussion.
The speed of rotational motion of a point fixed to the surface of the Earth is $v \ll c$, so there are no relativistic field-mixing effects for which we need to account; we need only relate the (Earth-fixed frame) longitude on the Earth's surface, $\tilde{\phi}$, to the azimuthal inertial co-ordinate $\phi$.
This is trivial:
\begin{align}
    \phi=\tilde\phi+2\pi f_dt,
    \label{eq:phaseRot}
\end{align}
where $f_d = (\text{sidereal day})^{-1}$.
As measured with respect to the inertial reference frame, the station at a fixed location $\tilde{\Omega} = (\tilde{\theta},\tilde{\phi})$ on the Earth's surface thus sees the magnetic field evolution
\begin{align}
    \bm{B}(\tilde{\Omega},t) = \bm{B}(\theta = \tilde{\theta},\phi=\tilde{\phi}+2\pi f_d t,t).
    \label{eq:Bevol}
\end{align}
The properties of the VSH are such that
\begin{align}
\bm{\Phi}_{1m}(\theta=\tilde{\theta},\phi=\tilde{\phi}+2\pi f_d t) = e^{2\pi if_dt} \bm{\tilde{\Phi}}_{1m}(\tilde{\Omega}),
\end{align}
where $\bm{\tilde{\Phi}}_{1m}(\tilde{\Omega})$ are the VSH as constructed by the observer using the body-fixed reference frame tied rigidly to the rotating Earth.%
\footnote{\label{ftnt:frames}%
    For the avoidance of any doubt as to the construction we intend: the Cartesian components of the VSH in the body-fixed frame are obtained using the exact same formal definitions as for the Cartesian components of the VSH in the inertial frame that are given in \appref{app:vectorSphericalHarmonics}, by replacing $ (\theta,\phi) \rightarrow  (\tilde{\theta},\tilde{\phi})$.
    The difference between the two constructions is of course that the Cartesian components in the body-fixed frame are defined with respect to a set of basis vectors that rotate in the inertial frame.
} %
In the body-fixed frame, which is, of course, the most convenient frame to use to compute fields measured at stations fixed to the surface of the rotating Earth, the observable signal at $\tilde\Omega=(\tilde{\theta},\tilde{\phi})$ is thus given by the real part of
\begin{align}
    \bm{B}(\tilde{\Omega},t)&=\sqrt{\frac\pi3}\varepsilon m_{A'}^2R \sum_{m=-1}^1A'_m\bm{\tilde{\Phi}}_{1m}(\tilde\Omega)e^{-i(m_{A'}-2\pi f_dm)t}.
    \label{eq:signal}
\end{align}

A comment on the temporal coherence of this signal [\eqref{signal}] is in order; see also \secref{sec:phenomenologyOverview}.
Thus far, we have assumed an exactly monochromatic oscillatory time dependence $\sim e^{-im_{A'}t}$ for the dark-photon background field; this dependence leads directly to the exactly monochromatic magnetic field signal $\sim e^{-im_{A'}t}$.
In reality, the dark-photon field is the vector sum of multiple plane-wave components that have both an average speed and a velocity dispersion on the order of $v_{\textsc{dm}}\sim 10^{-3}$.
As a result, the dark-photon field can be treated as essentially monochromatic only on timescales up to the coherence time $T_\text{coh}\sim 2\pi/(m_{A'} v_{\textsc{dm}}^2) \sim 10^6 T_{\text{osc}}$, where $T_{\text{osc}}$ is the dark-photon oscillation period (see, e.g., \citeR[s]{Graham:2013gfa,Chaudhuri:2014dla}).
For our mass range of interest, we have $T_\text{coh}\sim 2$--$45\,$yr.
Therefore, as written, \eqref{signal} is applicable for times $t\lesssim T_\text{coh}$; both the temporal phase and polarization of the signal will be randomized on timescales $\gtrsim T_\text{coh}$.

Because $T_{\text{coh}} > 1\,\text{yr}$, the motion of the Earth around the Sun takes place within the same coherence patch of the dark-photon field, and so we expect side-bands in the signal at frequencies $f = f_0 \pm 1/(\text{yr})$ where $f_0 = m_{A'}/(2\pi)$. 
Moreover, again because $T_{\text{coh}}>1\,\text{yr}$, even given a single coherence time worth of data, these side-bands are in principle resolvable outside the intrinsic $\Delta f / f_0 \sim v_{\textsc{dm}}^2 \sim 10^{-6}$ width of the main signal at $f=f_0$.
However, it is straightforward to see that the amplitude of the side-bands is much smaller than the amplitude of the signal at $f=f_0$: because the spatial gradients of the field are only probed by the Earth's motion around the Sun over length-scales $\sim \text{AU}$, while the dark-photon field has $\mathcal{O}(1)$ fractional spatial gradients only on length-scales $\sim 1 / (m_{A'} v_{\textsc{dm}})$, the fractional side-band amplitude can be estimated as $\sim ( 1\,\text{AU} ) \times (m_{A'} v_{\textsc{dm}}) \lesssim 5\times 10^{-2}$ for $m_{A'}\lesssim 7\times 10^{-17}\,\text{eV}$, assuming $v_{\textsc{dm}}\sim10^{-3}$.
Additional side-bands at $f = f_0 \pm 1/(\text{day})$ would appear owing to the rotation of the Earth causing the individual stations to probe the dark-photon field gradients, but they are even more severely suppressed by $\sim R\times (m_{A'}v_{\textsc{dm}})$; note that this is separate from the rotational effects on the vectorial orientation of the signal that are accounted for at \eqrefRange{phaseRot}{signal}.

\subsection{Earth model 2: Interplanetary medium as boundary}
\label{sec:generalEarthTheory}

In this subsection, we consider our second, less idealized model for the electrical environment near the Earth, in which we discard the assumption from \secref{sec:earthTheory} that the ionosphere is an idealized spherical surface on which the active mode is damped effectively.
Instead, the model is now as follows:
we continue to take the inner boundary of the region of interest for the computation of the dark-matter-induced magnetic field signal to be a spherical ball of infinite conductivity slightly interior to the surface of the Earth, which effectively damps the active mode.
The outer boundary of the region of interest is, however, now taken to be the aspherical magnetopause (see \secref{sec:magnetosphere}), which marks the onset of the interplanetary medium where the plasma frequency is high and the active mode is damped effectively.

We account for the asphericity of the outer boundary in our computation for this model, but we show that it does not significantly impact the signal so long as $m_{A'}L \ll 1$, where $L$ is the characteristic radial distance to the magnetopause, as measured from the center of the Earth.
In the worst case scenario, the magnetopause can extend as far as $L\sim200R$ (with $R$ still the Earth radius) in the direction downwind of the Earth with respect to the flow of the solar wind.
Since we consider $m_{A'} \lesssim 7\times 10^{-17}\,$eV, then at worst we have $m_{A'} L\lesssim 0.5$, which might be slightly marginal at this upper end of our mass range with this worst-case value of $L$.
In the best-case scenario, the magnetopause is only $L\sim 10R$ distant in the upwind direction; then $m_{A'}L \lesssim 0.02 \ll 1$ throughout our mass range of interest.
As such, we work in the $m_{A'}L \ll 1$ limit, as it applies over the majority of our mass range, and in all but the worst-case assumption about the value of $L$ that should be used. 

The result of our computation of the leading-order magnetic field in this section will show that the TM contribution (in inertial coordinates) is still given precisely by \eqref{pre_signal} [which is easily modified to account for rotation to obtain \eqref{signal}], but that there are additional TE contributions to the leading-order magnetic field.
However, as these TE contributions involve different VSH components as compared to the TM contributions [cf.~\eqref[s]{BTE} and (\ref{eq:BTM})], they can be distinguished from each other globally, and it suffices to search for the TM signal.

We further argue at the end of this section that our calculation here captures all the relevant physics, and that our result is insensitive to the details of any additional varying conductive regions in the gap between the surface of the Earth and the magnetopause.

The argument in this subsection will proceed as follows.
First, we show that regardless of the shape of the boundaries of effective shields for the active components in this second model now under consideration, the same leading-order electric field result derived for the first model in \secref{sec:earthTheory} applies, up to corrections at $\mathcal{O}(x_0^2)$ where $x_0=m_{A'}L$ with $L$ defined as above.
We recall that this is a TM-type electric field, with a leading term at $\mathcal{O}(x_0^0)$, and we will show that the leading TE-type contributions could only possibly appear at $\mathcal{O}(x_0^2)$.

Second, we examine the implications of these realizations for the magnetic field.
Performing a consistent perturbative expansion of the fields in powers of $x_0$ and applying Maxwell's equations, we show that the relative power counting for TM and TE field modes behaves differently: for integer $n$, there is a fixed relationship between the TE electric field at $\mathcal{O}(x_0^n)$ and the TE magnetic field at $\mathcal{O}(x_0^{n-1})$, while a TM electric field at $\mathcal{O}(x_0^n)$ has a fixed relationship to a TM magnetic field at $\mathcal{O}(x_0^{n+1})$.
As such, because the leading electric field is TM and $\mathcal{O}(x_0^0)$, it uniquely fixes the leading TM magnetic field in this second model at $\mathcal{O}(x_0^1)$ to again be the same as that which was found for the first model, \eqref{pre_signal}. 
However, the leading TE part of the magnetic field, necessarily at $\mathcal{O}(x_0^1)$ because the leading TE electric field can only appear at $\mathcal{O}(x_0^2)$, is not fixed by this argument since it requires knowledge of the TE electric field at $\mathcal{O}(x_0^2)$, which we will not compute.
However, for the reason noted in the previous paragraph, there is actually not a strong need to find this part of the magnetic field: it can be distinguished globally from the TM mode, and the latter can be searched for alone.

Before continuing to the computation proper, we note that since we will still be interested in calculating the magnetic field as measured at locations on the surface of the Earth, which we model as a sphere of fixed radius $R$, it is appropriate to continue to work in spherical coordinates and employ VSH decompositions of the electric and magnetic fields, even though the outer boundary of the region of interest is no longer spherical in this model.

As in \secref{sec:earthTheory}, we decompose the homogeneous electric and magnetic fields into TE and TM contributions
\begin{align}
    \bm{E}_\text{hom}&=\bm{E}_\text{TE} + \bm{E}_\text{TM},\\
    \bm{B}_\text{hom}&=\bm{B}_\text{TE} + \bm{B}_\text{TM},
\end{align}
whose forms are defined by \eqrefRange{ETE}{BTM}, although the coefficient functions $f_{\ell m}$ and $g_{\ell m}$ of course differ in principle in this case as compared to those in \secref{sec:earthTheory}.
Let us define $\xi=r/L$, where $L$ is taken to be the largest radial distance from the center of the Earth to the magnetopause ($L\sim 200 R$), and fix the same value of $L$ in the definition of the scale $x_0 = m_{A'}L$, so that $m_{A'}r=x_0\xi$.
Since the largest radial dimension in the problem is $L$, we have $\xi\lesssim 1$ for all relevant locations interior to the magnetopause, while $x_0$ is a fixed small parameter (see discussion above) that we can use as the parameter in formal power series expansions of the functions $f_{\ell m}$ and $g_{\ell m}$: 
\begin{align}
    f_{\ell m}(m_{A'} r)&\equiv \sum_{n=0}^{\infty} x_0^nf_{\ell m}^{(n)}(\xi), \label{eq:flmExp}\\
    g_{\ell m}(m_{A'} r)&\equiv \sum_{n=0}^{\infty} x_0^ng_{\ell m}^{(n)}(\xi).\label{eq:glmExp}
\end{align}
Substituting \eqref[s]{flmExp} and (\ref{eq:glmExp}) into \eqrefRange{ETE}{BTM} yields a formal power series for the fields. 
For notational simplicity, let $\bm E_\text{TE}^{(n)}$ and $\bm E_\text{TM}^{(n)}$ be fields defined to have the same forms as those given in \eqref[s]{ETE} and (\ref{eq:ETM}), respectively, but with the following replacements made: $f_{\ell m}\rightarrow f_{\ell m}^{(n)}$, $g_{\ell m} \rightarrow g_{\ell m}^{(n)}$, and $m_{A'}r \rightarrow \xi$ (note: we do \emph{not} mean $m_{A'}r\rightarrow x_0\xi$; we account for powers of $x_0$ separately below).
Similarly, let $\bm B_\text{TE}^{(n)}$ and $\bm B_\text{TM}^{(n)}$ be defined with the same replacements to the expressions appearing at \eqref[s]{BTE} and (\ref{eq:BTM}), respectively.
The formal power series expansions of the fields can then be written as 
\begin{align}
    \bm E_\text{TE}&=\sum_{n=0}^{\infty} x_0^n \bm E_\text{TE}^{(n)},\\
    \bm B_\text{TM}&=\sum_{n=0}^{\infty} x_0^n \bm B_\text{TM}^{(n)}.
\end{align}

Similar na\"ive manipulations would yield
\begin{align}
    \bm E_\text{TM}&\stackrel{?}{=}\sum_{n=0}^{\infty} x_0^{n-1} \bm E_\text{TM}^{(n)}, \label{eq:ETMnaive} \\
    \bm B_\text{TE}&\stackrel{?}{=}\sum_{n=0}^{\infty} x_0^{n-1} \bm B_\text{TE}^{(n)},\label{eq:BTEnaive}
\end{align}
where $\bm{E}_\text{TM}^{(n)}$ is determined completely by taking a derivative of $\bm{B}_\text{TM}^{(n)}$, and $\bm{B}_\text{TE}^{(n)}$ is determined completely by taking a derivative of $\bm{E}_\text{TE}^{(n)}$; see \eqref[s]{ETM} and (\ref{eq:BTE}).
However, \eqref[s]{ETMnaive} and (\ref{eq:BTEnaive}) would appear to allow for TM electric fields and TE magnetic fields at $\mathcal{O}(x_0^{-1})$, arising from the $n=0$ terms.
But because the coefficients in $f_{lm}$ and $g_{lm}$ that are fixed by boundary conditions can have at most one power of $\varepsilon m_{A'}$ arising directly from the Lagrangian couplings [cf.~e.g., \eqref[s]{am} and (\ref{eq:bm})],%
\footnote{\label{ftnt:scalingArgument}%
    This scaling is clear from the background current approach (see \appref{app:effectiveCurrent}): the dark-photon-sourced background current is $J \sim \varepsilon m_{A'}^2 A'$ which, in the long-wavelength ($m_{A'}L\ll 1$) limit, sources an inhomogeneous electric field $E\sim m_{A'}^{-1} J \sim ( \varepsilon m_{A'} ) A'$. 
    This inhomogeneous field fixes all of the homogeneous parts of the solution via boundary conditions; since electric field superposition is linear, all homogeneous field components thus have a single power of $( \varepsilon m_{A'} )$.
    Any additional powers of $m_{A'}$ must appear with a length-scale $\sim m_{A'}L = x_0$.
} %
any physical field component $\propto x_0^{-1}$ would have a piece that either diverges or fails to go to zero as $m_{A'} \rightarrow 0$; however, it is a well-known fact~\cite{Holdom:1985ag}, and clear from the interaction-basis Lagrangian, \eqref{interactionBasisLagrangianLinear}, that all physical effects of the dark photon must decouple as $m_{A'} \rightarrow 0$ for fixed $A_m'$. 
As a result, it must be the case that $\bm{E}_\text{TM}^{(0)}=0$ and $\bm{B}_\text{TE}^{(0)}=0$, and so the correct expressions are:
\begin{align}
    \bm E_\text{TM}&=\sum_{n=0}^{\infty} x_0^{n} \bm E_\text{TM}^{(n+1)},\\
    \bm B_\text{TE}&=\sum_{n=0}^{\infty} x_0^{n} \bm B_\text{TE}^{(n+1)}.
\end{align}

Synthesizing this, the full homogeneous electric and magnetic fields contributions at $\mathcal{O}(x_0^n)$ for $n=0,1,\ldots$ are given by
\begin{align}
    \bm E_\text{hom}^{(n)}&=\bm E_\text{TE}^{(n)}+\bm E_\text{TM}^{(n+1)},\\
    \bm B_\text{hom}^{(n)}&=\bm B_\text{TE}^{(n+1)}+\bm B_\text{TM}^{(n)}.
\end{align}

We will now argue that the leading-order homogeneous electric field is the same as one derived in \secref{sec:earthTheory}.
Recall that in the background-current approach that is applicable in the limit $\varepsilon \ll 1$, a dark-photon field can be treated as an effective current $\bm{J}_{\text{eff}}$ that sources an inhomogeneous electric field component that is given in Earth-centered inertial spherical coordinates by
\begin{align}
    \bm E_\text{inh}=\sqrt{\frac{4\pi}3}i\varepsilon m_{A'}\sum_{m=-1}^1A'_m(\bm Y_{1m}+\bm\Psi_{1m})e^{-im_{A'}t}.
\end{align}
Nothing about that argument depends on the geometry of the boundaries of the lower atmospheric `cavity'.

Where the geometry of the cavity does enter is in fixing the homogeneous part of the electric field by virtue of the boundary conditions that, for boundaries where the active mode is efficiently damped, fix the components of the electric field in the tangent plane to the boundary surface to be zero everywhere on that surface.
We argue that it is actually possible to find a homogeneous field solution that not only cancels the in-tangent-plane components of $\bm E_\text{inh}$ at the boundaries of the region of interest, but (up to sub-leading corrections) actually cancels $\bm E_\text{inh}$ \emph{everywhere inside the cavity, including on the boundaries, regardless of the boundary geometry}. 
Because this is \emph{one solution} to Maxwell's equations that satisfies the boundary conditions, uniqueness theorems then dictate that it is \emph{the only solution}, at least up to sub-leading corrections.

Let us see how this works: using the VSH curl properties in \appref{app:vectorSphericalHarmonics}, it is easy to see that $\nabla\times\left( r \bm\Phi_{1m}\right) = - 2 ( \bm Y_{1m}+\bm\Psi_{1m} )$; therefore, $\bm E_\text{inh}$ can be written as
\begin{align}
    \bm E_\text{inh}&=-\sqrt{\frac\pi3}i\varepsilon m_{A'} A'_m\sum_{m=-1}^1\nabla\times\left( r \bm\Phi_{1m}\right)e^{-im_{A'}t} \\
    &= -\sqrt{\frac\pi3}i\varepsilon A'_m\sum_{m=-1}^1\nabla \times\left( x_0 \xi \bm\Phi_{1m}\right)e^{-im_{A'}t}. \label{eq:imhomogAlt}
\end{align}
This form of the inhomogeneous solution is suggestive of an $\ell = 1$ TM electric field solution to the \emph{homogeneous} equation, which suggests that we may be able to arrange the cancellation noted above using an $\ell = 1$ homogeneous TM electric field.

To make this precise, we must return to the TM electric field definition at \eqref{ETM}, and recall that we have more information about the function $g_{\ell m}$ appearing in that definition than simply its power series expansion in terms of $x_0$ given at \eqref{glmExp}.
In particular, $g_{\ell m}$ must be a linear combination of spherical Bessel and spherical Neumann functions, which we can write as [cf.~\eqref{fmDefn} for $\ell =1$]
\begin{align}
    g_{\ell m}(x) \equiv a_{\ell m} j_{\ell}(x) + b_{\ell m} x_0^{2\ell+1} y_{\ell}(x).
\end{align}
Taking $x = m_{A'}r = x_0 \xi$, expanding the coefficients as
\begin{align}
    a_{\ell m} &= \sum_{p=0}^{\infty} x_0^p\cdot a_{\ell m}^{(p)}, &
    b_{\ell m} &= \sum_{p=0}^{\infty} x_0^p\cdot b_{\ell m}^{(p)},
\end{align}
and using standard power-series expansions%
\footnote{\label{ftnt:series}%
    Specifically, we use the series expansion for the cylindrical Bessel function $J_\nu$ given at Eq.~8.402 in \citeR{GradshteynRyzhik} for $|\text{arg}\, z<\pi|$, 
    \begin{align*}
        J_\nu(z) &= z^\nu \sqrt{\frac{2}{\pi}}  \sum_{k=0}^\infty C_{\nu}^{k} z^{2k}; & 
        C_{\nu}^{k} &\equiv \frac{\sqrt{\pi}(-1)^k 2^{-2k-\nu-1/2}}{\Gamma[k+1]\Gamma[\nu+k+1]},
    \end{align*}
    along with the definitions~\cite{ArfkenWeber} $j_{\ell}(x) \equiv \sqrt{\pi/(2x)}\, J_{\ell+1/2}(x)$ and $y_{\ell}(x) \equiv (-1)^{\ell+1} \sqrt{\pi/(2x)}\, J_{-\ell-1/2}(x)$.
} %
for the $j_{\ell}$ and $y_{\ell}$, it is reasonably straightforward to show that 
\begin{align}
    &g_{\ell m}(m_{A'}r)\label{eq:glmFull}\\
    &= x_0^{\ell} \sum_{k=0}^{\infty} \sum_{p=0}^{\infty} x_0^{p+2k}
      \lb[
        a_{\ell m}^{(p)} c^{a}_{\ell k} \xi^{\ell+2k} 
        + b_{\ell m}^{(p)} c^{b}_{\ell k}  \xi^{-\ell-1+2k} 
\rb],\nonumber
\end{align}
where $c^{\{a,b\}}_{\ell k}$ are known numerical coefficients.%
\footnote{\label{ftnt:CaCb}%
    These can be read directly from the series expansion and definitions in footnote \ref{ftnt:series} and are $c^{a}_{\ell k} \equiv C_{\ell +1/2}^{k}$ and $c^{b}_{\ell k} \equiv C_{-l-1/2}^{k}$.
} %

We need to read off one specific result from \eqref{glmFull}:
\begin{align}
    g_{1m}^{(1)}(\xi) &= \frac{a_{1m}^{(0)}}{3} \xi - b_{1m}^{(0)} \xi^{-2};
\end{align}
since $g_{1m}^{(1)}$ contains a term that is $\propto \xi$, we can indeed exactly cancel $\bm E_\text{inh}$ everywhere at order $x_0^0$ using an $\ell = 1$ $\bm{E}_{\text{TM}}^{(1)}$ electric field for which we have set [cf.~\eqref[s]{am} and (\ref{eq:bm})]
\begin{align}
    a_{1m}^{(0)} &= \sqrt{3\pi} i\varepsilon m_{A'} A_m', & b_{1m}^{(0)} &= 0,\label{eq:coeffSoln0}
\end{align}
where we included a factor of $m_{A'}$ in $a_{1m}^{(0)}$ that arises from the leading $m_{A'}^{-1}$ in the definition at \eqref{ETM}.

In other words, the leading-order homogeneous electric field that exactly cancels the inhomogeneous field everywhere in the cavity volume (including on the boundary surfaces) is given by 
\begin{align}
    \bm E_\text{hom}^{(0)}&= \bm{E}_{\text{TM}}^{(1)}
    = \sqrt{\frac\pi3}i\varepsilon A'_m\sum_{m=-1}^1\nabla \times\left( r \bm\Phi_{1m}\right)e^{-im_{A'}t}.\label{eq:Ehom}
\end{align}
Moreover, $g_{\ell m}^{(1)}(\xi)=0$ for all $\ell\neq1$; and since no $\bm{E}_{\text{TE}}^{(0)}$ field was required, $f_{\ell m}^{(0)} = 0$ for all $\ell,m$.
To avoid confusion, we emphasize that this $\bm E_\text{hom}^{(0)}$ field means that the \emph{total} electric field $\bm{E_}{\text{tot}} = \bm{E}_\text{inh} + \bm{E}_\text{hom}$ has no term $\propto x_0^0$: it has been exactly canceled out between the inhomogeneous and homogeneous parts of the solution.

It remains to understand the order in $x_0$ at which corrections to this leading-order homogeneous electric field appear.
Irrespective of the boundary geometry, it is entirely consistent with the boundary conditions and the existing lower-order field solution $\bm{E}_{\text{hom}}^{(0)}$ that was required to cancel the inhomogeneous field, to set $a_{\ell m}^{(1)} = b_{\ell m}^{(1)} = 0$ in \eqref{glmFull} for all $\ell, m$, which sets $\bm{E}_{\text{TM}}^{(2)}=0$, and also to then independently set $\bm{E}_{\text{TE}}^{(1)} = 0$ such that $\bm{E}_{\text{hom}}^{(1)} = 0$ everywhere, including on the boundary.
Formally, the reason that this is possible is that the formal power series at \eqref{glmFull} skips orders of $x_0$ in the sum over $k$ (i.e., $\sum_k x_0^{2k} [\,\cdots]$ appears); this fact ultimately arises from a property of the spherical Bessel function power series expressions (and thus also holds for the cognate $f_{\ell m}$ functions).

It is not, however, consistent to simply zero out the higher-order corrections at $\bm{E}_{\text{hom}}^{(2)}$: we already know that an $\bm{E}_{\text{TM}}^{(3)}$ component to the homogeneous solution exists by virtue of the fact that $a_{1 m}^{(0)} \neq 0$.%
\footnote{\label{ftnt:detail}%
    This is because the $p=0$, $k=1$, $\ell=1$ term in \eqref{glmFull} depends on $a_{1 m}^{(0)}$ and contributes to $g_{1 m}^{(3)}$, which fixes $\bm{E}_{\text{TM}}^{(3)}$.
} %
There are, however, other electric field contributions at $\mathcal{O}(x_0^2)$ [e.g., the TM modes from the terms in \eqref{glmFull} with $p=1,\ell=2,k=0$ or $p=0,\ell=3,k=0$; and also TE modes] that are available to satisfy the boundary condition that the full $\mathcal{O}(x_0^2)$ field components in the tangent plane to the boundary must be canceled on all the cavity boundaries.
For general boundary geometries that lack spherical symmetry, engineering that cancellation will require both TM and TE fields at multiple $\ell,m$ values.
We do not attempt to calculate these corrections in closed form (the problem is in general analytically intractable); for reasons to become clear below, it suffices for us to have shown that $\bm{E}_{\text{hom}}^{(1)} = 0 = \bm{E}_{\text{TM}}^{(2)} = \bm{E}_{\text{TE}}^{(1)}$, but we note that the next-order fields are generally non-zero.

Let us now understand the implications of these observations for the magnetic field.
We have seen that $\bm E_\text{TM}^{(1)}$ is given by \eqref{Ehom}, and that $\bm{E}_{\text{TM}}^{(0)}$, $\bm E_\text{TE}^{(0)}$, and $\bm E_\text{TE}^{(1)}$ must vanish.
The leading order contribution to the magnetic field will thus be
\begin{align}
    \bm B_\text{hom}^{(1)}=\bm B_\text{TE}^{(2)}+\bm B_\text{TM}^{(1)}.
\end{align}
Since $\bm{E}_{\text{TM}}^{(1)}$ fixes $\bm{B}_{\text{TM}}^{(1)}$ uniquely and we know the form of $\bm{E}_{\text{TM}}^{(1)}$, we can compute
\begin{align}\label{eq:asphersignal}
    \bm B_\text{TM}^{(1)}=\sqrt{\frac\pi3}\varepsilon m_{A'}\xi\sum_{m=-1}^1A'_m\bm\Phi_{1m}e^{-im_{A'}t}.
\end{align}

In general, $\bm B_\text{TE}^{(2)}$ is non-zero for all except the simplest spherically symmetric boundary geometries because $\bm{E}_{\text{TE}}^{(2)}\neq0$ generally, and this gives another contribution to $\bm B_\text{hom}^{(1)}$.
However, TE and TM magnetic fields have different spatial patterns globally, and so it is in principle possible to distinguish these contributions to the signal with sufficient sampling of the field.
We are satisfied that \emph{a} signal given by \eqref{asphersignal} exists regardless of the boundary geometry; because it can be distinguished from any possible additional signal that may or may not appear depending on the geometry of the boundary, a search can target the signal \eqref{asphersignal} independent of the boundary-shape-dependent additional field.

Reintroducing the factor of $x_0$ from the power-series expansion, and evaluating \eqref{asphersignal} at $r=R$, we find that the TM part of the $\mathcal{O}(x_0)$ magnetic field is
\begin{align}
    \bm{B}(\Omega,t)&=\sqrt{\frac\pi3}\varepsilon m_{A'}^2R\sum_{m=-1}^1A'_m\bm{\Phi}_{1m}(\Omega)e^{-im_{A'}t},
    \label{eq:pre_signal_aspherical}
\end{align}
which is, of course, identical to \eqref{pre_signal}, and will in turn thus lead to a final form of the leading TM part of the signal that is identical to \eqref{signal}; the calculation of \secref{sec:earthTheory} thus indeed gives the correct $\bm\Phi_{1m}$ components to leading order.

Importantly, the only place that $R$ enters in \eqref{pre_signal_aspherical} comes from the location where the magnetic field is measured, and not from the location of the inner boundary.
This implies that our result is relatively insensitive to the details of the interior conductivity profile of the Earth presented in \secref{sec:conductivityEarthInterior}.
In other words, regardless of what one considers the appropriate inner boundary (e.g., the Earth's surface, upper mantle, or lower mantle), the length-scale that appears in the leading-order magnetic result when evaluated on the surface of the Earth will still be the radius of the Earth.

Finally, we argue that the signal we have computed using the simplified model that we considered in this subsection would not be modified in geometries with more complicated conductivity profiles.
In particular, as long as the surface at which the magnetic field is measured lies in a vacuum,%
\footnote{\label{ftnt:finiteconductivity}%
    In \appref{app:finiteConductivty}, we consider finite conductivity effects, and find that the leading order result \eqref{BfiniteFinal} is not affected by a homogeneous isotropic conductivity at the measurement surface, so long as the skin-depth in the air gap is much longer than the radius of the Earth.
    We have also verified that a radially varying conductivity near the measurement surface does not affect the argument in this subsection.
    Therefore, even this assumption can be relaxed.
} %
all that is necessary to know is that the total electric field is order $\mathcal O(x_0^2)$ at this surface.
Given this condition, the leading-order homogeneous electric field at the measurement surface will still be given by \eqref{Ehom}, and the rest of the argument carries through. 
Such a condition should generically be expected for the physical case of interest, since the total electric field is known to vanish deep within the Earth and deep in the interplanetary medium.
These locations are separated by sub-wavelength scales $L\ll m_{A'}^{-1}$, and so we should generically expect the total electric field to grow at most quadratically in $x_0=m_{A'}L$ between them. 
Therefore, in particular, the total electric field at the Earth's surface, where we measure our signal, should be order $\mathcal O(x_0^2)$.

We note that there are some caveats to this argument.
There are many complicated details of the conductivity/electrical environment of the atmosphere and magnetosphere which we have not explicitly considered, which could in principle give rise to resonance effects that would allow the electric field to ring up in the gap, thereby invalidating the assumption that it is order $\mathcal O(x_0^2)$ at the surface of the Earth.
The lowest-frequency cavity resonances of the Earth--ionosphere cavity---the so-called Schumann resonances~\cite{Schumann+1952+149+154,bliokh1980schumann,Satori2009}---are well-studied; the lowest observed resonance appears at $f_{\textsc{s}} \sim8\,\text{Hz}$~\cite{bliokh1980schumann} [this is approximately $f_{\textsc{s}} \sim 1/ (2\pi R) \gg 1/(2\pi m_{A'})$], which lies well above the upper end of our frequency range of interest.
Moreover, the cognate lowest-frequency mode which one could imagine occurring in the Earth--magnetopause cavity would be at a frequency $f_{\textsc{E--M}} \sim 1/ (2\pi L ) \sim 1/ (400 \pi R )\sim 3\times 10^{-2}\,$Hz, which also still lies (marginally) above our frequency range of interest.

Obtaining a resonance at a frequency corresponding to a Compton wavelength larger than the geometrical size of a cavity requires elements to the cavity to act as an effective high-$Q$ lumped-element circuit.
Certain physical magnetohydrodynamic (MHD) processes do give rise to effective lumped-element behavior: the Ionospheric Alfv\'en Resonator (IAR) induces resonances in the 0.1--10\,Hz range~\cite{BELYAEV1990781,yahnin2003morphology,bosinger2004fine,Boesinger:2008eth,sanfui2016studies,Nose:2017wdf,NokesMSc}, while MHD ringing of the entire magnetospheric cavity can induce resonances in the mHz range~\cite{Troitskaya:1967efh,Pilipenko:1990dsg,Hughes:1994ouh,sanfui2016studies} (the Alfv\'en speed is $v_A \ll c$).
However, the conditions required for the existence of these resonances show strong diurnal variation, and their effects are also quite strongly spatially varying; moreover, they are not typically high-$Q$ resonances.
We therefore find it unlikely that the naturally occurring and noisy electromagnetic environment near the Earth could conspire to achieve a sufficiently strong, stable, and persistent resonance condition in our frequency range of interest so as to invalidate our modeling.
Furthermore, the existence of such a strong resonance with sufficient spatial and temporal overlap with our signal so as to be problematic would undoubtedly make itself known in the magnetic field data we have analyzed in~\citeR{Fedderke:2021qva}; we find no evidence for this.

In summary, we are reasonably confident that no strong resonances exist within the interesting range of frequencies where we search for a signal (see \secref{sec:experimentalSearch}) that could lead to large electric fields [i.e., $> \mathcal{O}(x_0^2)$] ringing up at the surface of the Earth.
As a result, we conclude that the model considered in this section appropriately captures the physics of interest, and that our signal is robust to neglected details of the near-Earth electrical environment.

\section{Search for signal in existing geomagnetic field data}
\label{sec:experimentalSearch}
The signal described in this work, \eqref{signal}, is a narrow-band oscillating magnetic field with a magnitude
\begin{align}
    B \sim 0.7\,\text{nG} \times \lb( \frac{\varepsilon}{10^{-5}} \rb)\times \lb( \frac{m_{A'}}{4\times 10^{-17}\,\text{eV}} \rb), \label{eq:signalAmplitude}
\end{align}
assuming the dark photon is all of the dark matter.%
\footnote{\label{ftnt:allofDM}%
    We take $A'=\sqrt{2\rho_{\textsc{dm}}}/m_{A'}$ with $\rho_{\textsc{dm}}= 0.3\,\text{GeV/cm}^3$.
} %
It exhibits a long coherence time and would appear in-phase across the entire surface of the Earth in unshielded magnetometers, with a specific vectorial spatial pattern.

A close-to-ideal experimental setup to detect such a signal would thus be a network of geographically dispersed, unshielded three-axis magnetometer stations that each measure the ambient magnetic field at the location of the station as a function of time, and report those time-stamped data over long periods of time.
Serendipitously, exactly such a network of detectors has been operating in this fashion for decades for the purposes of, among other things, geophysical metrology: the SuperMAG Collaboration~\cite{Gjerloev:2009wsd,Gjerloev:2012sdg,SuperMAGwebsite} (see also \citeR[s]{Mann:2008wrb,Chi:2013sdn,Engebretson:1995gna,Yumoto:2001sfb,Tanskanen:2009wdg,Love:2013fbq,Lichtenberger:2013adg,1992NASCP3153..321H,INTERMAGNETmanual}) maintains a public database~\cite{SuperMAGwebsite} of three-axis magnetic field time series data taken with a one-minute time resolution (`cadence') at $\mathcal{O}(10^2)$ stations---dispersed across every continent, and on islands in most of the major oceans---since the 1970s; these data are reported in a common format, and with common reference system conventions.

\begin{figure*}[t]
\includegraphics[width=0.9\textwidth]{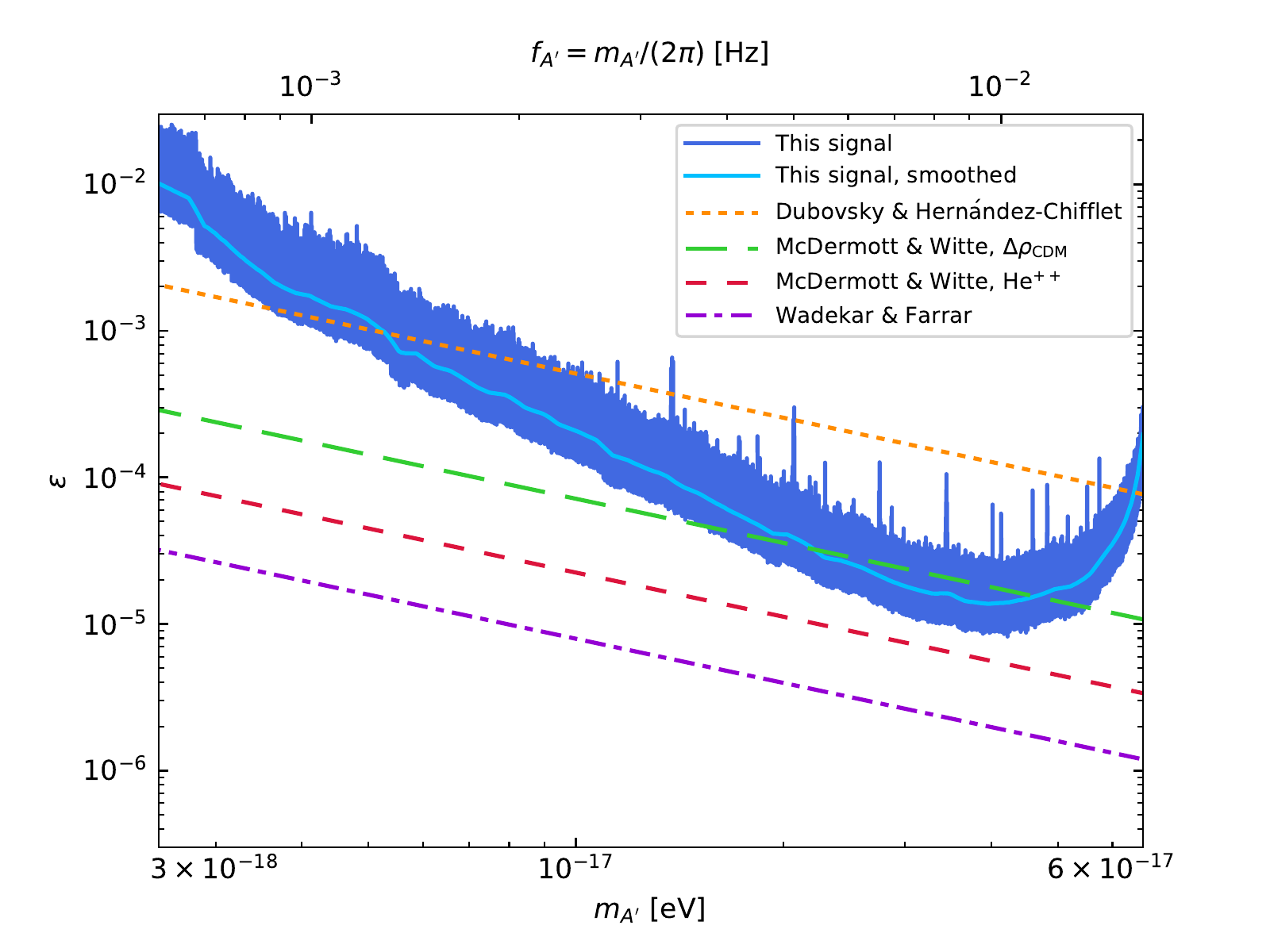}
\caption{\label{fig:limit}%
    The 95\%-credible upper limits on the kinetic mixing parameter $\varepsilon$ that result from the experimental search outlined in \secref{sec:experimentalSearch} (and detailed in our companion paper~\cite{Fedderke:2021qva}) are shown as the dark blue line as a function of the dark-photon mass $m_{A'}$ (the line appears as a band owing to the density of frequencies at which limits are plotted, and fluctuations of the limits from one frequency to the next).
    The sliding average of these limits over nearby frequencies are shown as a lighter blue line as a guide to the eye. 
    These limits assume that the dark photon is all of the dark matter.
    Also shown are a variety of existing astrophysical limits arising from heating effects of dark-photon dark matter on gas in various astrophysical settings: the ionized interstellar medium in the Milky Way (dotted orange line)~\cite{Dubovsky:2015cca}; the intergalactic medium around helium reionization (short-dashed red line, labeled `$\text{He}^{++}$')~\cite{McDermott:2019lch}; and gas in the Leo T dwarf galaxy (dot-dashed purple line)~\cite{Wadekar:2019xnf}.
    We also show a DM-depletion limit from non-resonant dark-photon--photon conversion~\cite{McDermott:2019lch} (long-dashed green line, labeled `$\Delta \rho_{\textsc{cdm}}$').
    }
\end{figure*}

In our companion paper~\cite{Fedderke:2021qva}, we undertake a detailed analysis of the SuperMAG data for the signal described by \eqref{signal}.
We summarize our approach and results here:
our analysis effectively projects the components of the three-axis magnetic field time series measurements from the $\mathcal{O}(\text{few} \times 10^2)$ individual stations'~measurements onto a small number of global time series variables that appropriately describe the VSH pattern of the signal, \eqref{signal}.
We then analyze these time series variables coherently across chunks of data of temporal duration equal to the signal coherence time, by first Fourier analyzing each such chunk, and then searching for narrow-band excess power in the frequency domain (this is equivalent to a matched-filter search in the time domain for a monochromatic signal).
Where relevant, we incoherently combine the results from multiple such coherence times, using a Bayesian analysis framework to take into account the stochastic fluctuations from one coherence time to the next of the magnitude and polarization state of dark-photon dark matter which \emph{on average} constitutes all of the local dark matter; this mildly degrades the signal sensitivity (see, e.g., \citeR[s]{Centers:2019dyn,roussy2021experimental,Lisanti:2021vij,Gramolin:2021mqv} for discussion of similar procedures applied to axions). 
Marginalizing over irrelevant signal parameters [including the spatial orientation (i.e., polarization state) of the dark-photon field, which we did not fix \emph{a priori}] and taking a reparametrization--invariant Jeffreys prior on the kinetic mixing parameter $\varepsilon$, we obtain the posterior distribution on $\varepsilon$ at each frequency at which we search over the range $6\times 10^{-4}\Hz \lesssim f_{0} \lesssim 2\times 10^{-2}\Hz$, roughly corresponding to $2\times 10^{-18}\eV \lesssim m_{A'} \equiv 2\pi f_0 \lesssim 7\times 10^{-17}\eV$.

As we report in \citeR{Fedderke:2021qva}, from these posteriors we identified $\mathcal{O}(30)$ na\"ive signal candidates (some of these candidates are visible by eye as narrow peaks above the dark blue exclusion band in \figref{fig:limit}).
However, on the basis of further careful resampling checks carried out on data subsets to test for the temporal constancy and/or spatial uniformity of these na\"ive signal candidates, we conclude that none of them constitute a robust dark-photon dark-matter signal candidate: we report no robust candidate signals of dark-photon dark matter in the SuperMAG dataset on the basis of our analysis~\cite{Fedderke:2021qva}.
SuperMAG has also recently released data taken with a higher cadence of one second, but over a shorter total time period (a little over ten years), and from a smaller total number of stations~\cite{SuperMAGwebsite}; it would be interesting to revisit this analysis with those data, as they would enable access to higher dark-photon masses (frequencies).

With no robust signal candidates identified in the one-minute-cadence dataset that we analyzed, we use the posterior distributions on $\varepsilon$ to extract 95\%-credible upper limits (local significance) on $\varepsilon$ as a function of $m_{A'}$; see \figref{fig:limit} and \citeR{Fedderke:2021qva} for our results.

Our exclusion results are competitive with, or complementary to, various astrophysical bounds on dark-photon dark matter~\cite{Dubovsky:2015cca,McDermott:2019lch,Wadekar:2019xnf},%
\footnote{\label{ftnt:otherRefs}%
    Bounds similar to those in \citeR{Wadekar:2019xnf} also appear in \citeR{Bhoonah:2018gjb} (with the exception of the much stronger bound explicitly marked as `preliminary' in the later reference that arises from a gas cloud of anomalously low, and somewhat disputed, temperature; see also the discussion in \citeR[s]{Farrar:2019qrv,Bhoonah:2018wmw}).
    Some questions have, however, been raised as to the validity of the bounds in \citeR{Bhoonah:2018gjb} owing to their being set using observations of gas clouds that are close to the center of the Milky Way and part of a large outflow of gas~\cite{Wadekar:2019xnf}, although these concerns were addressed in a note added in \citeR{Bhoonah:2018gjb}.
    We take no position on this point of debate and note only that the strongest non-preliminary bounds in \citeR{Bhoonah:2018gjb} are similar, within an $\mathcal{O}(1)$ factor, to those in \citeR{Wadekar:2019xnf}.
}\up{,}%
\footnote{\label{ftnt:limitupdate}%
    Per \citeR{WadekarPrivate2021}, the limits in the published version of \citeR{Wadekar:2019xnf} are slightly weaker than those shown in earlier \texttt{arXiv} versions of the latter, owing in part to updates to published gas metallicities for Leo T.
} %
and represent direct terrestrial laboratory exclusions of dark-photon dark-matter parameter space that are not subject to significant astrophysical uncertainties.
Future cosmological bounds from 21\,cm observations in this mass range are also expected to be highly competitive~\cite{Kovetz:2018zes}; however, in light of the uncertainty around the EDGES global 21\,cm anomaly~\cite{Bowman:2018sdg}, we do not display these limits in \figref{fig:limit}.
Moreover, future analyses of the higher-cadence SuperMAG data, as well as possible future experiments looking for our signal at even higher frequencies, could allow access to regions of parameter space significantly beyond current constraints.

We also note that our limits are set under the assumption that the dark-photon dark matter has $\rho_{\textsc{dm}} = 0.3\,\text{GeV/cm}^3$ on average, and a velocity dispersion in the vicinity of the Earth of $\sim 10^{-3}$, as for a standard halo model (SHM; see, e.g., \citeR{Evans:2018bqy}).
If the local DM abundance has stream-like structures (see, e.g., \citeR[s]{Myeong:2017skt,myeong2018shards,Lancaster_2019,Malhan_2018,Meingast_2019,OHare:2019qxc}), then these limits could be too conservative.
In a stream, the DM velocity dispersion is typically smaller than in the SHM, narrowing the signal width; moreover, the DM abundance could be boosted by such streams.
Both of these effects would make a signal more detectable.

\section{Conclusion}
\label{sec:conclusion}

The dark photon is an interesting and well-motivated dark-matter candidate over a wide mass range.
Direct laboratory probes of ultralight dark-photon dark matter, however, often suffer severe signal suppression by ratios of laboratory length-scales to the (much larger) dark-matter Compton wavelength. 
As such, most existing constraints on the lightest region of parameter space rely on astrophysical observations.
In this work, we have presented a novel terrestrial signature of ultralight dark-photon dark matter that exploits the size of the Earth itself in order to alleviate this usual length-scale suppression.
We have shown that there exists a coherently oscillating, quasi-monochromatic magnetic field signal of dark-photon dark matter, \eqref{signal}, that has a specific global vectorial spatial pattern, and which is detectable near the surface of the Earth in unshielded magnetometer data. 
This signal would be present in principle for any dark photon in the mass range $10^{-21}\,\text{eV}\lesssim m_{A'}\lesssim 3\times 10^{-14}\,\text{eV}$ (see \secref{sec:conductivitySummary}).

The signal we have presented is obtained utilizing a somewhat simplified model of the electrical conductivity environment near the Earth, in which we model the poorly conductive lower atmosphere around the surface of the Earth as a region of vacuum sandwiched between two layers of material that efficiently absorb ordinary photons: the conductive Earth below, and either the ionosphere or the interplanetary medium above.
This geometry is reminiscent of the conductor geometries employed in laboratory-scale direct-detection experiments, such as ADMX and DM Radio (which bound higher-mass dark photons), in order to mediate the generation of an observable magnetic-field signal of dark-photon dark matter.
Those signals are suppressed by $m_{A'}L$, where $L$ is the characteristic geometrical size of the experiment.
One might have thus expected that the length-scale suppression in our scenario would involve the scale-height of the lower atmospheric conductivity gap $h \sim 10^2$\,km; we have, however, shown that the relevant length-scale that enters in our signal is the much larger radius of the Earth, $R \sim 6\times 10^3$\,km.
This observation significantly enhances the amplitude of the magnetic field signal.
We have also shown that our signal prediction is robust to whether the ionosphere acts as a conductive shield, and we have argued that it is insensitive to many of the detailed complexities of the real near-Earth environment that are elided in our modeling.

We presented in this work the results of a search for the magnetic field signal \eqref{signal}, the details of which appear in our companion paper~\cite{Fedderke:2021qva}.
This analysis employs a large public dataset of geomagnetic field measurement data maintained by the SuperMAG Collaboration~\cite{SuperMAGwebsite,Gjerloev:2009wsd,Gjerloev:2012sdg}, which comprise a time series of unshielded three-axis magnetic field measurements made at $\mathcal O(10^2)$ stations widely dispersed around the surface of the Earth, with a one-minute time resolution, beginning in the 1970s.
Our analysis finds no robust dark-photon dark-matter signal candidates, and as such is used to place upper limits on the kinetic mixing parameter $\varepsilon$ as a function of the mass of the dark-photon dark matter in the ultralight region of parameter space.
In particular, these data allow us to present limits in the mass range $2\times 10^{-18}\,\text{eV}\lesssim m_{A'} \lesssim 7\times 10^{-17}$\,eV.
These limits are shown in \figref{fig:limit}, and are complementary to existing astrophysical constraints in this mass range.

The SuperMAG Collaboration is currently in the process of releasing data with higher time resolution (one second, instead of one minute); we defer to future work an analysis of that dataset, which would extend the sensitivity of the search to higher masses (frequencies), and presumably strengthen the existing exclusion limits (absent a signal detection).
We also note that in principle our limits appear to improve relative to existing astrophysical gas-heating bounds moving to higher frequencies; this strongly motivates additional experimental exploration of this signal at frequencies above the range we have considered in this work, as this approach may allow access to parameter space that is technically difficult to probe with laboratory-scale experiments, $f \lesssim \text{kHz}$~\cite{Chaudhuri:2014dla}.
It may also be worthwhile exploring whether better sensitivity could be achieved with improved magnetometers.
Finally, it would be interesting to consider the cognate signal that would be expected to appear for axion-like (i.e., ALP) dark matter.%
\footnote{\label{ftnt:ArielArza}%
    We acknowledge ongoing discussions with Ariel Arza on this point.
} %

\acknowledgments

We thank Surjeet Rajendran, Dmitry Budker, and Alex Sushkov for enlightening conversations at early stages of this project.

M.A.F.~thanks the Berkeley Center for Theoretical Physics at the University of California Berkeley and Lawrence Berkeley National Laboratory for their long-term hospitality during which the earliest stages of this work were completed. 

M.A.F., P.W.G.,~and S.K.~were supported by DOE Grant No.~DE-SC0012012, NSF Grant No.~PHY-2014215, the Heising-Simons Foundation Grants No.~2015-037 and No.~2018-0765, DOE HEP QuantISED Award No.~100495, and the Gordon and Betty Moore Foundation Grant No.~GBMF7946.
This work was also supported by the U.S.~Department of Energy, Office of Science, National Quantum Information Science Research Centers, Superconducting Quantum Materials and Systems Center (SQMS) under contract No.~DE-AC02-07CH11359.
D.F.J.K.~was supported by NSF Grant No.~PHY-1707875 as well as the Simons and Heising-Simons Foundations. 
S.K.~was also supported by NSF Grant No.~DGE-1656518.

Some of the computing for this project was performed on the Sherlock cluster. 
We thank Stanford University and the Stanford Research Computing Center for providing computational resources and support that contributed to these research results.

We gratefully acknowledge the SuperMAG Collaboration for maintaining and providing the database of ground magnetometer data that were analyzed to present the limits in \figref{fig:limit} and \citeR{Fedderke:2021qva}, and we thank Jesper W.~Gjerloev for helpful correspondence regarding technical aspects of the SuperMAG data.
SuperMAG receives funding from NSF Grants No.~ATM-0646323 and No.~AGS-1003580, and NASA Grant No.~NNX08AM32G S03.

We acknowledge those who contributed data to the SuperMAG Collaboration: 
INTERMAGNET, Alan Thomson; 
CARISMA, PI Ian Mann; 
CANMOS, Geomagnetism Unit of the Geological Survey of Canada; 
The S-RAMP Database, PI K.~Yumoto and Dr.~K.~Shiokawa; 
The SPIDR database; AARI, PI Oleg Troshichev; 
The MACCS program, PI M.~Engebretson; 
GIMA; 
MEASURE, UCLA IGPP and Florida Institute of Technology; 
SAMBA, PI Eftyhia Zesta; 
210 Chain, PI K.~Yumoto; 
SAMNET, PI Farideh Honary; 
IMAGE, PI Liisa Juusola; 
Finnish Meteorological Institute, PI Liisa Juusola; 
Sodankylä Geophysical Observatory, PI Tero Raita; 
UiT the Arctic University of Norway, Troms\o\ Geophysical Observatory, PI Magnar G.~Johnsen; 
GFZ German Research Centre For Geosciences, PI J\"urgen Matzka; 
Institute of Geophysics, Polish Academy of Sciences, PI Anne Neska and Jan Reda; 
Polar Geophysical Institute, PI Alexander Yahnin and Yarolav Sakharov; 
Geological Survey of Sweden, PI Gerhard Schwarz; 
Swedish Institute of Space Physics, PI Masatoshi Yamauchi; 
AUTUMN, PI Martin Connors; 
DTU Space, Thom Edwards and PI Anna Willer; 
South Pole and McMurdo Magnetometer, PIs Louis J.~Lanzarotti and Alan T.~Weatherwax; 
ICESTAR; 
RAPIDMAG; 
British Antarctic Survey; 
McMac, PI Dr.~Peter Chi; 
BGS, PI Dr.~Susan Macmillan; 
Pushkov Institute of Terrestrial Magnetism, Ionosphere and Radio Wave Propagation (IZMIRAN); 
MFGI, PI B.~Heilig; 
Institute of Geophysics, Polish Academy of Sciences, PI Anne Neska and Jan Reda; 
University of L’Aquila, PI M.~Vellante; 
BCMT, V.~Lesur and A.~Chambodut; 
Data obtained in cooperation with Geoscience Australia, PI Marina Costelloe; 
AALPIP, co-PIs Bob Clauer and Michael Hartinger; 
SuperMAG, PI Jesper W.~Gjerloev; 
Data obtained in cooperation with the Australian Bureau of Meteorology, PI Richard Marshall.

We thank INTERMAGNET for promoting high standards of magnetic observatory practice~\cite{INTERMAGNETwebsite}.

\appendix

\section{Photon--dark-photon dynamics}
\label{app:systemBehavior}

In this Appendix, we will give a short review of some of the underlying theoretical issues at play for the kinetically mixed dark photon, discussing basis choices, and the propagating eigenmodes of the EM-photon--dark-photon system in regions of high and low conductivity as well as their relationship to the vacuum mass eigenstates of the system and the interacting eigenstates that couple to EM-charged matter.
See also \citeR[s]{Nelson:2011sf,Graham:2014sha,Chaudhuri:2014dla,Raffelt:1996wa}.

Consider a massive dark photon, $(A_{\textsc{k}}')_\mu$, that is kinetically mixed with the ordinary photon of electromagnetism, $(A_{\textsc{k}})_\mu$, with kinetic mixing parameter $\varepsilon$, which we assume to be small $(\varepsilon \ll 1)$ [cf.~\eqref{kineticallyMixedLagrangian}]:%
\footnote{\label{ftnt:basisLabel}%
    Because we will in this Appendix at least initially be discussing various basis choices with which to write \eqref{kineticallyMixedLagrangianApp}, and the relationships between these basis choices, we have written an explicit subscript~`$\textsc{k}$' on the kinetically mixed basis states.
}%
\up{,}%
\footnote{\label{ftnt:epsilonLimit}%
    If $\varepsilon = \pm 1$, then one can `complete the square' on the kinetic terms such that only one linear combination $\kin{F} \mp \primekin{F}$, respectively, has a kinetic term while the other linear combination has no kinetic term and is thus not a propagating mode.
    Therefore, even absent the assumption $\varepsilon \ll 1$, there is still a limited range of values in which $\varepsilon$ can be varied continuously away from $\varepsilon = 0$ and yield a theory with two independent propagating eigenmodes: $-1<\varepsilon<1$. 
    This restriction will also become manifest when we examine the interaction and mass bases.
} %
\begin{align}
    \LL &\supset - \frac{1}{4} \kin{F}_{\mu\nu}\kin{F}^{\mu\nu} - \frac{1}{4} \primekin{F}_{\mu\nu} \primekin{F}^{\mu\nu} \nl
    + \frac{\varepsilon}{2} \kin{F}_{\mu\nu} \primekin{F}^{\mu\nu} + \frac{1}{2} m_{A'}^2 \primekin{A}_\mu \primekin{A}^{\mu} \nl
    - J_{\textsc{em}}^\mu \kin{A}_\mu, \qquad \text{\footnotesize[kinetically mixed basis]}
    \label{eq:kineticallyMixedLagrangianApp}
\end{align}
where $F^{(\prime)}_{\mu\nu} \equiv \partial_{\mu} A^{(\prime)}_{\nu} - \partial_{\nu} A^{(\prime)}_{\mu}$ are the respective field strength tensors.
We refer to the basis in which \eqref{kineticallyMixedLagrangianApp} is written as the `kinetically mixed' basis.

\subsection{Basis choices (vacuum)}
\label{app:basisChoices}

While the kinetically mixed basis is convenient to write the Lagrangian from a theoretical perspective (because it makes manifest the `vector portal' nature of the coupling), it does not make the phenomenology of the system readily apparent.
Of course, while the physics is invariant to the basis choice, different basis choices are convenient for different applications and, at the level of the Lagrangian, there are two such common alternative bases employed to write \eqref{kineticallyMixedLagrangian}: the (vacuum) mass basis and the interaction basis.
The (vacuum) mass basis is reached via the non-unitary field redefinition
\begin{align}
    \begin{pmatrix}
    A_{\textsc{m}} \\[1ex]
    A_{\textsc{m}}'  
    \end{pmatrix}
    &= \begin{pmatrix}
        1 & -\varepsilon \\[1ex]
        0 & \sqrt{1-\varepsilon^2}
    \end{pmatrix}
    \begin{pmatrix}
   A_{\textsc{k}} \\[1ex]
   A_{\textsc{k}}'
    \end{pmatrix},
    \label{eq:kineticTOmass}
\end{align}
in terms of which we have
\begin{align}
    \LL &\supset - \frac{1}{4} \mass{F}_{\mu\nu}\mass{F}^{\mu\nu} - \frac{1}{4} \primemass{F}_{\mu\nu} \primemass{F}^{\mu\nu} \nl
    + \frac{1}{2} \frac{m_{A'}^2}{1-\varepsilon^2} \primemass{A}_\mu \primemass{A}^{\mu} \qquad \text{\footnotesize[vacuum mass basis]} \nl
    - J_{\textsc{em}}^\mu \lb[ \mass{A}_\mu + \dfrac{\varepsilon}{\sqrt{1-\varepsilon^2}} \primemass{A}_\mu \rb].  
    \label{eq:massBasisLagrangianApp}
\end{align}
It is clear that the (vacuum) mass basis modes are the propagating (i.e., momentum) eigenmodes in vacuum: the massless mode $\mass{A}$ and the massive mode $\primemass{A}$ are independent if $J_{\textsc{em}} = 0$.
However, a linear combination of the two vacuum mass basis modes couples to EM charges; the mass basis states are thus not interaction eigenstates.

The interaction basis is reached by a different non-unitary field redefinition:
\begin{align}
    \begin{pmatrix}
    A_{\textsc{i}} \\[1ex]
    A_{\textsc{i}}'  
    \end{pmatrix}
    &= \begin{pmatrix}
        \sqrt{1-\varepsilon^2} & 0 \\[1ex]
    -\varepsilon & 1 
    \end{pmatrix}
    \begin{pmatrix}
   A_{\textsc{k}} \\[1ex]
   A_{\textsc{k}}'
    \end{pmatrix},
    \label{eq:kineticTOinteraction}
\end{align}
in terms of which we have
\begin{align}
    \LL &\supset - \frac{1}{4} \inter{F}_{\mu\nu}\inter{F}^{\mu\nu} - \frac{1}{4} \primeinter{F}_{\mu\nu} \primeinter{F}^{\mu\nu} \nl
    + \frac{1}{2} m_{A'}^2 \lb[\begin{array}{l}
            \primeinter{A}_\mu \primeinter{A}^{\mu} + \dfrac{2\varepsilon}{\sqrt{1-\varepsilon^2}} \inter{A}^\mu \primeinter{A}_\mu \\[1ex]
            + \dfrac{\varepsilon^2}{1-\varepsilon^2} \inter{A}_\mu \inter{A}^\mu
            \end{array} \rb] \nl
    - \dfrac{1}{\sqrt{1-\varepsilon^2}} J_{\textsc{em}}^\mu \inter{A}_\mu.  \qquad \text{\footnotesize[interaction basis]}
    \label{eq:interactionBasisLagrangianApp}
\end{align}
It is clear that the interaction basis modes are the interaction eigenstates: the `interacting mode' $\inter{A}$ couples to EM charges; the `sterile mode' $\primeinter{A}$ does not.
However, the presence of the mass-mixing terms in \eqref{interactionBasisLagrangianApp} makes clear that the interaction basis states are not propagation eigenstates in vacuum.

The relationship between the vacuum mass basis (propagating eigenstates in vacuum) and the interaction basis (interaction eigenstates) is given by the unitary transformation
\begin{align}
    \begin{pmatrix}
    A_{\textsc{i}} \\[1ex]
    A_{\textsc{i}}'  
    \end{pmatrix}
    &= \begin{pmatrix}
        \sqrt{1-\varepsilon^2} & +\varepsilon \\[1ex]
    -\varepsilon & \sqrt{1-\varepsilon^2}
    \end{pmatrix}
    \begin{pmatrix}
   A_{\textsc{m}} \\[1ex]
   A_{\textsc{m}}'
    \end{pmatrix},
    \label{eq:massTOinteraction}
\end{align}
which shows explicitly that the propagation and interaction eigenstates do not coincide in the presence of kinetic mixing.

Note also that it is common in the literature for all the results shown in this subsection to be written with all effects at $\mathcal{O}(\varepsilon^2)$ omitted; we have retained them here for completeness.

\subsection{Equations of motion and in-medium effects}
\label{app:EoMandInMedium}

So far, we have demonstrated that the vacuum mass-basis states $A_{\textsc{m}}$ and $A'_{\textsc{m}}$ are indeed the vacuum propagation (i.e., momentum) eigenstates, but that these states differ from the interaction eigenstates.
As we will demonstrate in this section, in a conducting medium with linear response, these statements must be modified.

Straightforward application of the Euler-Lagrange equations to the kinetically mixed basis Lagrangian \eqref{kineticallyMixedLagrangianApp} yields the equations of motion
\begin{align}
    \partial_\mu\big[ \kin{F}^{\mu\nu} - \varepsilon \primekin{F}^{\mu\nu} \big] &= J_{\textsc{em}}^{\nu},  \label{eq:eomFkin1} \\
    \partial_\mu\big[ \primekin{F}^{\mu\nu} - \varepsilon \kin{F}^{\mu\nu} \big] &+ m_{A'}^2 \primekin{A}^{\nu} = 0.   \label{eq:eomFkin2}
\end{align}
Because the local $U(1)$ gauge transformation $\kin{A}_\mu \rightarrow \kin{A}_\mu + \partial_\mu \Lambda$ (for any function $\Lambda$) remains a good symmetry of \eqref{kineticallyMixedLagrangian} [assuming a conserved EM current, $\partial_\mu J^{\mu}_{\textsc{em}} = 0$], we are still free to assume the Lorenz gauge condition $\partial_\mu \kin{A}^\mu = 0$.
Equivalently, \eqref[s]{eomFkin1} and (\ref{eq:eomFkin2}) contain only $F_{\textsc{k}}$ and not $A_{\textsc{k}}$, and $F_{\textsc{k}}$ is, of course, invariant to this gauge transformation.
On the other hand, there is no gauge freedom associated with the other $U(1)$ that is broken by the explicit mass term in \eqref{kineticallyMixedLagrangianApp}; however, applying $\partial_\nu$ to both sides of \eqref{eomFkin2} and recalling that $\partial_{\nu}\partial_{\mu}$ is symmetric on its indices while both field strength tensors $(F_{\textsc{k}}^{(\prime)})^{\mu\nu}$ are anti-symmetric yields the on-shell Proca consistency condition $m_{A'}^2[ \partial_\mu \primekin{A}^\mu ] = 0$.
Since $m_{A'}\neq0$ by assumption in this work, we must have $\partial_\mu \primekin{A}^\mu = 0$ on shell.

Therefore, we have
\begin{align}
    \partial^2 \big[ \kin{A}^{\mu} - \varepsilon \primekin{A}^{\mu} \big] &= J_{\textsc{em}}^{\mu},  \label{eq:eomAkin1} \\
    \partial^2 \big[ \primekin{A}^{\mu} - \varepsilon \kin{A}^{\mu} \big] &+ m_{A'}^2 \primekin{A}^{\mu} = 0,   \label{eq:eomAkin2} \\
    \partial_\mu \kin{A}^\mu &=0, \\
    \partial_\mu \primekin{A}^\mu &=0.
\end{align}

We will from this point assume that the fields $(A^{(\prime)}_{\textsc{k}})^\mu$ are plane waves:%
\footnote{\label{ftnt:notation}%
    To avoid a proliferation of notation, we write the field value at $x=0$ with the same symbol as we have up until now used to denote the field value at a general location $x$; namely, $(A^{(\prime)}_{\textsc{k}})^\mu$.
} %
\begin{align}
    (A^{(\prime)}_{\textsc{k}})^\mu(x) = (A^{(\prime)}_{\textsc{k}})^\mu e^{-ik_\nu x^\nu}.
\end{align}
Moreover, let us now consider these equations in a medium with a linear response, and no other free charge $(J_{\text{free}}^\nu = 0$).
Because the interaction eigenstate $\inter{A}$ is proportional to $A_{\textsc{k}}$ [indeed, ignoring $\mathcal{O}(\varepsilon^2)$ terms, they coincide; see \eqref{kineticTOinteraction}], and because charges in the medium will respond to only the interacting mode, we should set $J_{\textsc{em}}^{\mu} = - \Pi^{\mu\nu} \kin{A}_\nu$, where $\Pi^{\mu\nu}$ is the self-energy tensor for the medium (note importantly that this relationship is imposed in 4-momentum space; the position-space analog is generally non-local~\cite{Raffelt:1996wa}).
This yields
\begin{align}
    (-\omega^2 + k^2 ) \big[ \kin{A}^{\mu} - \varepsilon \primekin{A}^{\mu} \big] + \Pi^{\mu\nu} \kin{A}_\nu &=0,  \label{eq:eomAkinLinResp1} \\
    (-\omega^2 + k^2 ) \big[ \primekin{A}^{\mu} - \varepsilon \kin{A}^{\mu} \big] + m_{A'}^2 \primekin{A}^{\mu} &= 0,   \label{eq:eomAkinLinResp2} \\
    k_\mu \kin{A}^\mu &=0,  \label{eq:eomAkinLinResp3}\\
    k_\mu \primekin{A}^\mu &=0. \label{eq:eomAkinLinResp4}
\end{align}

The polarization tensor in a homogeneous medium can be written as a sum over the mode self-energies $\pi_a$ and mode projection operators $P_a^{\mu\nu}$~\cite{Raffelt:1996wa}:
\begin{align}
    \Pi^{\mu\nu} &\equiv \sum_{a=1,2,L} \pi_a P_a^{\mu\nu}; &
    P_a^{\mu\nu} &\equiv - (e_a)^\mu (e_a^{*})^{\nu},
\end{align}
where $a=1,2,L$ labels the two 3-transverse modes [i.e., $k_i (e_{\{1,2\}})^i =0$] and one 3-longitudinal mode [$(e_L)^i \propto k^i$], respectively, and $e_a^\mu$ are the corresponding orthonormal polarization 4-vectors, normalized%
\footnote{\label{ftnt:metricSignature}%
    We use the $(+,-,-,-)$ metric sign convention.
} %
such that $(e_a)^\mu (e^*_{b})_{\mu} = -\delta_{ab}$, and obeying the 4-transversality condition $k_\mu (e_a)^\mu = 0$.
Assuming that the medium does not distinguish between the transverse modes (as it could do if it were, e.g., birefringent or otherwise anisotropic), we can write the two transverse-mode self-energies as $\pi_1=\pi_2 = \pi_T$.
Similarly, decomposing
\begin{align}
    (A^{(\prime)}_{\textsc{k}})^\mu = \sum_a (A^{(\prime)}_{\textsc{k}})^a(e_a)^\mu,
\end{align}
we have
\begin{align}
    (-\omega^2+k^2) \big[ \kin{A}^{a} - \varepsilon \primekin{A}^{a} \big] + \pi_a \kin{A}^a &=0,  \label{eq:eigenEqn1} \\
    (-\omega^2+k^2) \big[ \primekin{A}^{a} - \varepsilon \kin{A}^{a} \big] + m_{A'}^2 \primekin{A}^{a} &= 0,   \label{eq:eigenEqn2}
\end{align}
where, here and throughout what follows, we have defined $k^2 = \bm{k}\cdot \bm{k}$ (we distinguish the contracted 4-vector $k$ as $k_\mu k^\mu$ where necessary).
The gauge [\eqref{eomAkinLinResp3}] and consistency [\eqref{eomAkinLinResp4}] conditions are automatically satisfied by construction of the polarization tensors as 4-transverse.

It remains to write expressions for the self-energies $\pi_a$. 
We will be interested in examining the behavior of these fields in media that can be considered to be ohmic conductors, where $\bm{J} = \sigma \bm{E}$.
Now,
\begin{align}
    \bm{E} &= - \partial_t \bm{A}_{\textsc{k}} - \bm{\nabla} A_{\textsc{k}}^0 \\
           &= +i \omega \bm{A}_{\textsc{k}} -i\bm{k} A_{\textsc{k}}^0 \\
           &= i\omega \sum_{a=1,2} \kin{A}^a \bm{e_a} + i \kin{A}^L [ \omega \bm{e_L} - \bm{k}(e_L)^0 ].
\end{align}
However, the 4-transversality condition imposes that $(e_L)^0 = \bm{k} \cdot \bm{e_L} / \omega$; moreover, since $\bm{e_L}\propto \bm{k}$, we have that $\bm{k} ( \bm{k} \cdot \bm{e_L} ) = k^2 \bm{e_L}$.
Therefore,
\begin{align}
    \bm{E} &= i\omega \sum_{a=1,2} \kin{A}^a \bm{e_a} + i \omega  [ 1 -  k^2 / \omega^2  ] \kin{A}^L \bm{e_L}, \label{eq:EfromA}\\
\Rightarrow \bm{J} &= i\omega\sigma \sum_{a=1,2} \kin{A}^a \bm{e_a} + i \omega \sigma  [ 1 -  k^2 / \omega^2  ] \kin{A}^L \bm{e_L} \\
    &\equiv - \sum_{a=1,2,L} \pi_a \kin{A}^a \bm{e_a}.
\end{align}
Therefore, we can read off 
\begin{align}
    \pi_T &= - i \omega \sigma, \label{eq:sigmaT} \\
    \pi_L &= - i \omega \sigma \lb[ 1 - k^2 / \omega^2 \rb].\label{eq:sigmaL}
\end{align}
Note that if we can consider non-relativistic modes $(k\ll \omega)$, we have $\pi_L \approx \pi_T$ and, moreover, there is no distinction between the $T$ and $L$ modes in terms of their relationships to the physical $\bm{E}$ field; cf.~\eqref{EfromA}.

This discussion, of course, also applies for an isotropic plasma with a number density $n$ of charge carriers of charge $Q=qe$ and mass $m$, and with a collision frequency $\nu \equiv \tau^{-1}$, such that the plasma frequency is $\omega_p^2 = 4\pi n q^2 \alpha / m$.
In this case, we simply take $\sigma$ to be a function of frequency: 
\begin{align}
   \sigma(\omega) &= \frac{\omega_p^2 \tau}{1-i\omega \tau} 
   = \frac{i\omega_p^2/\omega}{1+i\nu / \omega }.\label{eq:plasmaRepl}
\end{align}
For $\omega \tau \ll 1 \Rightarrow \nu/\omega \gg 1$, collisions dominate and such a plasma behaves as a DC conductor with conductivity $\sigma = \omega_p^2\tau$; on the other hand, for $\omega \tau \gg 1 \Rightarrow \nu/\omega \ll 1$, the plasma is effectively collisionless and we can replace $\sigma \rightarrow i\omega_p^2/\omega$ in the self-energies \eqref[s]{sigmaT} and (\ref{eq:sigmaL}), and elsewhere throughout this Appendix.%
\footnote{\label{ftnt:noExtraSolutions}%
    Note that because we solve the dispersion relation eigenvalue problems in the following subsections [\appref[ces]{app:transverseCase} and \ref{app:longitudinalCase}] for the momentum eigenvalues $k=k(\omega)$ and the corresponding momentum eigenmodes, whether $\sigma$ is a function of frequency has no influence on whether additional solutions to the eigenvalue equations exist.
} %
Generally however, for a plasma with $\nu \neq0$, the full replacement implied by \eqref{plasmaRepl}, $\sigma \rightarrow (i\omega_p^2/\omega) / \lb( 1+i\nu / \omega \rb)$, is required: for instance, damping effects for some modes are $\propto \nu$ (see \appref[ces]{app:transverseCase} and \ref{app:longitudinalCase}).

We will now discuss the transverse [\appref{app:transverseCase}] and longitudinal [\appref{app:longitudinalCase}] cases in turn.

\subsubsection{Transverse case}
\label{app:transverseCase}

Substituting \eqref{sigmaT} into \eqref{eigenEqn1}, and considering the resulting equation along with \eqref{eigenEqn2}, we have an eigenvalue problem for the propagating (i.e., momentum) modes, which can be cast in the form $k^2 \vec X = M(\omega) \vec X$, where the column vector $\vec X$ has components $\vec X = [ \kin{A}^T , \primekin{A}^T ]$; i.e., \eqref[s]{eigenEqn1}, (\ref{eq:eigenEqn2}), and (\ref{eq:sigmaT}) specify the transverse dispersion relations.
This is easily solved using standard linear algebra techniques%
\footnote{\label{ftnt:diagonalization}%
    Recall that if the matrix $M$ has linearly independent eigenvectors, then the matrix $U$ that is formed with columns that are equal to these eigenvectors is invertible and diagonalizes $M$ by a similarity transformation: $U^{-1}MU = D$, where $D$ is a diagonal matrix with the eigenvalues on the diagonal, ordered in the same sense as the columns of $U$.
    Since $k^2 \vec{X} = M \vec{X} = U D U^{-1}\vec{X}$ we have $ k^2 [ U^{-1} \vec{X} ] = D [ U^{-1} \vec{X}]$, and so the eigenstates are given by the components of $U^{-1} \vec{X}$.
} %
to yield the two eigenvalues $k_{T\{1,2\}}^2$ that are, correct to $\mathcal{O}(\varepsilon^2)$, given by (see also Appendix 3 of \citeR{Graham:2014sha}, but note a difference in our sign convention---we use $\omega$ of opposite sign---and method of derivation here in terms of the kinetically mixed basis)
\begin{align}
    k_{T1}^2 &= \omega^2 + i\omega\sigma - \frac{\varepsilon^2 \omega^2 \sigma^2}{i\omega \sigma + m_{A'}^2 }\label{eq:kT1Sq}\\
    &\approx   \omega^2 + i\omega\sigma, \label{eq:dispersion1}\\[3ex]
    k_{T2}^2 &= \omega^2 - m_{A'}^2 - \frac{\varepsilon^2 m_{A'}^4}{i\omega \sigma + m_{A'}^2 }\label{eq:kT2Sq}\\
    &\approx \omega^2 - m_{A'}^2 + i\omega\sigma \frac{ \varepsilon^2 m_{A'}^4}{ (\omega \sigma)^2 + m_{A'}^4 }.\label{eq:dispersion2}   
\end{align}
These correspond to the transverse propagation eigenstates, all correct to $\mathcal{O}(\varepsilon)$, given in the various basis sets by~\cite{Graham:2014sha}
\begin{align}
    A^T_1 &= \kin{A}^T - \frac{\varepsilon m_{A'}^2}{i\omega \sigma + m_{A'}^2 } \primekin{A}^T \label{eq:mode1kin}\\
    &= \inter{A}^T - \frac{\varepsilon m_{A'}^2}{i\omega \sigma + m_{A'}^2 } \primeinter{A}^T \label{eq:mode1inter}\\
    &=  \mass{A}^T + \frac{i \varepsilon  \omega \sigma }{i\omega \sigma + m_{A'}^2 } \primemass{A}^T, \label{eq:mode1mass}\\[3ex]
    A^T_2 &= \primekin{A}^T - \frac{i \varepsilon \omega\sigma }{i\omega \sigma + m_{A'}^2 } \kin{A}^T \label{eq:mode2kin}\\
    &=\primeinter{A}^T + \frac{ \varepsilon m_{A'}^2 }{i\omega \sigma + m_{A'}^2 } \inter{A}^T \label{eq:mode2inter}\\
    &= \primemass{A}^T - \frac{i \varepsilon \omega\sigma }{i\omega \sigma + m_{A'}^2 } \mass{A}^T. \label{eq:mode2mass}
\end{align}
As noted above, the expressions \eqrefRange{kT1Sq}{mode2mass} also apply in the case of a plasma, under the replacement rule implied by \eqref{plasmaRepl}: $\sigma \rightarrow i \omega_p^2/\omega \times ( 1+i\nu/\omega )^{-1}$, where $\omega_p$ and $\nu$ are the plasma and collision frequencies, respectively; see the comment in footnote \ref{ftnt:noExtraSolutions}.

In the conductor case, two limits are interesting: $\sigma\ll m_{A'}^2/\omega $ (poor conductor, or no medium) and $\sigma \gg m_{A'}^2 / \omega$ (good conductor).
Note that if we consider the case $\omega^2 \sim m_{A'}^2$ such that the second momentum mode is non-relativistic ($k_{T2}^2 \ll \omega^2$), the condition for a good conductor simplifies to $\sigma \gg m_{A'}$, while `vacuum' means $\sigma \ll m_{A'}$.

In the poor conductor limit, we find $k_{T1}^2 = \omega^2$ and $k_{T2}^2 = \omega^2-m_{A'}^2$, along with $A_{1}^T = \kin{A}^T - \varepsilon \primekin{A}^T = \mass{A}^T$ and $A_{2}^T = \primekin{A}^T = \primemass{A}^T$, correct up to terms at $\mathcal{O}(\varepsilon^2)$.
This, of course, is exactly the expected result: the vacuum mass-basis modes are the propagating momentum modes, and their dispersion relations are correct for massless and massive modes, respectively.

The good-conductor limit yields $k_{T1}^2 \approx \omega^2 + i \omega \sigma$ and $k_{T2}^2 \approx \omega^2 - m_{A'}^2 + i \varepsilon^2 m_{A'}^4/(\sigma \omega)$; in both expressions, terms parametrically suppressed compared to those shown have been omitted.
We also have $A_{1}^T = \kin{A}^T = \inter{A}^T$ and $A_{2}^T = \primekin{A}^T - \varepsilon \primekin{A}^T = \primeinter{A}^T$, where these two expressions correct up to omitted terms at $\mathcal{O}(\varepsilon^2)$.
This is an extremely important result: the in-medium propagation eigenstates in a good conductor are the \emph{interaction}-basis states, not the vacuum mass-basis states.

The interacting state has a complex momentum eigenvalue $k_{T1} \approx \sqrt{\omega \sigma} e^{i\pi/4}$, leading to an exponential damping factor
\begin{align}
&\propto \exp\lb[- i (k_{T1})_\mu x^\mu\rb] \nonumber\\
&= \exp\lb[ - i\omega t + i (\bm{\hat{k}} \cdot \bm{x}) \sqrt{\omega \sigma / 2} \rb] \nl \times \exp\lb[  - (\bm{\hat{k}} \cdot \bm{x}) \sqrt{\omega \sigma / 2} \rb]. 
\end{align}
That is, the interacting mode field \emph{amplitude} damps over length-scales $\delta_{T1} \sim (\omega \sigma/2)^{-1/2} \ll m_{A'}^{-1}$, which we refer to as the skin-depth [note that various conventions exist for skin-depth in the literature, largely depending on whether they are defined for the power (i.e., Poynting flux) or for the field amplitude; the various definitions differ from ours by $\mathcal{O}(1)$ numerical factors].

On the other hand, the sterile state has the usual dispersion relation for a massive mode with mass $m_{A'}$, with the addition only of a highly suppressed imaginary component.
Extracting a damping length requires some care: 
consider that if $\omega^2 -m_{A'}^2 \approx m_{A'}^2 v_{\textsc{dm}}^2$ with $v_{\textsc{dm}} \sim 10^{-3}$, then the dispersion relation reads
$k_{T2}^2 \approx m_{A'}^2 v_{\textsc{dm}}^2 + i\varepsilon^2m_{A'}^3/\sigma$.
If $\varepsilon^2m_{A'}/\sigma \gg v_{\textsc{dm}}^2$, then we can approximate $k_{T2}^2 \approx i\varepsilon^2m_{A'}^3/\sigma$, which would give $k_{T2} = \varepsilon m_{A'}e^{i\pi/4}\sqrt{m_{A'}/\sigma}$, leading to a skin-depth of $\delta_{T2}^{(1)} \sim \sqrt{2\sigma/m_{A'}}/( \varepsilon m_{A'}) \sim \delta_{T1} \times \sigma/(\varepsilon m_{A'})$, assuming $\omega \sim m_{A'}$.
However, whenever $\varepsilon^2m_{A'}/\sigma \ll v_{\textsc{dm}}^2$ the real term in the dispersion relation dominates over the complex one in magnitude, and in solving for the real ($k_R)$ and imaginary ($k_I$) parts of $k$ in the dispersion relation we must instead estimate $k_R \approx \sqrt{ \omega^2 - m_{A'}^2} \approx m_{A'} v_{\textsc{dm}}$, and $2k_Rk_I \approx \varepsilon^2m_{A'}^3/\sigma \Rightarrow k_I \approx \varepsilon^2m_{A'}^2/(2\sigma v_{\textsc{dm}})$, implying a damping length $\delta_{T2}^{(2)} \sim 1/k_I \sim (2\sigma v_{\textsc{dm}})/( \varepsilon^2m_{A'}^2) \sim (v_{\textsc{dm}}/\varepsilon) \sqrt{2\sigma/m_{A'}}\times \delta_{T2}^{(1)}$.
Because $\delta_{T2}^{(2)}>\delta_{T2}^{(1)}$ whenever the condition for the validity of the estimate leading to the former is satisfied (i.e., $\varepsilon^2m_{A'}/\sigma \ll v_{\textsc{dm}}^2$), we should thus instead take the combined result%
\footnote{\label{ftnt:bothRegimes}%
    We possibly access both regimes in various locations: we have $\varepsilon \lesssim 10^{-2}$, $m_{A'} \lesssim 10^{-16}\,$eV, and $\sigma$ in highly conductive layers that is as large as $\sigma \sim 10^{-2}$\,eV in the ionosphere, so $\varepsilon^2m_{A'}^3/\sigma \lesssim 10^{-18} m_{A'}^2 \ll v_{\textsc{dm}}^2 m_{A'}^2$.
    On the other hand, in the lower atmosphere we also have other regions where $\sigma \sim m_{A'}$, so there we have $\varepsilon^2m_{A'}^3/\sigma \sim 10^{-4} m_{A'}^2 \gg v_{\textsc{dm}}^2 m_{A'}^2$ at its largest.
} %
\begin{align}
    \delta_{T2} \sim \text{max}\lb[ 1 , (v_{\textsc{dm}}/\varepsilon) \sqrt{2\sigma/m_{A'}} \rb] \times \sqrt{2\sigma/m_{A'}}/( \varepsilon m_{A'});
    \label{eq:deltaT2}
\end{align}
in either case, this is an extremely long length-scale.

Similar qualitative observations hold for the case of a nearly collisionless plasma with a high plasma frequency ($\omega_p\gg\omega \sim m_{A'}\gg\nu$), which is the case of physical relevance in the interplanetary medium (see discussion in \secref{sec:interplanetaryMedium}).
In that case, in the non-relativistic limit $k^2 \ll \omega^2 \sim m_{A'}^2 \ll \omega_p^2$, we have $k_{T1} \approx i \omega_p$, leading to a very short active-mode damping length $\delta_{T1} \sim \omega_p^{-1} \ll m_{A'}^{-1}$.
In the same series of limits, we have $k_{T2}^2 \approx \omega^2 - m_{A'}^2 + i\varepsilon^2 m_{A'}^3 \nu/\omega_p^2$, where terms parametrically suppressed compared to those shown have been omitted.
Note that for an exactly collisionless plasma ($\nu=0$) there is no damping of this mode [see also Eq.~(23) of \citeR{Dubovsky:2015cca}]: the omitted terms in the expression for $k_{T2}^2$ are all real and positive in this limit.

Extracting the damping length for $\nu \neq 0$ again requires the same degree of care as needed to obtain the skin-depth for the $T2$ mode in a conductor.
However, because the mathematical structure of the expression for $k_{T2}^2$ for the plasma is identical to that for the conductor, requiring only the parametric replacement $\sigma \rightarrow \omega_p^2/\nu$, we can immediately write down the damping length for this mode in a plasma from \eqref{deltaT2}: 
\begin{align}
    \delta_{T2,p} &\sim \text{max}\lb[ 1 , (v_{\textsc{dm}}/\varepsilon) \sqrt{2\omega_p^2/(m_{A'}\nu)} \rb] \nl \quad\times \sqrt{2\omega_p^2/(m_{A'}\nu)}/( \varepsilon m_{A'});
    \label{eq:deltaT2p}
\end{align}
this is again an extremely long length-scale.

\subsubsection{Longitudinal case}
\label{app:longitudinalCase}

The longitudinal case requires some care in interpretation.
Formally, we may proceed as for the transverse case: substituting \eqref{sigmaL} into \eqref{eigenEqn1}, and considering it along with \eqref{eigenEqn2} we again obtain a system of equations that can be cast into an eigenvalue problem for the propagation eigenstate(s).
Proceeding na\"ively, we find two solutions:
\begin{align}
    k_{L1}^2 &= \omega^2 - m_{A'}^2 \frac{\sigma - i \omega}{\sigma - i (1-\varepsilon^2) \omega}\label{eq:kL1Sq} \\
    &= \omega^2 - m_{A'}^2 \frac{\sigma^2 + (1-\varepsilon^2) \omega^2}{\sigma^2 + (1-\varepsilon^2)^2\omega^2} \nl
    + i \frac{ \varepsilon^2 m_{A'}^2\sigma\omega }{ \sigma^2 + (1-\varepsilon^2) \omega^2 }\\
    &\approx \omega^2 - m_{A'}^2 + i \frac{ \varepsilon^2 m_{A'}^2\sigma\omega }{ \sigma^2 + \omega^2 } ,\\[3ex]
    k_{L2}^2 &= \omega^2,\qquad \text{\footnotesize[unphysical]}
\end{align} 
with corresponding eigenmodes
\begin{align}
    A_{1}^{L} &= \primekin{A}, \label{eq:A1L}\\[2ex]
    A_{2}^{L} &= \kin{A} - \frac{\varepsilon\omega}{\omega+i\sigma}\primekin{A}. \qquad \text{\footnotesize[unphysical]}\label{eq:A2L}
\end{align} 
While the first of these solutions (eigenvalue $k_{L1}^2$ and eigenmode $A_{1}^{L}$) is physical, the second solution (eigenvalue $k_{L2}^2$ and eigenmode $A_{2}^{L}$) is not.
There are any number of ways to see this, but the most straightforward is to note that in the limit $\omega^2 = k^2$, the longitudinal polarization vector $e^\mu_L = \lb( k/\sqrt{ k_\nu k^\nu }  , \,\bm{\hat{k}}\, \omega /\sqrt{ k_\nu k^\nu }  \rb)^\mu$ formally diverges: this is actually symptomatic of the fact that it is not possible to find a normalizable polarization vector $e_L^\mu$ that simultaneously satisfies $k_\mu e_L^\mu = 0$, $e_{L}^\mu (e^*_{L})_\mu = -1$, $e_L^i \propto k^i$, and $\omega^2 = k^2$. 
Attempting to impose all of these conditions leads to a logical contradiction.
The longitudinal mode is thus not physical if $\omega^2 = k^2$.

A corollary of this observation is that we must assume $\omega \neq -i\sigma$, to avoid $k_{1L}^2=\omega^2$ from \eqref{kL1Sq}.
Assuming that $\omega$ is real is, of course, natural in this situation, but this condition has non-trivial implications for the plasma case; see discussion at end of this section.

Because there is only one propagating eigenmode, the other degree of freedom in the system must be fixed by a constraint. 
Indeed, examining the equation that results from substituting \eqref{sigmaL} into \eqref{eigenEqn1}, we find that it reads
\begin{align}
    (-\omega^2 + k^2)\big[ \lb( 1+ i \sigma/\omega\rb) \kin{A}^L - \varepsilon \primekin{A}^L \big] &= 0;
\end{align}
demanding that this is solved for $\omega^2 \neq k^2$ leads to the constraint
\begin{align}
    \kin{A}^L = \varepsilon \primekin{A}^L \frac{\omega}{ \omega + i \sigma };
    \label{eq:constraint}
\end{align}
this, of course, enforces that the spurious mode vanishes identically, $A_{2}^{L} = 0$, as expected since we have assumed that the solution we are seeking has $\omega^2 \neq k^2$, and thus must be orthogonal to the (spurious) mode with the (spurious)  eigenvalue $\omega^2=k^2$.
Note that in the $\sigma = 0$ vacuum limit, $A_{2}^{L} = \mass{A}^L$, so this constraint sets the massless longitudinal mode in vacuum to zero.
This of course is expected because that mode does not actually exist: any $L$ mode must have a non-zero $\mu=0$ component since $e_L^0 \neq 0$ [i.e., $\mass{A}^0 \neq 0$], but we know that in vacuum such a component can be removed by a residual restricted gauge transformation (this is the massless mode, so the gauge symmetry is unbroken): $\mass{A}_\mu \rightarrow \mass{A}_\mu + \partial_\mu \Lambda$ with $\partial^2 \Lambda = 0$ by the choice $\partial_0 \Lambda = - \mass{A}_0$.
This is also consistent with counting of degrees of freedom: a system of one massless and one massive photon should, in vacuum, have only 5 physical degrees of freedom.
Note that, on the other hand, $A_{1}^L = A_{\tsc{m}}'$ exactly.

As a result of the constraint, we have, at $\mathcal{O}(\varepsilon)$, the following interaction basis relationships to the propagating longitudinal mode:%
\footnote{\label{ftnt:twoModesAreOne}%
    Of course, there is only 1 degree of freedom, so relating it to a basis of two modes is slightly odd.
    Nevertheless, because the interacting mode in the interaction basis is the only part of the system that couples to charges, this exercise is useful to understand how the propagating mode interacts with charges.
} %
\begin{align}
    \inter{A}^L &\approx \kin{A}^L =  \varepsilon \primekin{A}^L \frac{\omega}{ \omega + i \sigma } \nonumber\\
    &= \varepsilon A_{1}^{L} \frac{\omega}{ \omega + i \sigma },\\[2ex]
    \primeinter{A}^L &= \primekin{A}^L - \varepsilon \kin{A}^L \approx  A_{1}^L .
\end{align}
The overlap of the propagating mode $A_{1}^{L}$ with the interacting mode $A_{\tsc{i}}^L$ shows that the former drives charges in vacuum ($\sigma=0$) at $\mathcal{O}(\varepsilon)$.
Moreover, we see that in a good conductor (here, defined as $\sigma \gg \omega$), the propagating mode has a highly suppressed overlap with the interacting mode: $\inter{A}^L\rightarrow - i \varepsilon A_{1}^{L} (\omega / \sigma)$; it is instead closely aligned with the sterile mode.
In the same limit, we have $k_{L1}^2 \rightarrow \omega^2 - m_{A'}^2 + i \varepsilon^2 m_{A'}^2 \omega / \sigma$, which shows that the damping of the propagating mode in a good conductor is again highly suppressed: for a non-relativistic mode with $\omega \sim m_{A'}$, the damping length $\delta_{L1}$ is given by the same expression as that for $\delta_{T2}$ shown at \eqref{deltaT2}, up to sub-leading corrections.
Of course, in vacuum, we have $k_{L1}^2 = \omega^2 - m_{A'}^2$ exactly, and the longitudinal dark-photon mode propagates without any damping.

It remains to discuss how the field $A_{1}^{L}$ is sourced by free currents, which we will do here by considering how it may be sourced by such currents flowing outside of a conductive medium ($\sigma = 0$).
To this end, consider solving \eqref{eomAkin1} for $\partial^2 \kin{A}^{\mu}$ and substituting into \eqref{eomAkin2}, which yields
\begin{align}
    \partial^2 \primekin{A}^\mu + \frac{m_{A'}^2}{1-\varepsilon^2} \primekin{A}^\mu &=  \frac{\varepsilon}{1-\varepsilon^2} J_{\tsc{em}}^\mu.
\end{align}
Dropping terms at $\mathcal{O}(\varepsilon^2)$, projecting onto the longitudinal polarization vector by contracting with $(e_L^*)_\mu$, we find
\begin{align}
    \partial^2 \primekin{A}^L + m_{A'}^2 \primekin{A}^L &= \varepsilon J^L_{\tsc{em}},
\end{align}
where $J^L_{\tsc{em}} \equiv -(e_L^*)_\mu J_{\tsc{em}}^\mu$.
But $\primekin{A}^L = A_{1}^L$, so this shows that $A_{1}^L$ is sourced at $\mathcal{O}(\varepsilon)$ by ordinary EM currents (specifically, the piece of the current proportional to the longitudinal polarization vector).

In sum then, we see that for the longitudinal case, only one mode propagates in a conductor.
In vacuum, this propagating mode coincides with the massive mass-basis mode, and is undamped.
In a perfect conductor, this propagating mode coincides with the sterile state in the interaction basis and in the perfect conductor limit is also undamped.
This field is also sourced at $\mathcal{O}(\varepsilon)$ by ordinary free EM currents flowing in vacuum, and it couples to test charges in vacuum at $\mathcal{O}(\varepsilon)$.

Finally, we comment on the situation in plasma.
Once again, we would make the replacement $\sigma \rightarrow i \omega_p^2/\omega \times ( 1+i\nu/\omega )^{-1}$, but the derivation above for the \emph{momentum} eigenvalues is unchanged: as long as $\omega \neq \omega_p$ (the cognate of the condition $\omega \neq -i\sigma$ above), there is only one propagating momentum eigenmode: $k_{L1}=k_{L1}(\omega)$.
Note, however, that the energy spectrum of the longitudinal excitations at fixed $k$, $\omega = \omega(k)$, always contains two modes: for $k \not\sim \omega_p$, there is one mode at $\omega^2 \sim k^2 + m_{A'}^2$, and a second, non-propagating mode at $\omega^2 \sim \omega_p^2$.
As we are interested in the response of a system with spatial profiles when driven by a monochromatic background dark-photon field at frequency $\omega \sim m_{A'} \ll \omega_p$, it is appropriate on physical grounds for us to consider the momentum $k$ to be a function of $\omega$: $k_{L1}=k_{L1}(\omega)$.

For a non-relativistic mode propagating in a nearly collisionless plasma with a high plasma frequency (i.e., assuming $\omega_p^2 \gg \omega^2 \sim m_{A'}^2 \gg \nu^2$), we have $k_{L1}^2 \approx \omega^2-m_{A'}^2 + i \varepsilon^2 m_{A'}^3\nu/\omega_p^2$ (terms parametrically suppressed compared to those shown have been omitted), leading to a damping length $\delta_{L1,p}$ which has the same expression as that for $\delta_{T2,p}$ displayed at \eqref{deltaT2p}.

\subsection{Effective current approach}
\label{app:effectiveCurrent}

As our computations in the \secref{sec:signal} rely on treating the sterile field $\primeinter{A}$ as an effective current source for the interacting field $\inter{A}$ in the interaction basis, we briefly explain the origin of that approach here.

In the interaction basis the equations of motion \eqref[s]{eomFkin1} and (\ref{eq:eomFkin2}) read, at $\mathcal{O}(\varepsilon)$,
\begin{align}
    \partial_\mu \inter{F}^{\mu\nu} &= J^\nu - \varepsilon m_{A'}^2 \primeinter{A}^\nu, \label{eq:interEOM1} \\
    \partial_\mu \primeinter{F}^{\mu\nu} + m_{A'}^2 \primeinter{A}^\nu &= -\varepsilon m_{A'}^2 \inter{A}^\nu. \label{eq:interEOM2}        
\end{align}

Suppose we perform a systematic formal perturbative expansion of the fields in powers of $\varepsilon$:
\begin{align}
    \inter{A} &\equiv \sum_{n=0}^\infty \varepsilon^n \inter{A}^{(n)},\\
    \primeinter{A} &\equiv \sum_{n=0}^\infty \varepsilon^n \primeinter{A}^{(n)}.
\end{align}
Substituting this expansion into \eqref[s]{interEOM1} and (\ref{eq:interEOM2}), treating the resulting equations as a formal power series in $\varepsilon$ to be satisfied by setting the coefficients of equal powers of $\varepsilon$  equal, and keeping only terms up to $\mathcal{O}(\varepsilon)$ in line with the terms retained in \eqref[s]{interEOM1} and (\ref{eq:interEOM2}), we have a system of four equations (assuming that $J^\nu \sim \varepsilon^0$):
\begin{align}
    \partial_\mu \inter{F}^{(0)\, \mu\nu} &= J^\nu, \label{eq:interEOM1exp1} \\
    \partial_\mu \primeinter{F}^{(0)\, \mu\nu} + m_{A'}^2 \primeinter{A}^{(0)\, \nu} &= 0, \label{eq:interEOM2exp1}   \\
    \partial_\mu \inter{F}^{(1)\, \mu\nu} &= - m_{A'}^2 \primeinter{A}^{(0)\, \nu}, \label{eq:interEOM1exp2} \\
    \partial_\mu \primeinter{F}^{(1)\, \mu\nu} + m_{A'}^2 \primeinter{A}^{(1)\, \nu} &= - m_{A'}^2 \inter{A}^{(0)\, \nu}. \label{eq:interEOM2exp2}   
\end{align}

The leading-order interacting and sterile solutions are unperturbed: $\inter{F}^{(0)}$ obeys the standard sourced Maxwell equations, \eqref{interEOM1exp1}; and $\primeinter{F}^{(0)}$ obeys the source-free Proca equations, \eqref{interEOM2exp1}.

In the presence of a non-zero background field $\primeinter{A} \equiv \primeinter{A}^{(0)}$ that obeys \eqref{interEOM2exp1}---e.g., the dark-photon dark-matter field---we see from \eqref{interEOM1exp2} that the leading impact on the observable field $\inter{F}$ is at $\mathcal{O}(\varepsilon)$, and can be computed by treating the background $\primeinter{A}$ field as an effective current source [cf.~the forms of \eqref[s]{interEOM1exp1} and (\ref{eq:interEOM1exp2})]: 
\begin{align}
    J_{\text{eff}}^\nu = - \varepsilon m_{A'}^2 \primeinter{A}^\nu.
\end{align}
Note that the effective current is also conserved:
\begin{align}
    \partial_\mu J_{\text{eff}}^\mu &= - \varepsilon m_{A'}^2 \partial_\mu \primeinter{A}^\mu \\
    &= - \varepsilon m_{A'}^2 \big( \partial_\mu \primekin{A}^\mu - \varepsilon \partial_\mu \kin{A}^\mu \big) \\
    &= 0,
\end{align}
since both kinetically coupled basis modes obey $\partial_{\mu} (A_{\tsc{k}}^{(\prime)})^\mu = 0$.
Therefore, for plane-wave $\primeinter{A}^\mu$, we have 
\begin{align}
    J_{\text{eff}}^0 = \frac{k}{\omega} \bm{\hat{k}} \cdot \bm{J}_{\text{eff}}.
\end{align}
For non-relativistic modes ($k \ll \omega$) then, the effective charge density $J_{\text{eff}}^0$ vanishes, and the effective current is simply a 3-current:
\begin{align}
    \bm{J}_{\text{eff}} = - \varepsilon m_{A'}^2 \bm{A}'_{\tsc{i}}.
\end{align}

Finally, note that in a EM-source-free region ($J^\nu=0$) with boundary conditions set such that the $A_{\tsc{i}}$ field would be zero if we set $\varepsilon=0$, the $A_{\tsc{i}}$ field remains zero at leading order when $\varepsilon\neq 0$: $\inter{A}^{(0)} = 0$. 
In this case, the back-reaction term on the RHS of \eqref{interEOM2exp2} vanishes, and the leading back-reaction on $A'_{\tsc{i}}$ is at $\mathcal{O}(\varepsilon^2)$.

\section{Finite conductivity effects}
\label{app:finiteConductivty}

In this Appendix, we repeat the calculation of \secref{sec:earthTheory} including the effect of nonzero, finite conductivities for the Earth, atmosphere, and assumed conductive ionosphere layer, in order to demonstrate that the result is unchanged.
For simplicity, we take the Earth and ionosphere to have the same conductivity $\sigma_1\gg m_{A'}$ and the atmosphere to have a conductivity $\sigma_2$; we will ignore spatial variation of the conductivity within each layer.
As in \secref{sec:earthTheory}, we will treat this calculation as a single-photon electromagnetism problem, where the effect of the dark photon is to source an inhomogeneous contribution to the observable electric field.
We will then compute the homogeneous contribution required to satisfy the appropriate boundary conditions.

The first step therefore becomes to determine what the inhomogeneous contributions inside the different conductors are.
In particular, since we will be interested in solving for the observable electric and magnetic fields, we want to know the contribution to the active component in the interaction basis.
Consider the case of transverse fields.
As described in \appref{app:EoMandInMedium}, there are two propagating modes inside a conductor, given in this basis by \eqref{mode1inter} and \eqref{mode2inter}.
Inverting these we can write the active and sterile components in the non-relativistic limit as
\begin{align}
    \begin{pmatrix}
        \inter{\bm{A}}^T\\[1ex]
        \primeinter{\bm{A}}^T
    \end{pmatrix}
    =
    \begin{pmatrix}
        1   &   \dfrac{\varepsilon m_{A'}^2}{i m_{A'}\sigma+m_{A'}^2}  \\[2ex]
        -\dfrac{\varepsilon m_{A'}^2}{i m_{A'}\sigma+m_{A'}^2}  &1
    \end{pmatrix}
    \begin{pmatrix}
        \bm{A}_1^T\\[1ex]
        \bm{A}_2^T
    \end{pmatrix}.
\end{align}
Deep inside the ionosphere, $\bm{A}_1^T=0$ since its dispersion relation has a large imaginary part.
Let $\bm{A}_2^T=\bm{A}'_0e^{-im_{A'}t}$ deep inside the ionosphere.
In the non-relativistic limit, this mode has highly suppressed spatial dependence and can be treated as a uniform field in the vicinity of the Earth; it will thus take this value everywhere in the ionosphere.
The boundary condition at the interface between the atmosphere and ionosphere, as well as at the interface between the atmosphere and Earth, will be that the components are continuous in the interaction basis.
Since $\bm{A}_2^T=\primeinter{\bm{A}}^T$ to leading order, then, in fact, $\bm{A}_2^T=\bm{A}'_0e^{-im_{A'}t}$ everywhere.
This implies that the active component is
\begin{align}
    \inter{\bm{A}}^T=\bm{A}_1^T+\frac{\varepsilon\bm{A}'_0}{\beta^2}e^{-im_{A'}t},
\end{align}
where $\beta^2=1+i\sigma/m_{A'}$.%
\footnote{\label{ftnt:plasmabeta}%
    This will be the expression for $\beta$ in a conductor.
    However this notation can also be generalized to plasmas (such as the interplanetary medium, cf.~\secref{sec:interplanetaryMedium}), in which case $\beta^2=1-\omega_p^2/m_{A'}^2$.
    The rest of the argument in this Appendix remains valid so long as $\beta$ has a large imaginary part in the plasma, which is true if $\omega_p\gg m_{A'}$.
} %
The corresponding observable electric field will then be
\begin{align}
    \bm{E}&=\bm{E}_\text{hom}+\bm{E}_\text{inh},\\
    \bm{E}_\text{hom}&=im_{A'}\bm{A}_1^T,\\
    \bm{E}_\text{inh}&=\frac{i\varepsilon m_{A'}\bm{A}'_0}{\beta^2}e^{-im_{A'}t}\label{eq:finite_inhom}.
\end{align}
We thus see that inside a conductor, the inhomogeneous field has an additional factor of $\beta^{-2}$ compared to the vacuum expression [cf.~\eqref{inh}].

With this notation, we can now solve for the homogeneous fields following a method similar to that in \secref{sec:earthTheory}.
The primary difference is that now we will solve for the electric field in the Earth and ionosphere, as well as the atmosphere.
As before, each region will have an inhomogeneous contribution given by \eqref{finite_inhom}, with the conductivity appearing in $\beta$ given by $\sigma_1$ or $\sigma_2$, as appropriate.
We solve for the homogeneous contribution by satisfying boundary conditions.
Because all conductivities are now finite, the appropriate boundary condition at the interfaces between the regions are that the parallel electric and magnetic fields%
\footnote{\label{ftnt:BC1}%
    Technically, it is the parallel magnetic $\bm{H}$ field that is continuous across the interface (assuming no free surface current is flowing on the surface interface).
    However, an isotropic ohmic conductivity is equivalent to an effective permittivity and not an effective permeability, which implies that the parallel magnetic $\bm{B}$ field is continuous.
} %
are continuous across the interface.
Additionally, as in \secref{sec:toyTheory}, we will require that the electric field is regular at the origin.
Finally, we will require that the homogeneous contribution is entirely outgoing at infinity (i.e., there is only a component proportional to $\exp[+i\beta m_{A'}r]$, not one proportional to $\exp[-i\beta m_{A'}r]$); this final condition has the interpretation that the active photon modes in the ionosphere can only be moving away from the Earth, and not toward it, since they are sourced by charge motion in the vicinity of the interface between the lower atmosphere and ionosphere. 
As in \secref{sec:earthTheory}, the only relevant modes will be the $\ell=1$ TM modes (because only $\ell =1$ modes appear in the background sterile field, and the boundaries are all assumed to be spherically symmetric in this computation).

\begin{widetext}
Given the above boundary conditions, we can write the homogeneous contribution as
\begin{align}
    \bm{E}_\text{hom}=
    \begin{cases}
        \sum_ma_m\lb[
        -\dfrac{2j_1(\beta_1m_{A'}r)}{m_{A'}r}\bm{Y}_{1m}
        -\lb(\beta_1j'_1(\beta_1m_{A'}r)+\dfrac{j_1(\beta_1m_{A'}r)}{m_{A'}r}\rb)\bm{\Psi}_{1m}\rb]
        e^{-im_{A'}t},  &   r<R \\[3ex]
        \sum_m\lb[
        -\dfrac{2g_m(\beta_2m_{A'}r)}{m_{A'}r}\bm{Y}_{1m}
        -\lb(\beta_2g'_m(\beta_2m_{A'}r)+\dfrac{g_m(\beta_2m_{A'}r)}{m_{A'}r}\rb)\bm{\Psi}_{1m}\rb]
        e^{-im_{A'}t},  &   R<r<R+h \\[3ex]
        \sum_md_m\lb[
        -\dfrac{2h^{(1)}_1(\beta_1m_{A'}r)}{m_{A'}r}\bm{Y}_{1m}
        -\lb(\beta_1h^{(1)\prime}_1(\beta_1m_{A'}r)+\dfrac{h^{(1)}_1(\beta_1m_{A'}r)}{m_{A'}r}\rb)\bm{\Psi}_{1m}\rb]
        e^{-im_{A'}t},  &   r>R+h, 
    \end{cases}
     \label{eq:Efinite}
\end{align}
where $\beta^2_i=1+i\sigma_i/m_{A'}$, $h^{(1)}_n = j_n + i y_n$ is the spherical Hankel function of the first kind, and
\begin{align}
    g_{m}(x)&=b_mj_1(x)+c_m x_0^3 y_1(x) \qquad
    [x_0 \equiv m_{A'}R].
\end{align}
Likewise, the total magnetic field will be given by 
\begin{align}
    \bm{B}=
    \begin{cases}
        -i\sum_ma_m\beta_1^2j_1(\beta_1m_{A'}r)\bm{\Phi}_{1m}
        e^{-im_{A'}t},  &   r<R \\[3ex]
        -i\sum_m\beta_2^2g_m(\beta_2m_{A'}r)\bm{\Phi}_{1m}
        e^{-im_{A'}t},  &   R<r<R+h  \\[3ex]
        -i\sum_md_m\beta_1^2h_1^{(1)}(\beta_1m_{A'}r)\bm{\Phi}_{1m}
        e^{-im_{A'}t},  &   r>R+h.
    \end{cases}
     \label{eq:Bfinite} 
\end{align}

The boundary conditions that the parallel electric and magnetic fields are continuous at $r=R$ and $r=R+h$ give four equations which determine $a_m,b_m,c_m,d_m$.
Again the general solution is complicated, and we examine only in the limit $|\beta_2m_{A'}R|\ll1\ll|\beta_1m_{A'}h|$.
The latter limit corresponds to the skin-depth in the atmosphere being much longer than $R$, while the former corresponds to the skin-depths in the ionosphere and Earth being much shorter than $h$.
In this limit, the solution becomes
\begin{align}
    a_m&=-\sqrt{\frac{4\pi}3}\frac{i\varepsilon m_{A'}A'_m}{\beta_1}(m_{A'}R)^2e^{i\beta_1m_{A'}R},\\
    b_m&=\frac{\sqrt{3\pi}i\varepsilon m_{A'}A'_m}{\beta_2^3}\lb[1+\frac{(\beta_2m_{A'}R)^2}5\lb(\frac{3+2Q[ h/R]}3+\frac{5iQ[h/R]}{3\beta_1m_{A'}h}-\frac{5i}{2\beta_1m_{A'}R}\rb)\rb],\\
    c_m&=-\frac{4\sqrt{\pi}}{15\sqrt{3}}i\varepsilon m_{A'}A'_m \lb(\beta_2m_{A'}R\rb)^2\lb(1+\frac{5i}{2\beta_1m_{A'}h}\rb)Q[h/R],\\
    d_m&=-\sqrt{\frac\pi3}\frac{i\varepsilon m_{A'}A'_m}{\beta_1}\lb[m_{A'}(R+h)\rb]^2e^{-i\beta_1m_{A'}(R+h)},
\end{align}
where
\begin{align}
    Q[x]=\frac{3(x+1)^3(x+2)}{2(x^2+3x+3)},
\end{align}
and $A'_m$ are defined as in \eqrefRange{Aplus}{Azero}.
The magnetic field in the atmosphere to leading order is then
\begin{align}
    \bm{B}(\Omega,t)=\sqrt{\frac\pi3}\varepsilon m_{A'}^2R\sum_mA'_m\bm{\Phi}_{1m}e^{-im_{A'}t};
    \label{eq:BfiniteFinal}
\end{align}
this agrees with \eqref{pre_signal}, and the rotation of the Earth can be accounted for just as in \secref{sec:earthTheory} to arrive at \eqref{signal}.
\end{widetext}

Note that \eqref{BfiniteFinal} does not depend on either of the conductivities in the problem.
In particular, note also that at no point did we actually assume $\sigma_2\ll m_{A'}$, so this condition on the atmospheric conductivity is not actually essential to our result.
Indeed, the same leading-order magnetic field is obtained even for $\sigma_2 \sim m_{A'}$, which is the physical case; see \figref{fig:conductivityProfile} and the discussion in \secref{sec:conductivityNearEarth}.

However, in order to show more fully that our solution here does match onto the solution given in the main text, consider the additional limit $\sigma_2 \ll m_{A'} \ll \sigma_1$: then $\beta_2\approx1$ and $\beta_1 \approx \sqrt{\sigma_1/m_{A'}}\,\exp[i\pi/4]$.
It follows that the forms of the solutions for $R<r<R+h$ at \eqref[s]{Efinite}~and~(\ref{eq:Bfinite}) match onto those at \eqref[s]{ETM}~and~(\ref{eq:BTM}), respectively.
Moreover, since $b_m \rightarrow \sqrt{3\pi}i\varepsilon m_{A'}A_m'$, the normalization of the magnetic field for $R<r<R+h$ agrees with that of \eqref{pre_signal}; cf.~\eqref{am}, but note that we have re-labeled the coefficient $a_m$ in the main text as $b_m$ in this Appendix.

Moreover, $a_m \propto \exp[-R\sqrt{\sigma_1 m_{A'}/2}]$, while $d_m \propto \exp[ + \sqrt{m_{A'} \sigma/2}\, (R+h) ]$. 
The electric and magnetic field solutions \eqref[s]{Efinite}~and~(\ref{eq:Bfinite}) for $0\leq r<R$ can thus be shown in the $\sigma_1 \gg m_{A'}$ limit to exhibit exponential damping suppressions moving into the inner conductor that go as $\sim \exp[ - \sqrt{\sigma_1 m_{A'}/2}\, ( R - r ) ]$; similarly, for $r>R+h$ in the same limit, they exhibit exponential damping suppressions moving into the outer conductor that go as $\sim \exp\{ - \sqrt{\sigma_1 m_{A'}/2}\, [ r  - (R+h) ] \}$.
Both of these results exhibit field-amplitude skin-depths $\delta \sim \sqrt{2/(\sigma_1 m_{A'})}$, in agreement with \appref{app:systemBehavior}.
In the limit of infinite conductivity $\sigma_1$, we thus recover our solutions in the main text exactly.

\section{Full coefficient expressions for solution \texorpdfstring{in \secref{sec:earthTheory}}{for Model 1}}
\label{app:fullCoeffModel1}

For completeness, we present the full solutions for the coefficients $a_m$ and $b_m$ which appear in the computation in \secref{sec:earthTheory}, which we only gave in the combined limits $m_{A'}R\ll1$ and $h\ll R$ in the main text.

The full solutions are
\begin{widetext}
\begin{align}
a_m &= 
-2 i \sqrt{\frac{\pi}{3}} A_m' \varepsilon m_{A'}^4  \nl
   \quad \times 
   \Big\{
        (h+R)^3 \lb[ 
                    \left(1-m_{A'}^2 R^2\right) \cos (m_{A'} R)+ (m_{A'} R) \sin(m_{A'} R) 
                \rb] 
        - [ R \leftrightarrow (h+R) ]
   \Big\} \nl 
   \quad \times 
   \Big\{ 
        \left[ 1 - m_{A'}^2 ( h^2 + hR + R^2 ) + m_{A'}^4 R^2 ( h+R )^2 \right] \sin (m_{A'} h) 
        - (m_{A'} h)\cos (m_{A'}h) \left[1 + m_{A'}^2 R (h+R)\right]  
    \Big\}^{-1};
\label{eq:amFull} \\
b_m & = -2 i \sqrt{\frac{\pi }{3}} A_m' \varepsilon m_{A'} R^{-3} \nl
    \quad \times \Big\{
   (h+R)^3 \left[ \left(1-m_{A'}^2 R^2\right) \sin (m_{A'}
   R)- (m_{A'} R) \cos (m_{A'} R)\right] - [ R \leftrightarrow (h+R) ] 
   \Big\} \nl 
   \quad \times \Big\{ \left[ 1 - m_{A'}^2 ( h^2 + hR + R^2 ) + m_{A'}^4 R^2 ( h+R )^2 \right] \sin (m_{A'} h) - (m_{A'} h)\cos (m_{A'}h)
   \left[1 + m_{A'}^2 R (h+R)\right]  \Big\}^{-1},
\label{eq:bmFull}
\end{align}
\end{widetext}
where $[ R \leftrightarrow (h+R) ]$ indicates repetition of the immediately preceding term, but with the replacements $R \rightarrow h+R$ and $h+R \rightarrow R$.

\section{Vector spherical harmonics}
\label{app:vectorSphericalHarmonics}
\allowdisplaybreaks

In this Appendix, we summarize our conventions for the VSH.
The VSH are defined in terms of the scalar spherical harmonics $Y_{\ell m}$ by the relations
\begin{align}\label{eq:VSHdef}
    \bm{Y}_{\ell m} &= Y_{\ell m}\rhat, &
    \bm{\Psi}_{\ell m} &= r\bm{\nabla} Y_{\ell m}, &
    \bm{\Phi}_{\ell m} &= \bm{r}\times\bm{\nabla} Y_{\ell m},
\end{align}
where $\rhat$ is the radial unit vector.
Thus $\bm{Y}_{\ell m}$ points radially, while $\bm{\Psi}_{\ell m}$ and $\bm{\Phi}_{\ell m}$ point tangentially to a constant-radius sphere.
Our conventions follow those of \citeR{Barrera_1985}, which differ slightly from those of \citeR{Jackson}: in particular, $\bm{\Phi}_{\ell m} = i\sqrt{\ell (\ell +1)}\bm{X}_{\ell m}$, where $\bm{X}_{\ell m}$ is the normalized VSH defined at Eq.~(9.119) in \citeR{Jackson}.

Our phase conventions, and some of the relevant VSH orthogonality and completeness properties are
\begin{align}
    \bm{Y}_{\ell,-m}&=(-1)^m\bm{Y}_{\ell m}^*,\label{eq:Yminus}\\
    \bm{\Psi}_{\ell,-m}&=(-1)^m\bm{\Psi}_{\ell m}^*,\label{eq:Psiminus}\\
    \bm{\Phi}_{\ell,-m}&=(-1)^m\bm{\Phi}_{\ell m}^*, \label{eq:Phiminus}
\end{align}
\begin{align}
    \bm{Y}_{\ell m}\cdot\bm{\Psi}_{\ell m}&=\bm{Y}_{\ell m}\cdot\bm{\Phi}_{\ell m}=\bm{\Psi}_{\ell m}\cdot\bm{\Phi}_{\ell m}=0,
\end{align}
\begin{align}
    \int d\Omega\,\bm{Y}_{\ell m}\cdot \bm{Y}_{\ell'm'}^*&=\delta_{\ell\ell'}\delta_{mm'},\\
    \int d\Omega\,\bm{\Psi}_{\ell m}\cdot\bm{\Psi}_{\ell'm'}^*&=\int d\Omega~\Phi_{\ell m}\cdot\Phi_{\ell'm'}^*\nonumber \\
	&=\ell(\ell+1)\delta_{\ell\ell'}\delta_{mm'},\\
    \int d\Omega\,\bm{Y}_{\ell m}\cdot\bm{\Psi}_{\ell'm'}^*&=\int d\Omega\,\bm{Y}_{\ell m}\cdot\bm{\Phi}_{\ell'm'}^*\nonumber \\
	&=\int d\Omega\,\bm{\Psi}_{\ell m}\cdot\bm{\Phi}_{\ell'm'}^*=0.
\end{align}

For any radially dependent function $f(r)$, the divergences and curls of the VSH are given by
\begin{align}
    \nabla \cdot \lb(f\bm{Y}_{\ell m}\rb)&=\lb( \frac{df}{dr} + \frac{2f}{r} \rb) Y_{\ell m},\label{eq:Ydiv}\\
    \nabla \cdot \lb(f\bm{\Psi}_{\ell m}\rb)&= -\ell(\ell+1) \frac{f}{r} Y_{\ell m},\label{eq:Psidiv}\\
    \nabla \cdot \lb(f\bm{\Phi}_{\ell m}\rb)&=0,\label{eq:Phildiv}\\[3ex]
    \nabla\times\lb(f\bm{Y}_{\ell m}\rb)&=-\frac fr\bm{\Phi}_{\ell m},\label{eq:Ycurl}\\
    \nabla\times\lb(f\bm{\Psi}_{\ell m}\rb)&=\lb(\frac{df}{dr}+\frac fr\rb)\bm{\Phi}_{\ell m},\label{eq:Psicurl}\\
    \nabla\times\lb(f\bm{\Phi}_{\ell m}\rb)&=-\frac{\ell(\ell+1)f}r\bm{Y}_{\ell m}-\lb(\frac{df}{dr}+\frac fr\rb)\bm{\Psi}_{\ell m},\label{eq:Phicurl}
\end{align}
with the Laplacians then being
\begin{align}
    \nabla^2\lb(f\bm{Y}_{\ell m}\rb)
        &=\lb(\frac1{r^2}\frac d{dr}\lb(r^2\frac{df}{dr}\rb)-\frac{(\ell(\ell+1)+2)f}{r^2}\rb)\bm{Y}_{\ell m}
        \nl\quad +\frac{2f}{r^2}\bm{\Psi}_{\ell m},\label{eq:Ylaplace}\\
    \nabla^2\lb(f\bm{\Psi}_{\ell m}\rb)
        &=\lb(\frac1{r^2}\frac d{dr}\lb(r^2\frac{df}{dr}\rb)-\frac{\ell(\ell+1)f}{r^2}\rb)\bm{\Psi}_{\ell m}
        \nl\quad +\frac{2\ell(\ell+1)f}{r^2}\bm{Y}_{\ell m},\label{eq:Psilaplace}\\
    \nabla^2\lb(f\bm{\Phi}_{\ell m}\rb)
        &=\lb(\frac1{r^2}\frac d{dr}\lb(r^2\frac{df}{dr}\rb)-\frac{\ell(\ell+1)f}{r^2}\rb)\bm{\Phi}_{\ell m}.\label{eq:Philaplace}
\end{align}

The explicit expressions for the VSH which are relevant to this work [see \eqref{signal}] are
\begin{align}
    \bm{Y}_{10}(\bm{r})&=\sqrt{\frac3{4\pi}}\cos\theta \rhat\label{eq:Y10},\\
    \bm{Y}_{11}(\bm{r})&=-\sqrt{\frac3{8\pi}}e^{i\phi}\sin\theta\rhat\label{eq:Y11},\\
    \bm{\Psi}_{10}(\bm{r})&=-\sqrt{\frac3{4\pi}}\sin\theta\thetahat\label{eq:Psi10},\\ 
    \bm{\Psi}_{11}(\bm{r})&=-\sqrt{\frac3{8\pi}}e^{i\phi}(\cos\theta\thetahat+i\phihat)\label{eq:Psi11},\\
    \bm{\Phi}_{10}(\bm{r})&=-\sqrt{\frac3{4\pi}}\sin\theta\phihat\label{eq:Phi10},\\
    \bm{\Phi}_{11}(\bm{r})&=\sqrt{\frac3{8\pi}}e^{i\phi}(i\thetahat-\cos\theta\phihat)\label{eq:Phi11},
\end{align}
where $\thetahat$ and $\phihat$ are unit vectors in the directions of increasing $\theta$ and $\phi$. 
The $m=-1$ harmonics can be obtained using \eqrefRange{Yminus}{Phiminus}.

Note that, as written here, if the coordinate system in question is aligned such that $+\bm{\hat{z}}$ points along the rotation axis of the Earth out of the Geographic North Pole, and the coordinate system is body fixed such that it co-rotates with the surface of the Earth, then the spherical coordinate $\phi$ coincides with the definition of longitude.
However, the spherical coordinate $\theta$ is not the latitude: $\theta$ increases from $\theta = 0$ at the Geographic North Pole (latitude $+90^\circ$), to $\theta = \pi/2$ on the Equator (latitude $0^\circ$), to $\theta = \pi$ at the Geographic South Pole (latitude $-90^\circ$).

\begin{figure*}[t]
\includegraphics[width=0.8\textwidth]{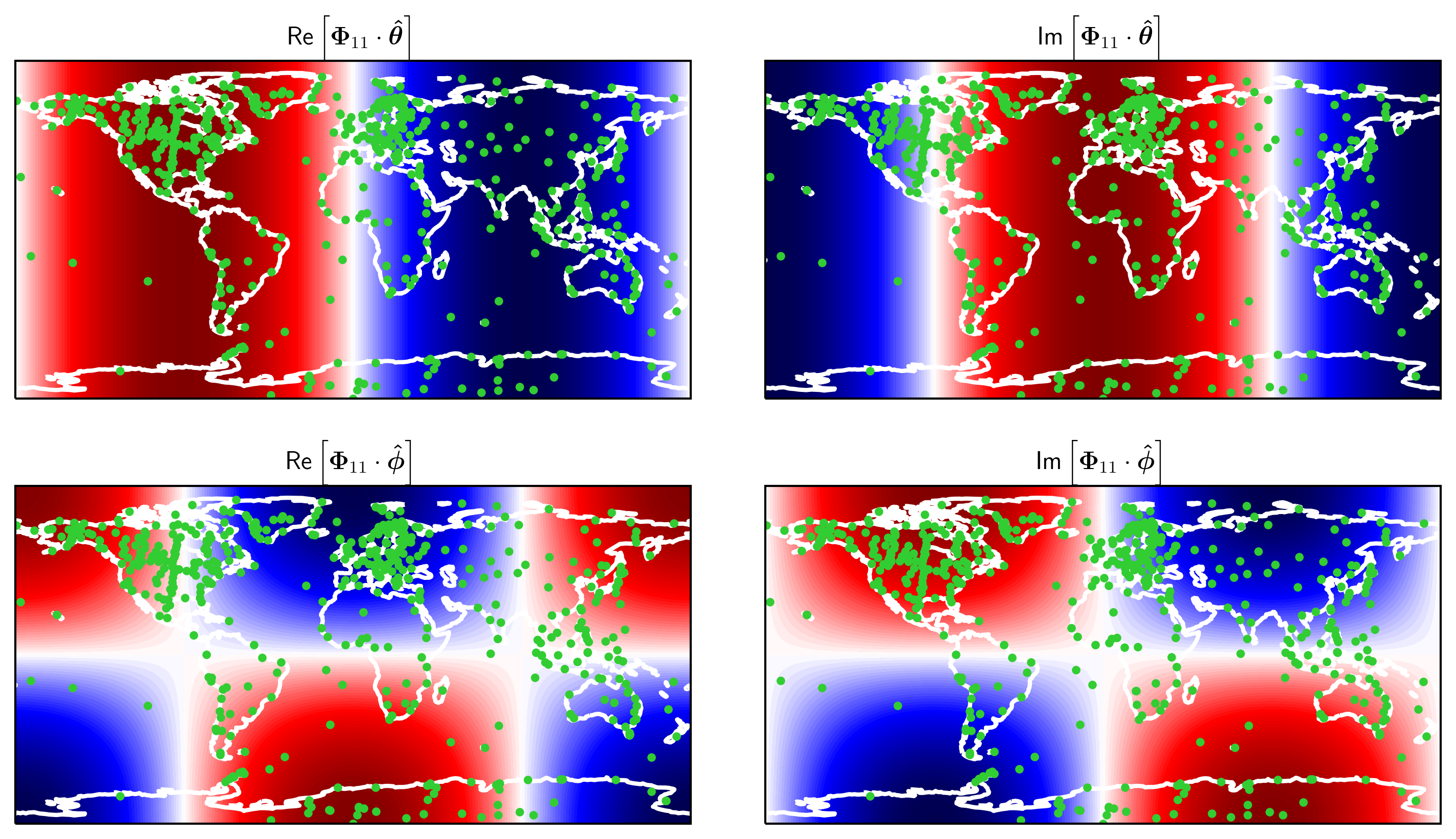}\\[1ex]
\includegraphics[width=0.4\textwidth]{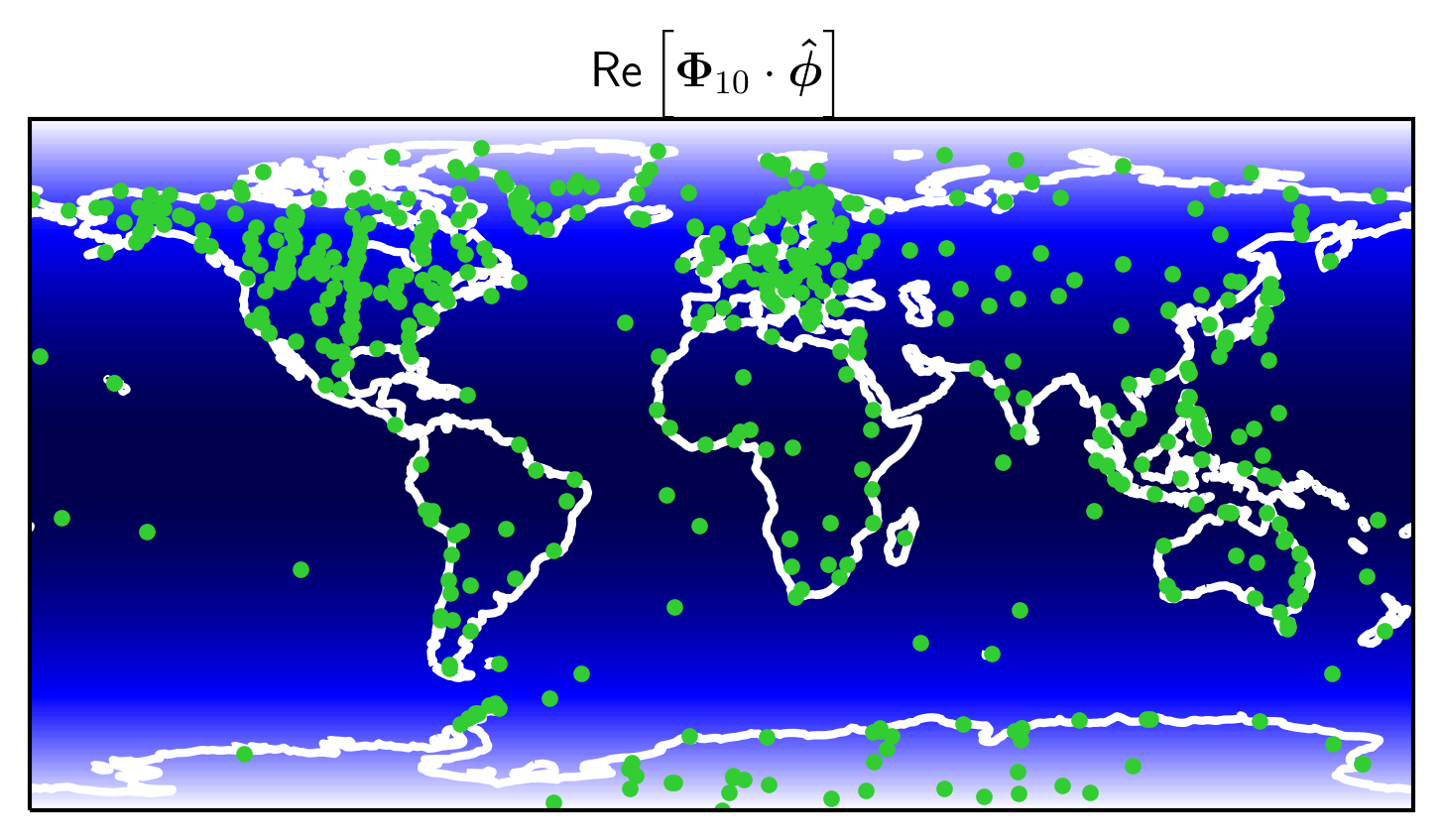}
\caption{\label{fig:VSHplots}%
    Shaded contour plots of the real and imaginary parts of all the non-zero $\bm{\hat{\theta}}$- and $\bm{\hat{\phi}}$-components of the vector spherical harmonics $\bm{\Phi}_{11}$ and $\bm{\Phi}_{10}$; the cognate plots for $\bm{\Phi}_{1,-1}$ can be read from those of $\bm{\Phi}_{11}$ using \eqref{Phiminus}.
    Red (blue) indicates positive (negative) values, with the color range for each plot independently normalized to span the range of values plotted.
    Overlaid are the outlines of the Earth's continents (white)~\cite{cartopy}. 
    The locations of the SuperMAG stations used in the analysis that is outlined in \secref{sec:experimentalSearch} (and which is the subject of \citeR{Fedderke:2021qva}) are shown as green points.
   }
\end{figure*}

\figref{fig:VSHplots} shows the real and imaginary components of the non-zero $\bm{\hat{\theta}}$- and $\bm{\hat{\phi}}$-components of $\bm{\Phi}_{11}$ and $\bm{\Phi}_{10}$, which are the relevant VSH that appear in the signal, \eqref{signal}.

Finally, we note that the Cartesian unit vectors can be written in terms of the VSH as
\begin{align}
    \xhat&=-\sqrt{\frac{2\pi}3}(\bm{Y}_{11}-\bm{Y}_{1,-1}+\bm{\Psi}_{11}-\bm{\Psi}_{1,-1}),\label{eq:xharm}\\
    \yhat&=\sqrt{\frac{2\pi}3}i(\bm{Y}_{11}+\bm{Y}_{1,-1}+\bm{\Psi}_{11}+\bm{\Psi}_{1,-1}),\label{eq:yharm}\\
    \zhat&=\sqrt{\frac{4\pi}3}(\bm{Y}_{10}+\bm{\Psi}_{10}).\label{eq:zharm}
\end{align}

\bibliographystyle{JHEP}
\bibliography{references.bib}

\end{document}